\documentclass[pdftex,twocolumn,epjc3_preprint,runningheads]{svjour3}

\usepackage[colorlinks,citecolor=blue,urlcolor=blue,linkcolor=blue,breaklinks=true]{hyperref}
\usepackage[dvipsnames]{xcolor}
\usepackage[table]{xcolor}
\definecolor{lightgray}{gray}{0.1}

\usepackage{verbatim}
\usepackage{enumitem}

\newcommand{\rownumber}{\texttt{\stepcounter{rownum}\arabic{rownum}}.}

\pdfoutput=1

\usepackage[T1]{fontenc}
\usepackage{lmodern}
\usepackage{calc}
\usepackage{graphicx}
\usepackage{booktabs}
\usepackage{textcomp}
\usepackage{xspace}
\usepackage{relsize}
\usepackage{amssymb}
\usepackage{amsmath}
\usepackage{listings}
\usepackage{microtype}
\usepackage{multirow}
\usepackage{tabularx}
\usepackage{array}
\usepackage{placeins}
\usepackage{cuted}
\usepackage{soul} 
\usepackage{fixltx2e}
\usepackage{slashed}
\usepackage{bm}
\usepackage[numbers,sort&compress]{natbib}
\usepackage[labelfont=bf,font=small]{caption}
\usepackage[skip=-2pt]{subcaption}
\usepackage[clockwise,figuresright]{rotating}
\usepackage{tikz}
\usepackage[normalem]{ulem}
\usepackage[utf8]{inputenc}
\usepackage{etoolbox}
\usepackage{siunitx}

\AfterEndEnvironment{strip}{\leavevmode}

\allowdisplaybreaks

\newcolumntype{L}{>{\raggedright\let\newline\\\arraybackslash\hspace{0pt}}X}
\newcolumntype{R}{>{\raggedleft\let\newline\\\arraybackslash\hspace{0pt}}X}
\newcolumntype{C}{>{\centering\let\newline\\\arraybackslash\hspace{0pt}}X}

\newcommand{\gambitinstitute}[1]{\expandafter\csname #1\endcsname\label{#1}}
\newcommand{\gi}[1]{\gambitinstitute{#1}\and}
\newcommand{\last}[1]{\gambitinstitute{#1}}

\makeatletter

\newcommand{\preprintnumber}[1]{\gdef\@preprintnumber{\begin{flushright}{#1}\end{flushright}}}

\g@addto@macro\bfseries{\boldmath}
\makeatother

\let\underscore\_
\renewcommand{\_}{\discretionary{\underscore}{}{\underscore}}

\makeatletter
\let\orgdescriptionlabel\descriptionlabel
\renewcommand*{\descriptionlabel}[1]{%
  \let\orglabel\label
  \let\label\@gobble
  \phantomsection
  \protected@edef\@currentlabel{#1}%
  \let\label\orglabel
  \orgdescriptionlabel{#1}%
}
\makeatother

\newcommand\postnewlinemarker{\hbox{\ensuremath{\hookrightarrow}}}
\newcommand\cpp[1]{{\lstinline!#1!}}  

\newcommand\yaml[1]{{\lstset{style=yaml}\lstinline!#1!\lstset{style=cpp}}}

\newcommand\term[1]{{\lstset{style=terminal}\lstinline!#1!\lstset{style=cpp}}}
\newcommand\termalt[1]{{\lstset{style=terminalalt}\lstinline!#1!\lstset{style=cpp}}}
\newcommand\fortran[1]{{\lstset{style=fortran}\lstinline!#1!\lstset{style=cpp}}}
\newcommand\py[1]{{\lstset{style=python}\lstinline!#1!\lstset{style=cpp}}}
\newcommand\customtilde{{\raisebox{0.2ex}{\scalebox{0.6}{\boldmath$\sim$}}}}
\newcommand\mathematica[1]{{\lstset{style=Mathematica}\lstinline!#1!\lstset{style=cpp}}}
\newcommand\guminline[1]{{{\lstset{style=gum}\lstinline!#1!}}}
\newcommand\textinline[1]{{{\lstset{style=text}\lstinline!#1!}}}

\def\be{\begin{equation}}
\def\ee{\end{equation}}
\def\ba{\begin{eqnarray}}
\def\ea{\end{eqnarray}}
\newcommand{\bea}{\begin{eqnarray}}
\newcommand{\eea}{\end{eqnarray}}

\lstnewenvironment{lstlistingyaml}{\lstset{style=yaml}}{\lstset{style=cpp}}
\lstnewenvironment{lstlistingterm}{\lstset{style=terminal}}{\lstset{style=cpp}}
\lstnewenvironment{lstlistingfortran}{\lstset{style=fortran}}{\lstset{style=cpp}}
\lstnewenvironment{lstcpp}{\lstset{style=cpp}}{\lstset{style=cpp}}
\lstnewenvironment{lstcppalt}{\lstset{style=cppalt}}{\lstset{style=cpp}}
\lstnewenvironment{lstcppnum}{\lstset{style=cppnum}}{\lstset{style=cpp}}
\lstnewenvironment{lstyaml}{\lstset{style=yaml}}{\lstset{style=cpp}}
\lstnewenvironment{lstgum}{\lstset{style=gum}}{\lstset{style=cpp}}
\lstnewenvironment{lstterm}{\lstset{style=terminal}}{\lstset{style=cpp}}
\lstnewenvironment{lsttermalt}{\lstset{style=terminalalt}}{\lstset{style=cpp}}
\lstnewenvironment{lsttext}{\lstset{style=text}}{\lstset{style=cpp}}
\lstnewenvironment{lstfortran}{\lstset{style=fortran}}{\lstset{style=cpp}}
\lstnewenvironment{lstpy}{\lstset{style=python}}{\lstset{style=cpp}}
\lstnewenvironment{lstmathematica}{\lstset{style=mathematica}}{\lstset{style=cpp}}

\newcommand{\tmpname}{}
\newcommand{\tmplistingname}{}
\makeatletter
\newif\ifATOlabelname
\lst@Key{labelname}{Listing}{\def\ATOlabelname{#1}\global\ATOlabelnametrue}
\makeatother
\lstnewenvironment{lstcpplabel}[1][]{
  \lstset{style=cpp,#1} 
  \ifATOlabelname
    \renewcommand{\tmpname}{\lstlistingname}
    \renewcommand{\tmplistingname}{\lstlistlistingname}
    \renewcommand{\lstlistingname}{\ATOlabelname}
    \renewcommand{\lstlistlistingname}{List of \lstlistingname s}
  \fi
}{
  \renewcommand{\lstlistingname}{\tmpname}
  \renewcommand{\lstlistlistingname}{\tmplistingname}
  \lstset{style=cpp}
}
\definecolor{solarized@base03}{HTML}{002B36}
\definecolor{solarized@base02}{HTML}{073642}
\definecolor{solarized@base01}{HTML}{586e75}
\definecolor{solarized@base00}{HTML}{657b83}
\definecolor{solarized@base0}{HTML}{839496}
\definecolor{solarized@base1}{HTML}{93a1a1}
\definecolor{solarized@base2}{HTML}{EEE8D5}
\definecolor{solarized@base3}{HTML}{FDF6E3}
\definecolor{solarized@yellow}{HTML}{B58900}
\definecolor{solarized@orange}{HTML}{CB4B16}
\definecolor{solarized@red}{HTML}{DC322F}
\definecolor{solarized@magenta}{HTML}{D33682}
\definecolor{solarized@violet}{HTML}{6C71C4}
\definecolor{solarized@blue}{HTML}{268BD2}
\definecolor{solarized@cyan}{HTML}{2AA198}
\definecolor{solarized@green}{HTML}{859900}
\definecolor{darkred}{HTML}{550003}
\definecolor{darkgreen}{HTML}{00AA00}
\definecolor{orchid}{HTML}{AF06F5}
\definecolor{lightgray}{gray}{0.8}

\newcommand\YAMLstringstyle{\footnotesize\color{solarized@green}\mdseries}
\newcommand\YAMLkeystyle{\footnotesize\color{solarized@blue}\ttfamily}
\newcommand\YAMLvaluestyle{\footnotesize\color{blue}\mdseries}
\newcommand\ProcessThreeDashes{\llap{\color{cyan}\mdseries-{-}-}}

\newcommand\CPPcommentstyle{\color{solarized@violet}\footnotesize\ttfamily}
\newcommand\CPPdirectivestyle{\color{solarized@magenta}\footnotesize\ttfamily}
\newcommand\termplainstyle{\footnotesize\ttfamily}

\newcommand\YAMLcommentstyle{\color{solarized@orange}\ttfamily}

\newcommand\processLongMacroDelimiter
{%
\CPPdirectivestyle%
\#define%
}

\lstdefinestyle{cpp}
{
  language=C++,
  basicstyle=\footnotesize\ttfamily,
  basewidth={0.53em,0.44em}, 
  numbers=none,
  tabsize=2,
  breaklines=true,
  escapeinside={@}{@},
  showstringspaces=false,
  numberstyle=\tiny\color{solarized@base01},
  keywordstyle=\color{solarized@orange},
  stringstyle=\color{solarized@red}\ttfamily,
  identifierstyle=\color{solarized@blue},
  commentstyle=\CPPcommentstyle,
  directivestyle=\CPPdirectivestyle,
  emphstyle=\color{solarized@green},
  frame=single,
  rulecolor=\color{solarized@base2},
  rulesepcolor=\color{solarized@base2},
  literate={~} {\customtilde}1,
  moredelim=*[directive]\ \ \#,
  moredelim=*[directive]\ \ \ \ \#
}

\lstdefinestyle{cppalt}
{
  language=C++,
  basicstyle=\footnotesize\ttfamily,
  basewidth={0.53em,0.44em}, 
  numbers=none,
  tabsize=2,
  breaklines=true,
  escapeinside={*@}{@*},
  showstringspaces=false,
  numberstyle=\tiny\color{solarized@base01},
  keywordstyle=\color{solarized@orange},
  stringstyle=\color{solarized@red}\ttfamily,
  identifierstyle=\color{solarized@blue},
  commentstyle=\CPPcommentstyle,
  directivestyle=\CPPdirectivestyle,
  emphstyle=\color{solarized@green},
  frame=single,
  rulecolor=\color{solarized@base2},
  rulesepcolor=\color{solarized@base2},
  literate={~}{\customtilde}1,
  moredelim=**[is][\processLongMacroDelimiter]{BeginLongMacro}{EndLongMacro} 
}

\lstdefinestyle{cppnum}
{
  language=C++,
  basicstyle=\footnotesize\ttfamily,
  basewidth={0.53em,0.44em}, 
  numbers=none,
  tabsize=2,
  breaklines=true,
  escapeinside={@}{@},
  numberstyle=\tiny\color{solarized@base01},
  showstringspaces=false,
  keywordstyle=\color{solarized@orange},
  stringstyle=\color{solarized@red}\ttfamily,
  identifierstyle=\color{solarized@blue},
  commentstyle=\CPPcommentstyle,
  directivestyle=\CPPdirectivestyle,
  emphstyle=\color{solarized@green},
  frame=single,
  rulecolor=\color{solarized@base2},
  rulesepcolor=\color{solarized@base2},
  literate={~} {\customtilde}1,
  moredelim=*[directive]\ \ \#,
  moredelim=*[directive]\ \ \ \ \#
}

\lstdefinestyle{python}
{
  language=Python,
  basicstyle=\footnotesize\ttfamily,
  basewidth={0.53em,0.44em},
  numbers=none,
  tabsize=2,
  breaklines=true,
  escapeinside={@}{@},
  showstringspaces=false,
  numberstyle=\tiny\color{solarized@base01},
  keywordstyle=\color{blue},
  stringstyle=\color{orange}\ttfamily,
  identifierstyle=\color{darkred},
  commentstyle=\color{purple},
  emphstyle=\color{green},
  frame=single,
  rulecolor=\color{solarized@base2},
  rulesepcolor=\color{solarized@base2},
  literate = {~}{\customtilde}1
             {\ as\ }{{\color{blue}\ as\ \color{black}}}3
             {.set}{{\color{black}.}{\color{darkred}set}}4
}

\lstdefinestyle{fortran}
{
  language=Fortran,
  basicstyle=\footnotesize\ttfamily,
  basewidth={0.53em,0.44em},
  numbers=none,
  tabsize=2,
  breaklines=true,
  escapeinside={@}{@},
  showstringspaces=false,
  numberstyle=\tiny\color{solarized@base01},
  keywordstyle=\color{blue},
  stringstyle=\color{orange}\ttfamily,
  identifierstyle=\color{Periwinkle},
  commentstyle=\color{purple},
  emphstyle=\color{green},
  morekeywords={and, or, true, false},
  frame=single,
  rulecolor=\color{solarized@base2},
  rulesepcolor=\color{solarized@base2},
  literate={~}{\customtilde}1
}

\lstdefinestyle{terminal}
{
  language=bash,
  basicstyle=\termplainstyle,
  numbers=none,
  tabsize=2,
  breaklines=true,
  escapeinside={@}{@},
  frame=single,
  showstringspaces=false,
  numberstyle=\tiny\color{solarized@base01},
  keywordstyle=\color{solarized@orange},
  stringstyle=\color{solarized@red}\ttfamily,
  identifierstyle=\color{black},
  commentstyle=\color{solarized@violet},
  emphstyle=\color{solarized@green},
  frame=single,
  rulecolor=\color{solarized@base2},
  rulesepcolor=\color{solarized@base2},
  morekeywords={gambit, cmake, make, mkdir, gum, python, wget, tar, cp, pippi, mpirun},
  deletekeywords={test},
  literate = {/gambit}{{/}{\color{black}}gambit}6
             {gambit/}{{\color{black}}gambit{/}}6
             {gum/}{{\color{black}}gum{/}}4
             {/include}{{/}{\color{black}}include}8
             {cmake/}{{\color{black}}cmake/}6
             {.cmake}{{.}{\color{black}}cmake}6
             {.gum}{{.}{\color{black}}gum}6
             {.tar}{{.}{\color{black}}tar}4
             {source/}{{\color{black}}source{/}}7
             { type}{{\color{black}}{}type}5
             {~}{\customtilde}1
             {math}{{\color{solarized@orange}}math}4
}

\lstdefinestyle{terminalalt}
{
  language=bash,
  basicstyle=\footnotesize\ttfamily,
  numbers=none,
  tabsize=2,
  breaklines=true,
  escapeinside={*@}{@*},
  frame=single,
  showstringspaces=false,
  numberstyle=\tiny\color{solarized@base01},
  keywordstyle=\color{solarized@orange},
  stringstyle=\color{solarized@red}\ttfamily,
  identifierstyle=\color{black},
  commentstyle=\color{solarized@violet},
  emphstyle=\color{solarized@green},
  frame=single,
  rulecolor=\color{solarized@base2},
  rulesepcolor=\color{solarized@base2},
  morekeywords={gambit, cmake, make, mkdir},
  deletekeywords={test},
  literate = {\ gambit}{{\ }{\color{black}}gambit}7
             {/gambit}{{/}{\color{black}}gambit}6
             {gambit/}{{\color{black}}gambit{/}}6
             {/include}{{/}{\color{black}}include}8
             {cmake/}{{\color{black}}cmake/}6
             {.cmake}{{.}{\color{black}}cmake}6
             {~}{\customtilde}1
}

\lstdefinestyle{text}
{
  language={},
  basicstyle=\footnotesize\ttfamily,
  identifierstyle=\color{black},
  numbers=none,
  tabsize=2,
  breaklines=true,
  escapeinside={*@}{@*},
  showstringspaces=false,
  frame=single,
  rulecolor=\color{solarized@base2},
  rulesepcolor=\color{solarized@base2},
  literate={~}{\customtilde}1
}

\lstdefinestyle{yaml}
{
  language=bash,
  escapeinside={@}{@},
  keywords={true,false,null},
  otherkeywords={},
  keywordstyle=\color{solarized@base0}\bfseries,
  basicstyle=\footnotesize\color{black}\ttfamily,
  identifierstyle=\YAMLkeystyle,
  sensitive=false,
  commentstyle=\YAMLcommentstyle,
  morecomment=[l]{\#},
  morecomment=[s]{/*}{*/},
  stringstyle=\YAMLstringstyle\ttfamily,
  moredelim=**[s][\YAMLkeystyle]{,}{:},   
  moredelim=**[l][\YAMLvaluestyle]{:},    
  morestring=[b]',
  morestring=[b]",
  literate =    {---}{{\ProcessThreeDashes}}3
                {>}{{\textcolor{solarized@red}\textgreater}}1
                {gtr}{\textgreater}1
                {grt}{\textgreater}1
                {|}{{\textcolor{solarized@red}\textbar}}1
                {\ -\ }{{\mdseries\color{black}\ -\ \negmedspace}}3
                {\}}{{{\color{black} \}}}}1
                {\{}{{{\color{black} \{}}}1
                {[}{{{\color{black} [}}}1
                {]}{{{\color{black} ]}}}1
                {~}{\customtilde}1,
  breakindent=0pt,
  breakatwhitespace,
  columns=fullflexible
}

\lstdefinestyle{gum}
{
  language=bash,
  escapeinside={@}{@},
  keywords={true,false,null,all},
  otherkeywords={},
  keywordstyle=\color{solarized@base02}\bfseries,
  basicstyle=\footnotesize\color{black}\ttfamily,
  identifierstyle=\color{solarized@magenta},
  sensitive=false,
  commentstyle=\color{solarized@cyan}\ttfamily,
  morecomment=[l]{\#},
  morecomment=[s]{/*}{*/},
  stringstyle=\footnotesize\color{solarized@base01}\mdseries\ttfamily,
  moredelim=**[l][\footnotesize\color{solarized@base02}\mdseries]{:},    
  morestring=[b]',
  morestring=[b]",
  literate =    {---}{{\ProcessThreeDashes}}3
                {grt}{{\textcolor{solarized@magenta}\textgreater}}1
                {gtr}{{\textcolor{solarized@base02}\textgreater}}1
                {/>}{{\textcolor{solarized@magenta}\textgreater}}1
                {/<}{{\textcolor{solarized@magenta}\textless}}1
                {lss}{{\textcolor{solarized@base02}\textless}}1
                {pls}{{\textcolor{solarized@magenta}+}}1
                {mns}{{\textcolor{solarized@magenta}-}}1
                {|}{{\textcolor{solarized@base02}\textbar}}1
                {\ -\ }{{\mdseries\color{solarized@base02}\ -\ \negmedspace}}3
                {\}}{{{\color{solarized@base02} \}}}}1
                {\{}{{{\color{solarized@base02} \{}}}1
                {[}{{{\color{solarized@base02} [}}}1
                {]}{{{\color{solarized@base02} ]}}}1
                {~}{\customtilde}1,
  breakindent=0pt,
  breakatwhitespace,
  columns=fullflexible
}

\lstdefinestyle{mathematica}
{
  language={Mathematica},
  basicstyle=\footnotesize\ttfamily,
  basewidth={0.53em,0.44em},
  numbers=none,
  tabsize=2,
  breaklines=true,
  postbreak=,
  escapeinside={@}{@},
  numberstyle=\tiny\color{black},
  showstringspaces=false,
  numberstyle=\tiny\color{solarized@base01},
  keywordstyle=\color{solarized@orange},
  stringstyle=\color{solarized@red}\ttfamily,
  identifierstyle=\color{solarized@orange}\ttfamily,
  commentstyle=\color{solarized@gray}\ttfamily,
  directivestyle=\color{solarized@orange}\ttfamily,
  emphstyle=\color{solarized@green},
  frame=single,
  rulecolor=\color{solarized@base2},
  rulesepcolor=\color{solarized@base2},
  literate={~} {\customtilde}1,
  moredelim=*[directive]\ \ \#,
  moredelim=*[directive]\ \ \ \ \#,
  mathescape=false
}

\newcommand{\doublecross}[2]{\hyperref[#2]{\textbf{#1}}}
\newcommand{\doublecrosssf}[2]{\hyperref[#2]{\textbf{\textsf{#1}}}}

\newcommand{\startglossary}{\section{Glossary}\label{glossary}Here we explain some terms that have specific technical definitions in \GB.\begin{description}}
\newcommand{\finishglossary}{\end{description}}

\newcommand{\bcode}{\begin{lstlisting}}
\newcommand{\ecode}{\end{lstlisting}}

\newcommand{\MSBar}{\overline{MS}}

\newcommand{\gambit}{\textsf{GAMBIT}\xspace}
\newcommand{\gambitversion}{2.7}
\newcommand{\gambitVer}{\gambit \textsf{\gambitversion}\xspace}

\newcommand{\colliderbit}{\textsf{ColliderBit}\xspace}

\newcommand{\specbit}{\textsf{SpecBit}\xspace}
\newcommand{\decaybit}{\textsf{DecayBit}\xspace}

\newcommand{\scannerbit}{\textsf{ScannerBit}\xspace}

\newcommand{\GB}{\gambit}

\newcommand{\buckfast}{\textsf{BuckFast}\xspace}

\newcommand{\pythia}{\textsf{Pythia}\xspace}
\newcommand{\pythiaeight}{\textsf{Pythia\,8}\xspace}

\newcommand{\heputils}{\textsf{HEPUtils}\xspace}

\newcommand{\rivet}{\textsf{Rivet}\xspace}

\newcommand{\contur}{\textsf{Contur}\xspace}

\newcommand\flexiblesusy{\FlexibleSUSY}
\newcommand\FlexibleSUSY{\textsf{FlexibleSUSY}\xspace}

\newcommand\SOFTSUSY{\textsf{SOFTSUSY}\xspace}

\newcommand\HDECAY{\textsf{HDECAY}\xspace}

\newcommand\SDECAY{\textsf{SDECAY}\xspace}
\newcommand\SUSYHIT{\textsf{SUSY-HIT}\xspace}

\newcommand\diver{\textsf{Diver}\xspace}

\newcommand{\sarah}{\textsf{SARAH}\xspace}

\newcommand\beq{\begin{equation}}
\newcommand\eeq{\end{equation}}

\newcommand{\subparagraph}{} 
\journalname{Eur.\ Phys.\ J.\ C}
\smartqed

\makeatletter
\patchcmd{\ttlh@hang}{\parindent\z@}{\parindent\z@\leavevmode}{}{}
\patchcmd{\ttlh@hang}{\noindent}{}{}{}
\makeatother

\usepackage{siunitx}
\usepackage{booktabs}

\usepackage[title]{appendix}

\usepackage{amsmath} 
\usepackage[displaymath, mathlines,switch]{lineno} 

\newcommand{\EWMSSM}{EWMSSM\xspace}
\newcommand{\GEWMSSM}{$\tilde{G}$-EWMSSM\xspace}

\renewcommand{\MSBar}{\ensuremath{\overline{\text{MS}}}}

\newcommand{\neu}[1]{\ensuremath{\tilde{\chi}_{#1}^0}}
\newcommand{\mneu}[1]{\ensuremath{m_{\neu{#1}}}}

\newcommand{\cha}[2][\pm]{\ensuremath{\tilde{\chi}_{#2}^{#1}}}
\newcommand{\mcha}[2][\pm]{\ensuremath{m_{\tilde{\chi}_{#2}^{#1}}}}

\definecolor{myviolet}{HTML}{7E57C2}

\newcommand{\gravitino}{\ensuremath{\tilde G}}
\newcommand{\mg}{\ensuremath{m_{3/2}}}

\begin{document}

\preprintnumber{gambit-physics-2026, OPENMAPP-26-04, ADP-26-13/T1310, MCNET-26-23}

\title{What has the LHC told us about the electroweakino sector of the Minimal Supersymmetric Standard Model?}

\author{The GAMBIT Collaboration: 
Peter Athron\thanksref{nnu} \and
Csaba Bal\'azs\thanksref{monash} \and
Andy Buckley\thanksref{glasgow} \and
Jon Butterworth\thanksref{ucl} \and
Christopher Chang\thanksref{oslo,a} \and
Andrew Fowlie\thanksref{xjtlu} \and
Tom\'as E.\ Gonzalo\thanksref{kitTTP,eumetsat} \and
Vinay Hegde\thanksref{baylor} \and
Ida-Marie Fauske Johansson\thanksref{oslo} \and
Adil Jueid\thanksref{kias} \and
Tore Klungland\thanksref{oslo} \and
Anders Kvellestad\thanksref{oslo,b} \and
Farvah Mahmoudi\thanksref{lyon, cernth, iuf} \and
Gregory D.\ Martinez\thanksref{ucla} \and
Holly Pacey\thanksref{oxford} \and
Tomasz Procter\thanksref{krakow} \and
Are Raklev\thanksref{oslo} \and
Roberto Ruiz de Austri\thanksref{ific} \and
Pat Scott\thanksref{gridsight} \and
Martin White\thanksref{adelaide} \and
Yang Zhang\thanksref{htu} \and
Pengxuan Zhu\thanksref{adelaide} 
}

\institute{
    \gi{nnu}
    \gi{monash}
    \gi{glasgow}
    \gi{ucl}
    \gi{oslo}
    \gi{xjtlu}
    \gi{kitTTP}
    \gi{eumetsat}
    \gi{baylor}
    \gi{kias}
    \gi{lyon}
    \gi{cernth}
    \gi{iuf}
    \gi{ucla}
    \gi{oxford}
    \gi{krakow}
    \gi{ific}
    \gi{gridsight}
    \gi{adelaide}
    \last{htu}
}

\thankstext{a}{c.j.chang@fys.uio.no}
\thankstext{b}{anders.kvellestad@fys.uio.no}

\date{Received: date / Accepted: date}

\maketitle




\begin{abstract}
We perform global fits of the electroweak sector of the Minimal Supersymmetric Standard Model (MSSM) using a comprehensive set of LEP searches, 34 Run 2 LHC searches, and 63 Run 2 LHC measurements. Scanning the bino, wino and Higgsino mass parameters, and the ratio of the Higgs vacuum expectation values, we find that for a light, bino \neu{1}, the mass of the next-to-lightest neutralino must be $\mneu{2} \gtrsim 760$\,GeV. While MSSM electroweakinos can explain individual excesses observed by ATLAS and CMS in searches targeting compressed spectra, we find no scenarios that fit these excesses simultaneously.
When we add a light gravitino, neutralinos are further excluded up to about 1\,TeV, though this depends on their composition; Higgsino-dominated $\neu{1}$ requires only $\mneu{1} \approx \mneu{2} \gtrsim 650$\,GeV. Lastly, the newer LHC searches and measurements exclude a low-mass region that was preferred in a previous study. This is the most complete summary of collider constraints on the electroweakino sector of the MSSM performed to date. 
\end{abstract}

\tableofcontents

\section{Introduction}
\label{sec:introduction}
At the dawn of the Large Hadron Collider (LHC) era in particle physics, there was great excitement regarding the prospects of discovering new physics. The instability of the weak scale with respect to quantum corrections from higher scales, the Weakly Interacting Massive Particle (WIMP) miracle explaining the measured relic abundance of dark matter, embedding the Standard Model (SM) gauge groups in grand unified theories, and some explanations of the matter anti-matter asymmetry and neutrino masses all motivate new physics at the weak or TeV scale.

Supersymmetric theories have the capacity to explain all of these problems of the SM that motivate new physics at the TeV scale; see, for example, the reviews in Refs.~\cite{Nilles:1983ge,Haber:1984rc,Martin:1997ns,Chung:2003fi,Feng:2013pwa}. Furthermore, the Minimal Supersymmetric Standard Model (MSSM) has a very rich phenomenology with many different signatures and is therefore often used in phenomenological investigations as a testing ground for the impact of experiments on new physics.  
However, early results from Run 1 already placed significant constraints on the masses of strongly interacting sparticles, which typically have the largest production cross-sections.  Subsequent runs pushed these constraints further, e.g.\ reaching $\sim1.8$\,TeV \cite{ATLAS:2018nud,CMS:2019zmd} for first and second generation squarks, 2.4\,TeV \cite{ATLAS:2022ckd, ATLAS:2022ihe, CMS:2021beq, CMS:2019ybf} for gluinos, and limits can be as high as 5\,TeV for some leptoquarks~\cite{CMS:2025iix}.   

Weakly interacting new physics is typically harder to constrain at the LHC, due to lower production cross-sections. Specific scenarios do nonetheless have significant limits, e.g.\ with mass limits on charginos reaching as high as 1.1\,TeV~\cite{ATLAS:2018ojr}. 
However, these limits tend to be highly model dependent and can be significantly weakened when moving away from the simplified model interpretations usually used to present search results in the experimental papers.

This issue was brought into sharp focus by the global fit of the electroweakino sector of the MSSM presented in Ref.~\cite{GAMBIT:2018gjo}. The electroweakino sector consists of four neutral fermions (neutralinos) and two charged fermions (charginos), and can be fully specified with just four parameters. In Ref.~\cite{GAMBIT:2018gjo} it was demonstrated that when robust reinterpretations of the available LHC searches are carried out and then combined in a statistically rigorous manner, the LHC searches available at that time implied no general constraint in the mass plane of the lightest chargino and lightest neutralino, viz.\ the ($m_{\cha{1}}$,\mneu{1}) mass plane. While there were strong constraints on particular slices of the parameter space, allowing all four parameters of the MSSM electroweakino sector to vary, there was always at least one choice of parameters orthogonal to the mass plane that enabled an escape from LHC searches.

Interest in electroweak new physics has also been driven by anomalies that have appeared in LHC searches. These are deviations from the SM predictions that are typically above $2 \sigma$, but below the large $5 \sigma$ threshold for discovery used in particle physics. In particular, there are a number of such anomalies in searches for electroweakinos \cite{ATLAS:2019lng,ATLAS:2021moa,CMS:2021edw,CMS:2024gyw,ATLAS:2025evx,ATLAS:2024tqe,CMS:2025myt}, which have received some attention in the literature. Due to the very large number of LHC searches, it is statistically inevitable that such anomalies will appear, and that many, if not all, of these will disappear over time as more data is collected. This has already been the case with earlier anomalies in this sector. Nonetheless, such anomalies could also be the first hints of new physics, and finding consistent explanations for the anomalies remains a valuable and useful exercise~\cite{Bagnaschi:2025ksk,Ellwanger:2024vvs,Agin:2024yfs,Martin:2024pxx,Agin:2025vgn,Hammad:2025wst,Araz:2025bww,Constantin:2025bqp}. In this paper, we will remain resolutely agnostic about the status of  anomalies, which are in any case included in our global fit along with null results, but we will comment on them in the presentation of our results.

Properly determining the true constraints on weakly interacting new physics in realistic models, and assessing the viability of various anomalies that have been identified, requires a rigorous statistical combination of the many different relevant results. As shown in Ref.\ \cite{GAMBIT:2023yih}, which studies the electroweakino sector of the MSSM extended by a light gravitino, seeing the complete picture can require combining not only the dedicated searches designed for the particular model, but also other more tangential LHC searches for new physics, in addition to assessing the impact of SM measurements on the parameter space~\cite{Butterworth:2016sqg,Buckley:2021neu}. 

Since the results of Refs.~\cite{GAMBIT:2018gjo} and \cite{GAMBIT:2023yih} were published, Run~2 of the LHC has provided a wealth of new information, and the LHC experiments have published many further searches that may close observed holes in exclusion limits. In this paper, we will therefore seek to reassess the status of LHC constraints on models with electroweakinos after the completion of the LHC Run 2 within the framework of the \gambit~\cite{gambit} global fitting tool. We will carry out our work in the context of the electroweakino sector of the MSSM, and the light gravitino extension of this model, updating and extending results in Refs.\ \cite{GAMBIT:2018gjo} and \cite{GAMBIT:2023yih}. Although Run 3 of the LHC has recently finished, the Run 2 results still constitute the public state of the art in terms of available data, making our study the most complete summary of constraints on the electroweakino sector of the MSSM performed to date.

To do this, we will combine the results from a total of 34 LHC direct searches and 63 separate SM measurements. The study will focus on collider constraints and remain
agnostic about the dark matter properties of the models investigated. We will begin by reviewing the two physics models in Sec.~\ref{sec:model}, before discussing the collider likelihoods used in Sec.~\ref{sec:collider likelihoods}. The details of our parameter scan are found in Sec.~\ref{sec:global_fit_setup}, and finally we present our results in Sec.~\ref{sec:results}. In the appendices, we provide more details on the LHC searches used and their implementation and validation in the \colliderbit~\cite{ColliderBit} module of \gambit, as well as the new functionality added to \colliderbit.

\section{Models}
\label{sec:model}

In this study, we consider two supersymmetric models involving the electroweakino sector while decoupling all the other states. In the first model, denoted by EWMSSM, we consider the electroweakino sector only, while in the second model ($\tilde{G}$-EWMSSM) we add an extra light gravitino. In this section, we briefly describe the particle content and associated parameters in these models.

\subsection{EWMSSM}
\label{sec:model_EWMSSM}
We first consider the electroweakino sector of the MSSM.  This sector contains SU(3) singlet fermions that are superpartners of the Higgs bosons (Higgsinos) and SU(2)$_W$ and U(1)$_Y$ gauge bosons (the winos and bino).  The Higgsinos are SU(2) doublets, $\tilde{H_u} = (\tilde{H}_u^+, \tilde{H}_u^0)^T$ with hypercharge $Y = 1/2$ and  $\tilde{H_d} = (\tilde{H}_d^0,  \tilde{H}_d^-)^T$ with $Y = -1/2$.  The bino ($\tilde{B}$) is an SU(2) singlet with $Y=0$, and the winos are an SU(2) triplet $( \tilde{W}^+, \tilde{W}^0,\tilde{W}^-)^T$. 

The neutralinos are formed from the mixing of the neutral states, while the charginos are formed from the mixing of the charged ones. Defining the neutral and charged states as $(\psi^0)^T = (\tilde{B}, \tilde{W}^0, \tilde{H}_d^0, \tilde{H}_u^0)$, $(\psi^\pm)^T = (\tilde{W}^+, \tilde{H}_u^+, \tilde{W}^-, \tilde{H}_d^-),$
the Lagrangian of the electroweakino sector can be written as 
\begin{equation}
{\cal L} = - \frac{1}{2} (\psi^0)^T M_N \psi^0 - \frac{1}{2} (\psi^\pm)^T M_C \psi^\pm + {\rm c.c.},
\end{equation}
with $M_N$ and $M_C$ being the mass matrices for the neutralino and chargino sectors, respectively. These are given at tree level by:
\begin{eqnarray}
    M_N = \left(\begin{array}{cccc}
       M_1  & 0   & -\frac{1}{2}c_\beta g_1\upsilon & \frac{1}{2}s_\beta g_1\upsilon   \\
       0    & M_2 & \frac{1}{2}c_\beta g_2 \upsilon  & -\frac{1}{2}s_\beta g_2\upsilon   \\ 
       -\frac{1}{2}c_\beta g_1\upsilon & \frac{1}{2}s_\beta g_1\upsilon & 0 & -\mu \\
       \frac{1}{2}c_\beta g_2 \upsilon  & -\frac{1}{2}s_\beta g_2\upsilon & -\mu & 0 \\
    \end{array} \right),
    \label{eq:MN}
\end{eqnarray}
\begin{eqnarray}
    M_C = \left(\begin{array}{cc}
      0   &  X^T \\
      X   &  0   \\
    \end{array}\right),~{\rm with}~ X = \left(\begin{array}{cc}
      M_2 & s_\beta\frac{g_2\upsilon}{\sqrt{2}}  \\ 
      c_\beta\frac{g_2\upsilon}{\sqrt{2}}  & \mu 
    \end{array}\right),
    \label{eq:MC}
\end{eqnarray}
where $g_1$ and $g_2$ are the coupling constants of the $U(1)_Y$ and $SU(2)_L$ gauge groups, respectively, $c_\beta \equiv \cos\beta$, $s_\beta \equiv \sin\beta$, and $\upsilon$ is the electroweak vacuum expectation value. Here $\tan\beta =\upsilon_2/\upsilon_1$ is a free parameter. 

The diagonalization of the neutralino mass matrix can be achieved via a unitary transformation
\begin{eqnarray}
    N^* M_N N^{-1} = {\rm diag}(\mneu{1}, \mneu{2}, \mneu{3}, \mneu{4}),
\end{eqnarray}
leading to real positive masses, where each state can be decomposed as
\begin{eqnarray}
    \neu{i} = N_{i1} \tilde{B} + N_{i2} \tilde{W}^0 + N_{i3} \tilde{H}_d^0 + N_{i4} \tilde{H}_u^0.
\end{eqnarray}
With only real entries in $M_N$ it is common to substitute the unitary diagonalization for orthogonal, leading to real, but not necessarily positive, mass eigenvalues. In this case, a sign on the mass must be kept track of, as it constitutes a rotation of the corresponding mass eigenstate field and has consequences for the neutralino interactions.

On the other hand, the chargino mass matrix can be diagonalized through a bi-unitary transformation, i.e.\
\begin{eqnarray}
    U^* X V^{-1} = {\rm diag}(m_{\cha{1}}, m_{\cha{2}}).
\end{eqnarray}
In the EWMSSM the lightest chargino is assumed to always be heavier than the lightest neutralino.

The electroweakinos interact with the SM through the usual gauge interactions for fermions according to their hypercharge and SU(2) representation, and also through the gaugino--Higgs--Higgsino gauge interactions.  They also have interactions that involve the sfermions, but here, since we will assume the sfermions, as well as the non SM-like Higgs states, are all much heavier and decoupled, they play no role in the phenomenology.

In summary, the phenomenology of the EWMSSM can be entirely described by four parameters
\begin{eqnarray}
    \{ M_1, \quad M_2, \quad \mu, \quad \tan{\beta} \}.
\end{eqnarray}
In this study, we restrict the model we consider to parameter regions where the lightest neutralino is heavier than 62.5\,GeV. This is to avoid the additional cost of having to simulate production of SM bosons which subsequently produce light electroweakinos through their decays. We therefore remain agnostic about the status of such low-mass scenarios, but note that they in general will be constrained by the invisible width of the SM-like Higgs boson and by direct searches at LEP.
As part of our model restriction we also require the lifetime of all the produced sparticles---except the lightest supersymmetric particle (LSP)---to be $\tau_0c<1$\,cm, as we do not simulate LHC searches for long-lived particles.

\subsection{\texorpdfstring{$\tilde{G}$-EWMSSM}{G-EWMSSM}}
\label{sec:gravitino_model}
The $\tilde{G}$-EWMSSM model has neutralinos and charginos formed from  Higgsinos, winos, and binos exactly as described for the EWMSSM model above. In addition, it also has a light gravitino $\tilde{G}$ as the lightest supersymmetric particle, motivated by gauge mediated supersymmetry breaking~\cite{Dine:1981gu,Dine:1981za,Dimopoulos:1981au,Alvarez-Gaume:1981abe,Nappi:1982hm,Dine:1993yw,Dine:1994vc,Dine:1995ag,Kolda:1997wt}.

The precise phenomenology of the $\tilde{G}$-EWMSSM model is controlled by the hierarchy of $M_1$, $M_2$ and $\mu$, with some additional effects coming from the value of $\tan\beta$. A chargino (\cha{1}) next-to-lightest supersymmetric particle (NLSP) will decay promptly to the gravitino and a (possibly off-shell) $W$ boson in the narrow regions of parameter space in which the chargino is the NLSP. The more generic possibility is a lightest neutralino NLSP (\neu{1}) decaying through the modes: $\neu{1}\rightarrow\{\gamma,Z,h\}\,\tilde{G}$. In the relevant limit, $\mg\ll m_{\{\tilde{\chi},Z,h\}}$, the decay widths take the form~\cite{Feng:2004mt,Covi:2009bk}:
\begin{linenomath*}
\begin{align}
    \Gamma(\neu{1}\rightarrow \gamma\tilde{G})&=|N_{11}c_W+N_{12}s_W|^{2}\,\mathcal{R}(\mneu{1})\,,
    \label{chitogammaG}
    \\
    \begin{split}
    \Gamma(\neu{1}\rightarrow Z\tilde{G})&=\left(|-N_{11}s_W+N_{12}c_W|^{2} \right.\\
    &\hspace{0.25cm}\left.+|-N_{13}c_\beta+N_{14}s_\beta|^{2}/2\right)\\
    &\hspace{1cm}\times
    \mathcal{C}(m_{Z},\mneu{1})\,\mathcal{R}(\mneu{1}),
    \end{split}
    \label{chitoZG}
    \\
    \begin{split}
    \Gamma(\neu{1}\rightarrow h\tilde{G})&=\frac{1}{2}|-N_{13}s_\alpha+N_{14}c_\alpha|^{2}\\
    &\hspace{1cm}\times\mathcal{C}(m_{h},\mneu{1})\,\mathcal{R}(\mneu{1})\,.
    \end{split}
    \label{chitohG}
\end{align}
\end{linenomath*}
Here, $\mg$ is the gravitino mass, $s_W$, $c_W$, $s_\alpha$ and $c_\alpha$ are the sines and cosines of the weak mixing angle $\theta_W$ and the mixing angle $\alpha$ between the $CP$-even neutral Higgs states, and
\begin{linenomath*}
\begin{align}
    \mathcal{R}(\mneu{1})=\frac{1}{48\pi M_{P}^{2}}\frac{\mneu{1}^{5}}{\mg^{2}}\,,
    \hspace{0.3cm}
    \mathcal{C}(m_{i},\mneu{1})=\Bigg(1-\frac{m_{i}^{2}}{\mneu{1}^{2}}\Bigg)^{4}\,,\nonumber
\end{align}
\end{linenomath*}
where $M_P$ is the Planck scale. Because of the small Planck scale-suppressed coupling to the gravitino, the heavy electroweakinos will dominantly decay to other electroweakinos.

In gauge mediated scenarios, experimental searches usually assume a gravitino mass of the order of a few eV, e.g.\ see the CMS search in Ref.~\cite{CMS:2019oou} which assumes 10\,eV. While the gravitino mass has a negligible effect on the collider kinematics, because of the strong dependence of the partial widths on the gravitino mass in $\mathcal{R}$, the mass choice can affect the decay channels of any particles that are nearly degenerate with the NLSP, specifically either the lightest chargino in the wino-dominated NLSP case, or the next-to-lightest neutralino and the lightest chargino in the Higgsino NLSP case. If there is little phase space available these particles have a competitive decay directly to the gravitino when the gravitino mass is sufficiently small, because the coupling increases as $1/\mg$~\cite{Fayet:1977vd}. On the other hand, the gravitino mass cannot be arbitrarily small since the limit $\mg\to 0$ is the supersymmetric limit and would require very light superpartners. The mild assumption of the heaviest superpartner having a mass of 3\,TeV requires $\mg>2\times10^{-3}$\,eV~\cite{Ambrosanio:1996jn}.

Since we want to remain as model independent as possible and avoid specifying the details of the supersymmetry breaking mechanism, we cannot connect the other superpartner masses explicitly to the gravitino mass. Instead, in this paper, we make the phenomenological choice of fixing the gravitino mass to either $\mg = 1$\,eV, $\mg = 1$\,meV or $\mg = 1$\,keV. We note here that the lowest gravitino mass is not entirely realistic given the discussion above, but it serves as a bookending of the model parameter space. We will later discuss how our results change with the assumed gravitino mass, and show that this tests a range of possible values that covers the interesting phenomenology.

\subsection{Mapping Lagrangian parameters to particle properties}
A one-loop calculation of the chargino and neutralino mass spectrum is performed using a \flexiblesusy~\cite{Athron:2014yba,Athron:2017fvs} spectrum generator, which in turn relies on \sarah~\cite{Staub:2008uz,Staub:2012pb} and some routines from \SOFTSUSY~\cite{Allanach:2001kg,Allanach:2009bv}. Decay widths are calculated using the \decaybit module. For final states not including a gravitino, these are obtained using \SUSYHIT \textsf{1.5} \cite{Djouadi:2006bz}, which is interfaced with the \SDECAY \cite{Muhlleitner:2003vg} and \HDECAY \cite{Djouadi:1997yw} packages. Decay widths to final states including a gravitino are available directly within \decaybit, based on the results contained in Refs.~\cite{Feng:2004mt,Covi:2009bk,Hasenkamp:2009zz}. The absence of corrections beyond one-loop order implies some uncertainty in the mapping between Lagrangian parameters and physical masses, but at scales well below the resolution of our results. For partial decay widths, the absence of higher-order corrections could have some impact in scenarios with small mass differences; however, we use the available state-of-the art tools.

\section{Collider likelihoods}
\label{sec:collider likelihoods}
In the following, we will perform frequentist global fits of both the \EWMSSM and \GEWMSSM, taking into account three different contributions to the likelihood arising from collider physics observations: direct LHC searches for specific new particles, model-independent LHC measurements that are sensitive to contamination from beyond-Standard Model physics processes, e.g.\ precision measurements of differential cross-sections, and limits on the cross-sections of electroweakino production from the Large Electron--Positron (LEP) collider. 

\subsection{LHC searches for new particles}
\label{sec:simulated_LHC_searches}
The most rigorous way of obtaining the likelihood of a given point in parameter space given LHC particle search results is to simulate sparticle production and decay, pass the results through a detector simulation, and then reproduce the statistical procedures utilised by the ATLAS and CMS experiments. This is performed using the \specbit, \decaybit and \colliderbit modules of \gambit~\cite{SDPBit,ColliderBit}. We include 20 ATLAS and 14 CMS searches from Refs.~\cite{ATLAS:2020pgy,ATLAS:2022hbt,ATLAS:2022zwa,ATLAS:2019wgx,ATLAS:2023lfr,ATLAS:2021jyv,ATLAS:2020aci,ATLAS:2024tqe,ATLAS:2019lng,ATLAS:2020syg,ATLAS:2021yqv,ATLAS:2021hza,ATLAS:2019lff,ATLAS:2021moa,ATLAS:2021yyr,ATLAS:2017avc,ATLAS:2019fag,ATLAS:2018nud,ATLAS:2022ckd,ATLAS:2018vzq,CMS:2021few,CMS:2022vpy,CMS:2021beq,CMS:2022sfi,CMS:2023xlp,CMS:2020bfa,CMS:2018xqw,CMS:2019zmd,CMS:2017gbz,CMS:2018fon,CMS:2017jrd,CMS:2019vzo,CMS:2020cpy,cms:2021cox-fix}. A short summary of each included search is provided in Appendix~\ref{app:searches}. 

The LHC electroweakino production at $\sqrt{s}=13$\,TeV is performed using the \pythiaeight generator~\cite{Sjostrand:2006za,Bierlich:2022pfr}, including its leading-order
plus leading-logarithm (LO+LL) production cross-sections for normalising signal yields. The use of LO+LL normalization typically reduces the predicted signal yield relative to next-to-leading-logarithmic accuracy (NLO+NLL) calculations, although the impact on the profile likelihood is process and spectrum dependent. The ATLAS and CMS detectors are simulated by the \gambit native \buckfast detector simulation~\cite{ColliderBit}, with the analysis cuts for each signal region provided by \colliderbit. As discussed in Refs.~\cite{GAMBIT:2023yih,EWMSSM,Cranmer:2021urp}, for many analyses it is not possible to perform a proper statistical combination of the various signal regions due to a lack of public information on how the background uncertainties are correlated between signal regions. In this case, a conservative approach is to include results only from the signal region that has the highest expected significance for each given point in the supersymmetry parameter space. 
Since electroweakino searches often have a small acceptance, there can be multiple competing signal regions, and the precise choice is subject to substantial Monte Carlo noise. To mitigate this issue, we generate a total of 16 million Monte Carlo events per parameter point. 

We define the likelihood function for each analysis as follows. If no signal region correlation information is publicly available for an analysis, we define the likelihood as

\begin{linenomath*}
\begin{align}
  \label{eq:search_1SR_like}
  \begin{split}
    \mathcal{L}_{\text{search}}^\text{1SR}(s_i, \gamma_i) 
    =& \left[ \frac{(s_i + b_i + \gamma_i)^{n_i} \, e^{-(s_i + b_i + \gamma_i)}}{n_i!} \right]\\
    & \times \frac{1}{\sqrt{2\pi}\sigma_i} e^{-\frac{\gamma_i^2}{2 \sigma_i^2}} \, , 
  \end{split}
\end{align}
\end{linenomath*}
where $s_i$, $b_i$ and $n_i$ are the expected signal yield, expected background yield and observed yield, respectively, for the signal region $i$ that has the best expected sensitivity. The uncertainty in the total predicted yield is treated by adding a Gaussian penalty term with a nuisance parameter $\gamma_i$, whose width $\sigma_i$ is the quadrature sum of the uncertainties of $s_i$ and $b_i$. At each parameter point, we profile $\mathcal{L}_{\text{search}}^\text{1SR}(s_i, \gamma_i)$ over $\gamma_i$:
\begin{linenomath*}
\begin{align}
  \mathcal{L}_{\text{search}}^\text{1SR}(s_i) \equiv \mathcal{L}_{\text{search}}^\text{1SR}(s_i, \hat{\hat{\gamma_i}}),
  \label{eq:search_1SR_like_profiled}
\end{align}
\end{linenomath*}
where $\hat{\hat{\gamma_i}}$ is the $\gamma_i$ value that maximises $\mathcal{L}_{\text{search}}^\text{1SR}(s_i,\gamma_i)$ for a given $s_i$.

For some searches, correlation information is publicly available. For example, a subset of CMS search results is published with a covariance matrix for the background uncertainties within the \emph{simplified likelihood} framework~\cite{Collaboration:2242860,Buckley:2018vdr}. A search with $n_\text{SR}$ signal regions is then described by the likelihood function
\begin{linenomath*}
\begin{align}
  \label{eq:search_simp_like}
  \begin{split}
    \mathcal{L}_{\text{search}}(\bm{s}, \bm{\gamma})
    =& \prod_{i=1}^{n_\text{SR}} \left[ \frac{(s_i + b_i + \gamma_i)^{n_i} \, e^{-(s_i + b_i + \gamma_i)}}{n_i!} \right]\\
    & \hphantom{\int} \times \frac{1}{\sqrt{\det2\pi\bm{\Sigma}}} e^{-\frac{1}{2} \bm{\gamma}^T \bm{\Sigma^{-1}} \bm{\gamma}}, 
  \end{split}
\end{align}
\end{linenomath*}
where $\bm{\Sigma}$ is the covariance matrix for the nuisance parameters $\gamma_i$ (with dimension $n_\text{SR} \times n_\text{SR}$). $\bm{\Sigma}$ is constructed by taking the covariance matrix provided by CMS and adding our signal uncertainties in quadrature along the diagonal. We then again profile $\mathcal{L}_{\text{search}}(\bm{s}, \bm{\gamma})$ over the nuisance parameters,
\begin{linenomath*}
\begin{align}
  \mathcal{L}_{\text{search}}(\bm{s}) \equiv \mathcal{L}_{\text{search}}(\bm{s}, \hat{\hat{\bm{\gamma}}}).
  \label{eq:search_simp_like_profiled}
\end{align}
\end{linenomath*} 

For some searches, ATLAS has published the information required to fully utilise all signal regions, through the \textit{full likelihood} framework~\cite{ATLAS:2019oik}. We have extended \colliderbit to make use of the ATLAS \textit{full likelihood} framework, and we use it to compute the likelihood contributions from the relevant searches where this information is available~\cite{ATLAS:2020pgy, ATLAS:2022zwa, ATLAS:2019wgx, ATLAS:2019lng, ATLAS:2021moa}.

A further issue is to what extent the same signal events risk being counted by two or more different searches within the same experiment. If there is significant overlap in the set of selected events between two searches, we can no longer treat these searches as statistically independent and simply add their log-likelihood contributions. The amount of expected overlap between two analyses will vary across the \EWMSSM and \GEWMSSM parameter spaces. Ideally, one would at each sampled parameter point use the simulated events to identify the optimal subset of analyses that can be treated as independent. However, this quickly becomes computationally impractical. As a pragmatic solution, we have instead performed a simulation-based investigation of analysis overlap for a selected set of benchmark parameter points, chosen as examples of different phenomenological scenarios contained within the \EWMSSM and \GEWMSSM models that are also expected to produce non-negligible LHC signal counts. We treat any analyses that are correlated at even a single one of these benchmark points as if correlated over the entire parameter scan.

\begin{figure*} 
  \centering
  \includegraphics[width=0.495\textwidth]{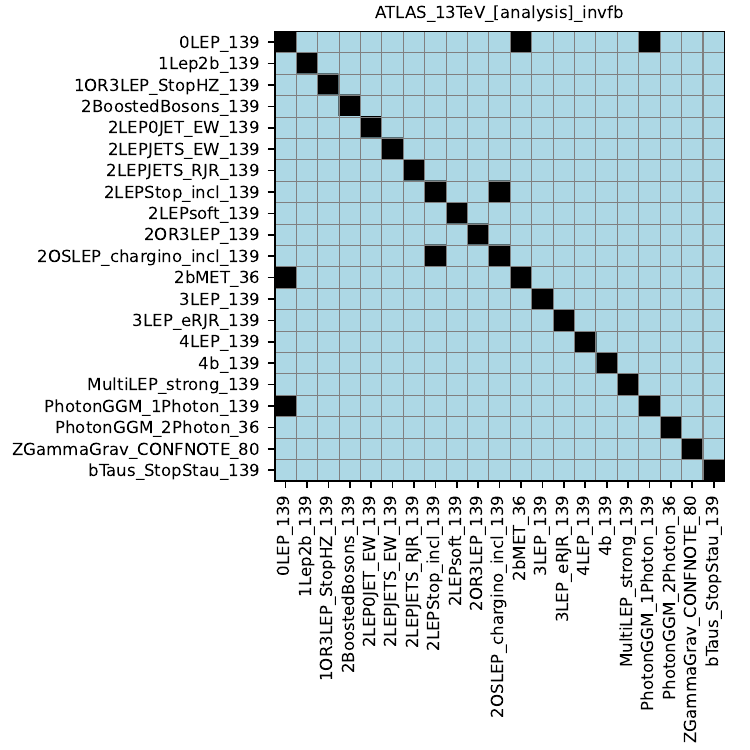}
  \includegraphics[width=0.495\textwidth]{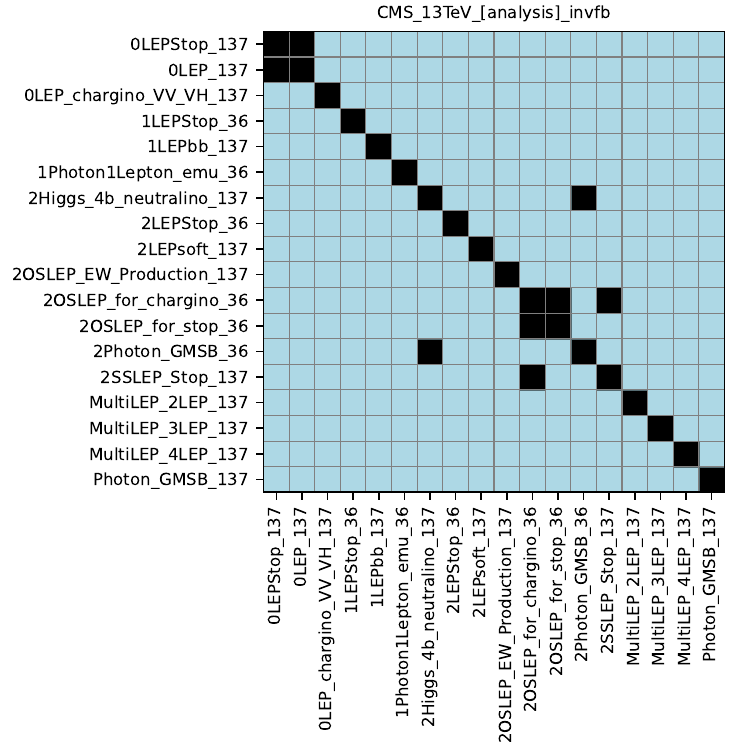}
  \caption{\label{fig:EWMSSM analysis correlations} Correlated analyses for ATLAS (left) and CMS (right) predictions for the \EWMSSM model. Black regions correspond to correlated analyses above a 5\% threshold. These are computed by finding whether any signal regions within a pair of analyses contain the same events, for each of a range of benchmark parameter samples.}
\end{figure*}

\begin{figure*} 
  \centering
  \includegraphics[height=0.495\textwidth]{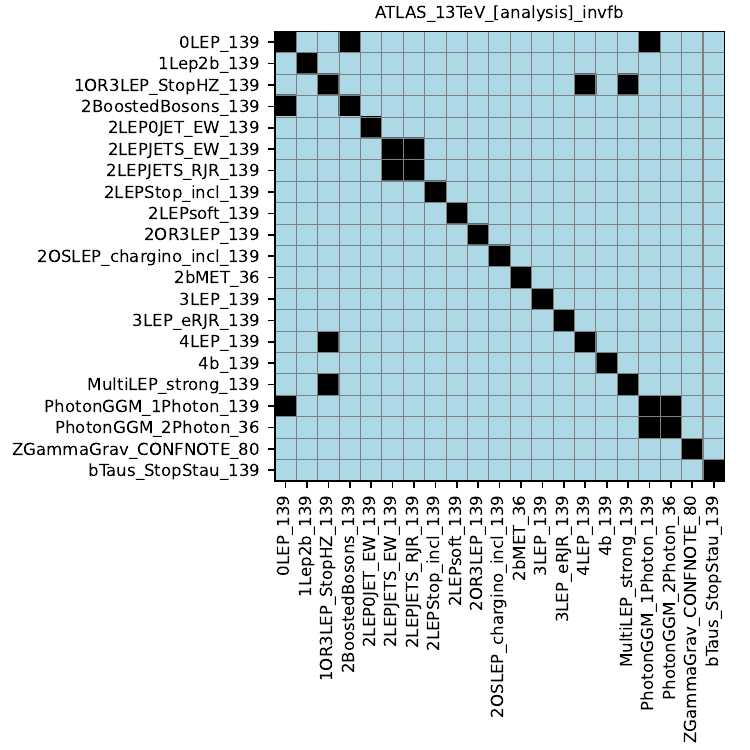}
  \includegraphics[height=0.495\textwidth]{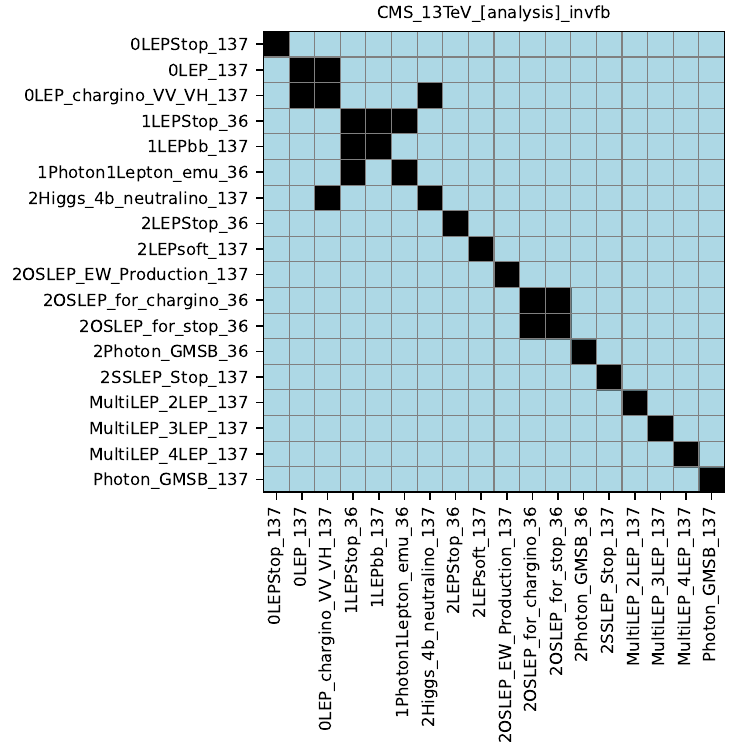}
  \caption{\label{fig:Gravitino analysis correlations} Correlated analyses for ATLAS (left) and CMS (right) predictions for the \GEWMSSM model. Black regions correspond to correlated analyses above a 5\% threshold. These are computed by finding whether any signal regions within a pair of analyses contain the same events, for each of a range of benchmark parameter samples.}
\end{figure*}

\colliderbit has been extended to allow direct output of a file containing an account of all analysis signal regions for a given model parameter point and whether a given signal event filled this signal region. Figs.~\ref{fig:EWMSSM analysis correlations} and~\ref{fig:Gravitino analysis correlations} show which analyses contain correlated signal regions.\footnote{These figures can be generated using the utility script \textsf{ColliderBit/scripts/generate\_correlation\_plots.py}.} Two analyses are claimed to be sufficiently correlated if the signal regions that produce the strongest expected exclusion at any of the benchmark parameter points correlate with each other above 5\%. This is similar to the approach followed in the TACO method~\cite{Araz:2022vtr}; however, it works at the analysis level rather than directly at the signal region level. For any set of analyses that fail this check, we choose which analyses to include in the total likelihood calculation by picking the combination of analyses that provides the greatest expected exclusion on a per-point basis. This follows our approach for handling signal region choice for any analysis without a correlation matrix. It is worth noting that this choice can potentially act as a source of per-point variation in the magnitude of the likelihood across regions of different analysis combinations, but the effect of this on any results presented here was found to be negligible.

The combined log-likelihood for all LHC searches is defined as 
\begin{linenomath*}
\begin{align}
 \label{eq:delta_lnlike_sum}
 \Delta \ln \mathcal{L}_\text{searches}(\bm{s}) = \sum_{j} \Delta \ln \mathcal{L}_j(\bm{s}),
\end{align}
\end{linenomath*}
where $\Delta \ln \mathcal{L}_j$ is the log-likelihood contribution from search $j$ defined as the difference with respect to the corresponding background-only ($\bm{s} = \bm{0}$) likelihood, 
\begin{linenomath*}
\begin{align}
 \label{eq:delta_lnlike}
 \Delta \ln \mathcal{L}_j(\bm{s}) = \ln \mathcal{L}_j(\bm{s}) - \ln \mathcal{L}_j(\bm{s} = \bm{0}).
\end{align}
\end{linenomath*}
If the combined likelihood $\Delta \ln \mathcal{L}_\text{searches}(\bm{s})$ is positive for a given parameter point, it means that the signal predictions for that point provide a better fit to the observed LHC data than the SM-only hypothesis. 

\subsection{Model-independent LHC measurements}\label{sec:contur}
In addition to direct searches for new particles, sparticle production at the LHC may contribute to the observed event yield in regions of phase space typically used for precision measurements. Such measurements are included in \colliderbit via an interface to the \contur~\cite{Butterworth:2016sqg,Buckley:2021neu} package. Simulated events are passed through analysis code from the \rivet~\cite{Bierlich:2019rhm} toolkit and projected into the fiducial phase space of the measured cross sections. 

The analyses available in \contur have detector effects unfolded bin-by-bin, which requires bin counts to be in the regime where a Poisson likelihood is indistinguishable from a $\chi^2$ test. In analogy with the LHC search likelihood, the $\chi^2$ is defined in terms of the log-likelihood difference between the ``signal-injection'' hypothesis and the SM null hypothesis (which amounts to assuming that the data yields are equal to the SM prediction). We thus define:
\begin{linenomath*}
\begin{align}
 \label{eq:contur_lnlike}
 \begin{split}
 \ln \mathcal{L}_\text{meas}(\bm{s}) 
 &= -\chi^2(\bm{s})/2\\
 &\equiv - \sum_{i \,\in\, \text{active bins}} \left[
 \frac{y_i^\text{s+b}(\bm{s}) - y_i^\text{obs}}{\Delta y_i}
 \right]^2 \Big/ 2
 \, ,
 \end{split}
\end{align}
\end{linenomath*}
where $y_i$ and $\Delta{y_i}$ are the bin values and uncertainties, respectively, and the sum is over all included measured bins. We use \contur in a mode which assumes the data is identical to the SM measurements, and so the log-likelihood difference is then simply 
$\Delta \ln \mathcal{L}_\text{meas}(\bm{s}) = \ln \mathcal{L}_\text{meas}(\bm{s}) - \ln \mathcal{L}_\text{meas}(\bm{s} = \bm{0}) = \ln \mathcal{L}_\text{meas}(\bm{s})$. This assumption does not permit the SM measurement likelihood to favour the model above the background hypothesis. Experimental uncertainties are assumed to be uncorrelated. To account
for statistical correlations between separate measurements, the measurements are divided based on the experiment, final states and run period into non-overlapping ``analysis pools''. The combined \contur likelihood is then defined using the most sensitive measurement from each pool.

Due to the need to generate and store additional events for \contur processing, we generate 100,000 extra events at each sampled parameter point in a post-processing run. The smaller number of events relative to the 16 million used in the main run is due to the much higher acceptances of LHC precision measurements compared to searches. 

We briefly discuss the impact of these measurements on the \GEWMSSM model in Sec.~\ref{sec:GEWMSSM_results}. For the \EWMSSM model, the profiled parameter samples we present in Fig.~\ref{fig:mN2mN1plane} are not strongly impacted by the contributions from LHC measurements, and so we do discuss these in detail. 

\subsection{Cross-section limits from LEP searches}\label{sec:lep}
For some parameter regions with low masses, LEP constraints on electroweakino production remain competitive with the more recent results from the LHC. We therefore include in our analysis LEP searches and measurements published as upper limits on particular electroweakino production cross-sections. 

A detailed description of our treatment of LEP results and the translation from cross section limits to likelihoods can be found in Refs.~\cite{ColliderBit,EWMSSM}. For this work, we have extended the available LEP searches in \colliderbit by an L3 multi-photon and missing energy search~\cite{L3:2003yon} that is particularly important for the \GEWMSSM model, see also Appendix~\ref{app:lep_extension}.

Note that some processes constrained by LEP can be interpreted in different ways depending on the model being studied. For example, a search for pair-produced charginos that both decay via $\tilde\chi^\pm \to \text{SM} + \tilde\chi$ can also be used to constrain the process $\tilde\chi^\pm \to \text{SM} + \tilde{G}$.

\section{Scan details}
\label{sec:global_fit_setup}

We perform all scans of the \EWMSSM and \GEWMSSM models using the \specbit \cite{SDPBit}, \decaybit \cite{SDPBit}, \colliderbit~\cite{ColliderBit} and \scannerbit \cite{ScannerBit} modules of the \GB~\textsf{2.7} global fit framework \cite{gambit}. The scans are performed with the differential evolution sampler \diver \textsf{1.0.5}~\cite{ScannerBit}, running in \textsf{jDE} mode (self-adaptive rand/1/bin evolution), which is based on Ref.~\cite{Brest06}. The number of points per \diver generation was 1000, with a tight convergence threshold of $10^{-12}$ to allow ample exploration of the best-fit regions. The final combined datasets consist of around 400k and 500k parameter samples for the \GEWMSSM and \EWMSSM models, respectively, and 30k--35k parameter samples for additional scans with varying gravitino mass.

For our scans of the \GEWMSSM model, the gravitino mass is not varied as a free parameter but, as discussed in Sec.~\ref{sec:gravitino_model}, it is instead fixed to a value of 1\,eV, with additional smaller scans at 1\,keV and 1\,meV to explore the full phenomenology. This means that, for both the \GEWMSSM and \EWMSSM models, the physics parameters of interest are the mass parameters $M_1$, $M_2$ and $\mu$, and the dimensionless $\tan\beta$
parameter. 

We utilise the \GB implementations of supersymmetric models in which the MSSM soft supersymmetry breaking Lagrangian parameters are defined at the scale $Q$, which we here set to be 3\,TeV. The most general model has 63 free parameters, comprising the gaugino masses $M_1$, $M_2$, and $M_3$, the trilinear coupling matrices $\mathbf{A}_u, \mathbf{A}_d$ and $\mathbf{A}_e$, the squared soft sfermion mass matrices $\mathbf{m}^\mathbf{2}_Q$, $\mathbf{m}^\mathbf{2}_u$, $\mathbf{m}^\mathbf{2}_d$, $\mathbf{m}^\mathbf{2}_L$ and $\mathbf{m}^\mathbf{2}_e$, and three additional parameters describing the Higgs sector. We set
all trilinear couplings to zero and set the diagonal entries of the squared soft sfermion mass matrices to be $Q^2$, with all off-diagonal entries set to zero. We assume the pseudo-scalar Higgs mass $m_A=5$\,TeV, and set the gluino mass $M_3$ to the same value, while the lightest Higgs mass is fixed to the measured value of the SM-like Higgs. These values are chosen to decouple all sparticles except for the electroweakinos---as well as the gravitino in the case of the \GEWMSSM model. The particular choices for these values do not affect our results once the decoupling regime is reached. 

Table~\ref{tab:parameters} summarises the scan range for each of the free parameters, along with the values of fixed parameters including relevant SM quantities. $M_2$ is restricted to positive values because physical results in these models are invariant under a global sign change for $M_1$, $M_2$ and $\mu$. By requiring that $|M_1|$, $M_2$ and $|\mu|$ are all below 1.5\,TeV, any point where \mneu{2}, and therefore also \mneu{3} and \mneu{4}, are approaching the upper mass bound, must necessarily be a mixed composition of some or all of \neu{2}, \neu{3} and \neu{4}. This is an acceptable compromise in order to allow us to explore the low mass region where the LHC has most sensitivity. To ensure adequate coverage of the parameter space, we combine the results of multiple scans that cover either the entire range of parameters or are restricted to positive/negative $M_1$, $M_2$ or $\mu$. These samples were combined with a single scan targeting a compressed mass range (retaining only parameter samples predicting $\mneu{2} < 800$\,GeV and $\mneu{1} < 150$ GeV), containing around 20k samples. 


\scannerbit applies a prior transformation for each scanned parameter, to allow the chosen scanning algorithm (in this case \diver) to operate within the unit hypercube. Both flat and logarithmic prior transformations were used for the $\tan\beta$ parameter.\footnote{The rejection of parameter samples where $\mneu{1} < 62.5$ (as described in Sec.~\ref{sec:model_EWMSSM}), combined with logarithmic priors, would prevent a sufficient number of accepted samples for \diver optimisation in the first set of sampling generations without careful tuning of the lower limits of $M_1$, $M_2$ and $\mu$. It is for this reason that $M_1$, $M_2$ and $\mu$ were not sampled with log priors.} Since our results are presented as profile likelihood maps, within the broader context of a frequentist statistical interpretation, the choice of prior transformations is only designed to influence the coverage and speed of convergence of the scans, not the final statistical interpretation. 

\begin{table}
\begin{center}
\begin{tabular}{l@{\ }c c}
\toprule
Parameter & Range/value &  Prior transf.\ \\
\toprule
$M_1(Q)$               & $[-1.5, 1.5]\,$TeV & flat  \\
$M_2(Q)$               & $[0, 1.5]\,$TeV  & flat  \\
$\mu(Q)$               & $[-1.5, 1.5]\,$TeV & flat  \\
$\tan\beta(m_Z)$       & $[1, 70]$       & log, flat     \\
$\mg$                  & $1\,$eV        & fixed         \\
\midrule
$Q$                    & $3\,$TeV & fixed \\
\midrule
$\alpha_s^{\MSBar}(m_Z)$ & $0.1181$ & fixed \\
Top quark pole mass      & $171.06\,$GeV & fixed \\
Higgs mass               & $125.09\,$GeV & fixed \\
\bottomrule
\end{tabular}
\caption{\label{tab:parameters} Ranges and prior transformations for the input parameters. }
\end{center}
\end{table}

\newpage

\section{Results}
\label{sec:results}
In this section, we present and discuss the profile likelihood maps resulting from the scans of our two models, starting with the \EWMSSM before moving on to the \GEWMSSM. 

\subsection{Global fit results for the \EWMSSM} \label{sec:EWMSSM_results}
\begin{figure*} 
  \centering
  \includegraphics[height=0.8\columnwidth]{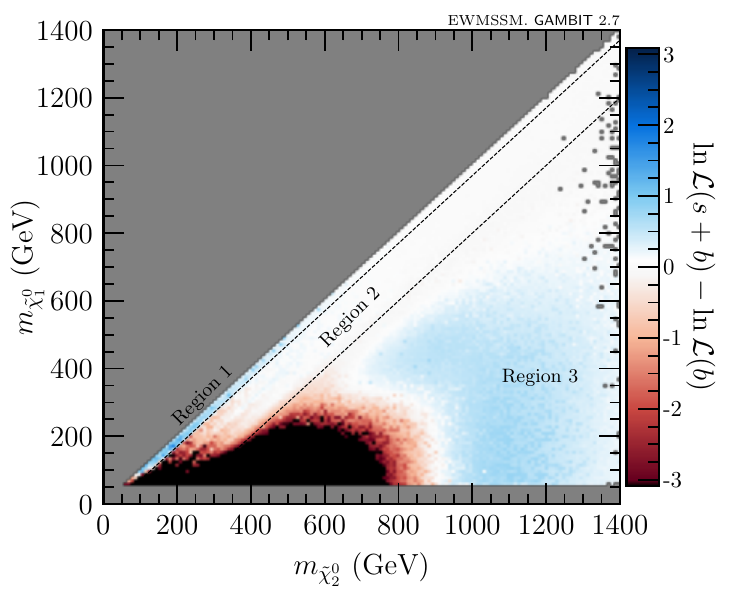}
  \includegraphics[height=0.8\columnwidth]{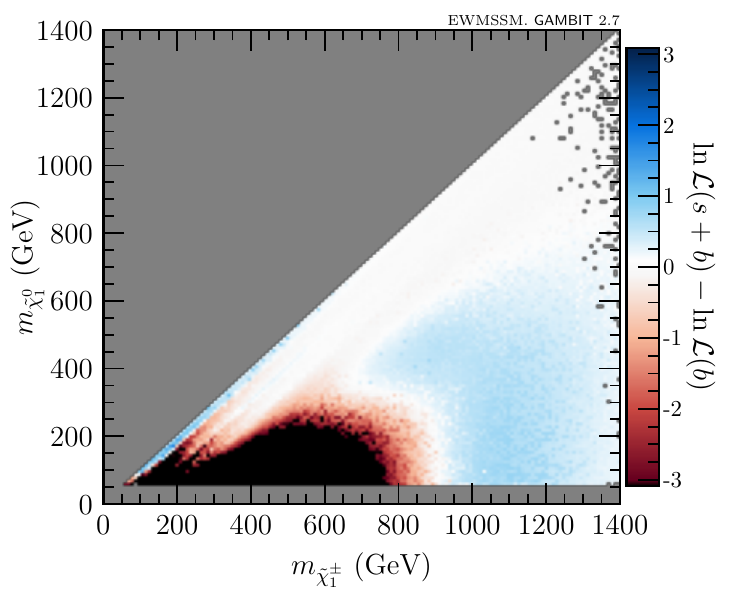}
  \caption{Profile likelihood for the EWMSSM model in the ($\mneu{2},\mneu{1}$) (left) and ($m_{\cha{1}},\mneu{1}$) (right) planes. Blue (red) regions are favoured (disfavoured) with respect to the background hypothesis, with black disfavoured at or beyond 2$\sigma$. Regions 1--3 as described in Sec.~\ref{sec:EWMSSM_results} are designated and separated by black lines.
  }
  \label{fig:mN2mN1plane}
\end{figure*}

In Fig.~\ref{fig:mN2mN1plane} we show the 2D profile likelihood in the ($\mneu{2},\mneu{1}$) and ($m_{\cha{1}},\mneu{1}$) planes, coloured according to the total log-likelihood difference $\Delta\ln\mathcal{L} = \ln\mathcal{L}(\bm{s}) - \ln\mathcal{L}(\bm{s}=\bm{0})$. Using the nominal Wilks' threshold for a 2$\sigma$ confidence region in 2D, we consider regions with a profiled log-likelihood difference $\Delta\ln\mathcal{L} \leq -3.09$ as excluded (black colour), when compared to the background-only hypothesis.\footnote{In interpreting these figures, we also remind the reader that parameter regions can end up as allowed ($\Delta\ln\mathcal{L} > -3.09)$ for two qualitatively different reasons. In the first case, none of the included searches/measurements are sufficiently sensitive to the scenario picked out by the profiling, as is typical e.g.\ at high masses. In the second case, multiple searches/measurements may be sensitive to the selected model point, but the log-likelihood penalty coming from observations in tension with the model is balanced by a positive log-likelihood contribution from other observations that prefer a non-zero signal.} 
Grey areas of the figures are regions where no parameter points were included in a given bin, such as where samples would lie outside scanned parameter ranges, or do not satisfy the model restrictions described in Sec.~\ref{sec:model}.

For a very light \neu{1}, the exclusion extends up to approximately 760\,GeV in both \mneu{2} and $m_{\cha{1}}$. This differs markedly from our previous \gambit global fit of the \EWMSSM based on 36\,fb$^{-1}$ of Run 2 LHC data, in which no region in the ($m_{\cha{1}},\mneu{1}$) plane was excluded.\footnote{See Fig.~6 of Ref.~\cite{EWMSSM}.} This indicates both the enhanced statistical power of the full Run 2 dataset, plus the ingenuity of LHC experimentalists in improving the sensitivity of searches. 

Unlike the simplified model limits presented by the ATLAS and CMS collaborations, which we discuss further below, our results allow for all the possible compositions of the electroweakinos accessible within the scanned \EWMSSM parameter space, with each bin of the profile likelihood plots featuring the combination that maximises the likelihood for those particular masses. It is therefore not trivial to unpack the effects of different choices for the electroweakino mixing. While we will here focus on the ($\mneu{2},\mneu{1}$) plane to discuss our results, it is important to keep in mind that for any given bin in the ($\mneu{2},\mneu{1}$) plane, the likelihood profiling might pick out parameter samples where the LHC sensitivity is determined as much by the masses and compositions of the two heavier neutralinos and the two charginos as by \mneu{1} and \mneu{2}. 

\begin{figure*}
    \centering
    \includegraphics[width=0.32\linewidth]{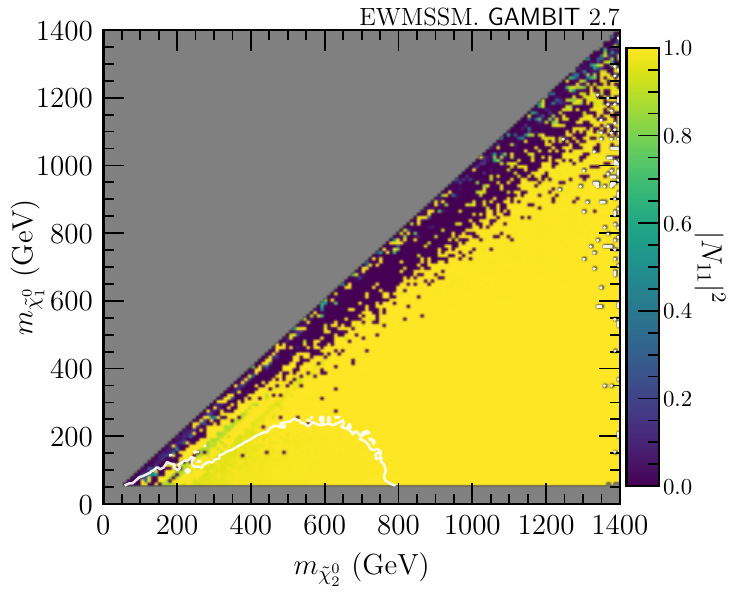}
    \includegraphics[width=0.32\linewidth]{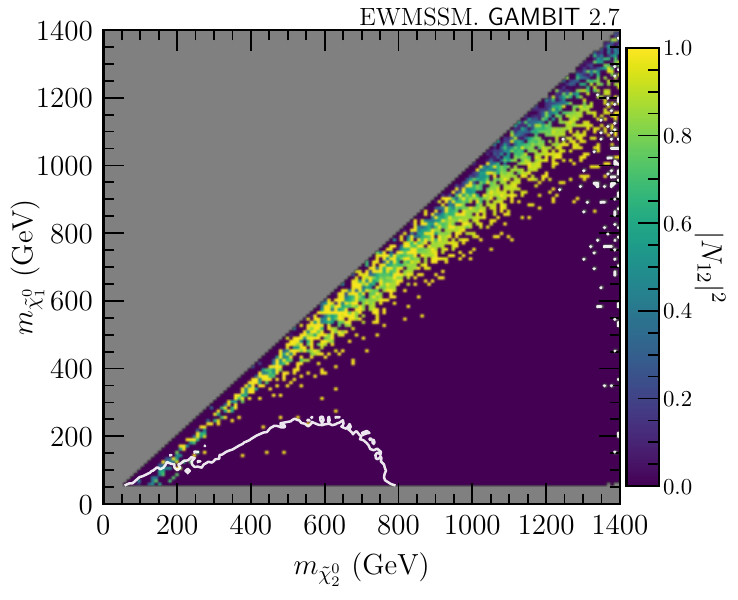}
    \includegraphics[width=0.32\linewidth]{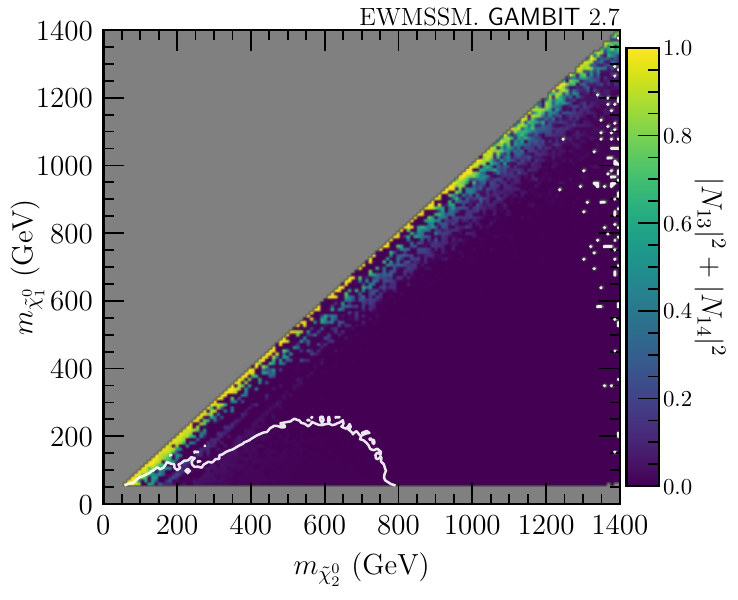}  \\ 
    \includegraphics[width=0.32\linewidth]{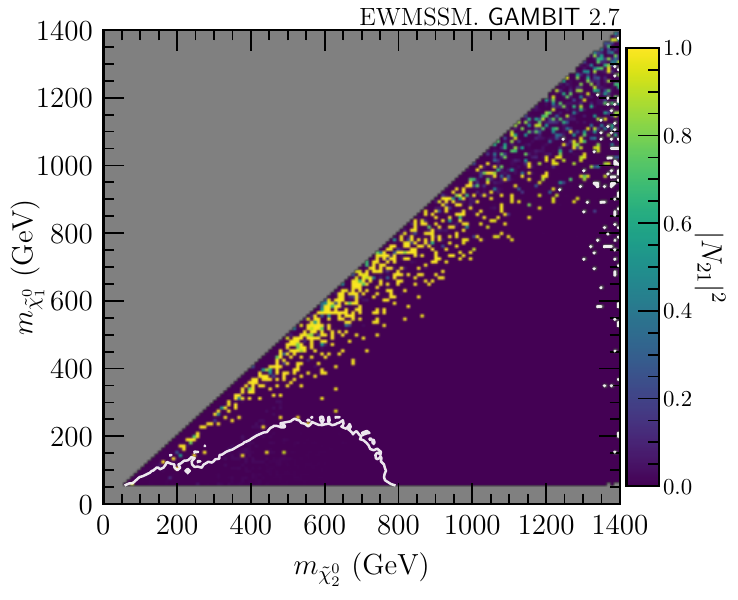}
    \includegraphics[width=0.32\linewidth]{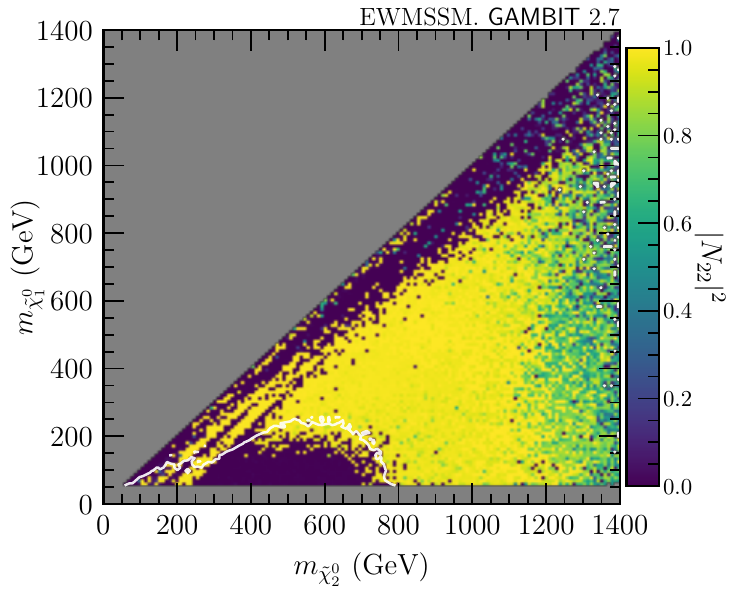}
    \includegraphics[width=0.32\linewidth]{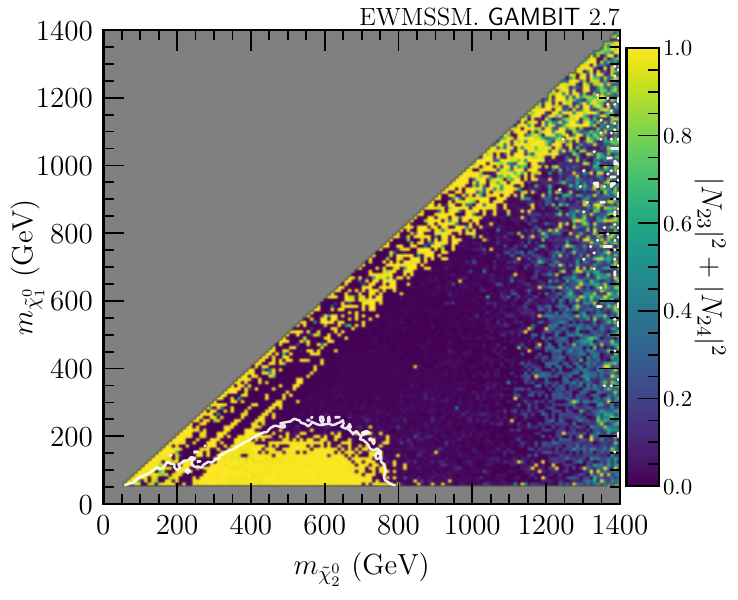}  \\
    \caption{Composition of the lightest neutralino (top) and second lightest neutralino (bottom) for the maximum likelihood parameter sample for each ($\mneu{2},\mneu{1}$) bin in the plane, as found by the profiling in Fig.~\ref{fig:mN2mN1plane} (left). The composition is indicated by colour scales for the bino fraction (left), wino fraction (middle) and Higgsino fraction (right) across the  ($\mneu{2},\mneu{1}$) plane. The white contour is the border of the 2$\sigma$ excluded region.  }
   \label{fig:mN2mN1planecomp}
\end{figure*}

To facilitate interpretation, we provide in Fig.~\ref{fig:mN2mN1planecomp} plots showing the bino (left), wino (middle) and Higgsino (right) mixture of the \neu{1} (top) and \neu{2} (bottom), for the parameter samples that make up our profile likelihood ``surface'' across the ($\mneu{2},\mneu{1}$) plane, i.e.\ the same parameter samples picked out by the profiling in Fig.~\ref{fig:mN2mN1plane} (top left).
Based on these composition plots we can see that the ($\mneu{2},\mneu{1}$) profile likelihood result can be viewed as consisting of three different non-excluded regions, as we move from small to large $\mneu{2} - \mneu{1}$ mass splitting:
\begin{itemize}[itemsep=\medskipamount]
    \item \textbf{Region 1}: The region along the diagonal, where either \neu{1} is mostly Higgsino, and, consequently, so is \neu{2}, leading to $\mneu{2} \approx \mneu{1}$, or \neu{1} is mostly bino with a wino-dominated \neu{2} close in mass.
    \item \textbf{Region 2}: The region with mass splitting $\mneu{2} - \mneu{1} \lesssim 200$\,GeV, in which the likelihood profiling selects parameter points from a range of different mixing scenarios, 
    most notably scenarios with bino-dominated \neu{1} 
    and wino-dominated \neu{2}, 
    scenarios where the \neu{1} is mostly bino with a non-negligible Higgsino component 
    and Higgsino-dominated \neu{2} (and \neu{3}),
    or scenarios where \neu{1} is a wino-Higgsino mixture and the \neu{2} either a bino-Higgsino mixture or mostly Higgsino.
    \item \textbf{Region 3}: The region at $\mneu{2} - \mneu{1} \gtrsim 200$\,GeV, where the \neu{1} is almost entirely bino. 
\end{itemize}

\begin{figure*} 
  \centering
  \includegraphics[height=0.56\columnwidth]{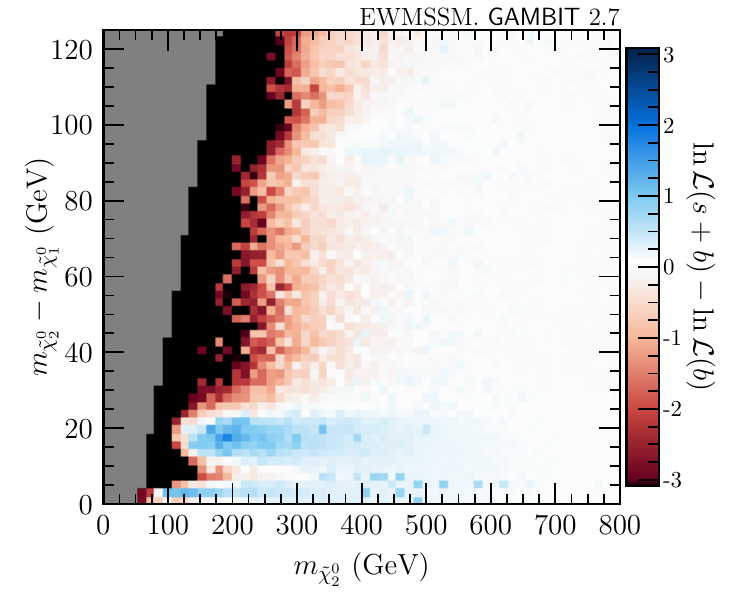}
  \includegraphics[height=0.56\columnwidth]{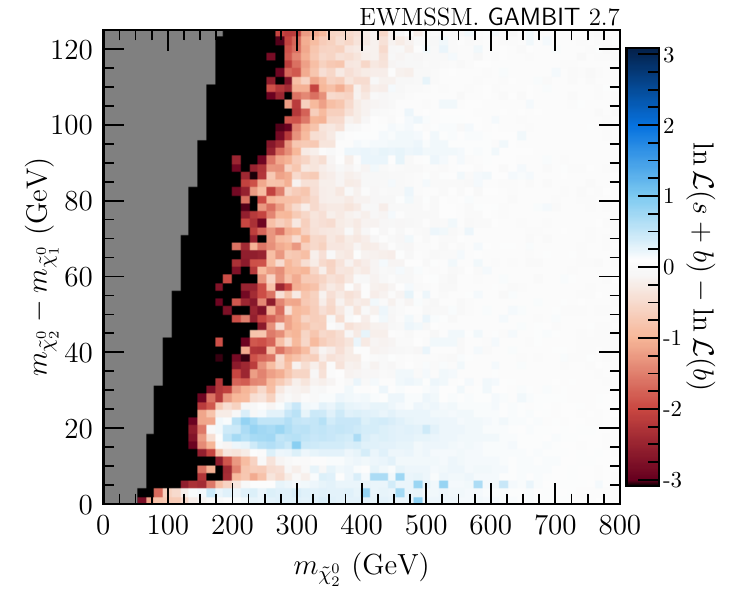}
  \includegraphics[height=0.56\columnwidth]{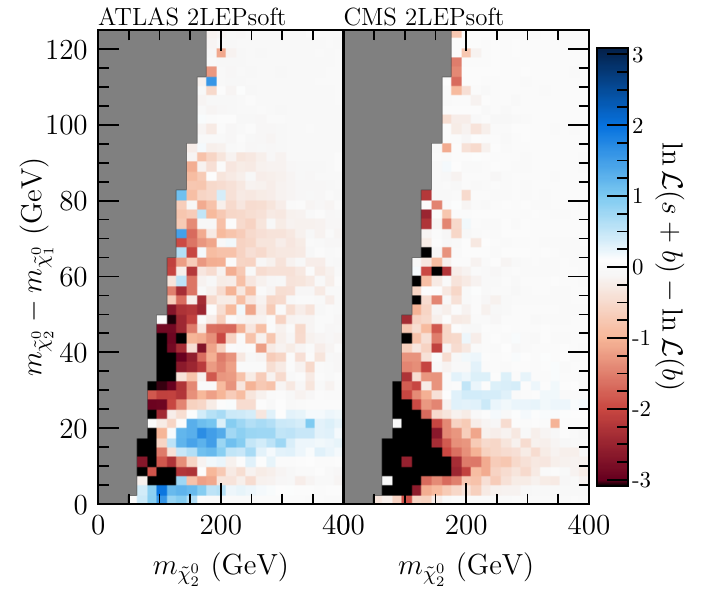}
  \caption{\label{fig:low mass excesses} Profile likelihood for the EWMSSM model focused on the compressed mass region. Results are shown with (middle) and without (left) including the CMS soft lepton search, where the combination without the CMS soft lepton search corresponds to our main scanned likelihood. The right panel shows the contributions from ATLAS (left half) and CMS (right half) soft-lepton searches to the total likelihood in the centre panel.}
  \label{fig:mN2mdiffplane}
\end{figure*}

For Region 1, along the diagonal in the ($\mneu{2},\mneu{1}$) plane, the most important searches for determining which scenarios end up as the least excluded low-mass scenarios in our combination are \texttt{ATLAS\_2LEPsoft\_139} and \texttt{CMS\_0LEP\_137}.\footnote{Here we use shorthand labels for the analyses, relative to the labels used in the \colliderbit code. For the full labels and corresponding ATLAS/CMS references, see Table~\ref{apptab:analist}.}
In the case of \texttt{ATLAS\_2LEPsoft\_139} there is a small, much discussed excess, e.g.~see Refs.~\cite{Bagnaschi:2025ksk,Ellwanger:2024vvs,Agin:2024yfs,Martin:2024pxx,Agin:2025vgn,Hammad:2025wst,Araz:2025bww,Constantin:2025bqp}, across multiple signal regions, leading to weaker than expected exclusion of scenarios with mass differences $\mneu{2} - \mneu{1} \sim 20$\,GeV and also for $\mneu{2} - \mneu{1} \lesssim 3$\,GeV. In our combined fit, we identify a corresponding weak preference for such mass splittings at low masses of the lightest neutralinos. This can most easily be seen as two horizontal bands of favoured points in Fig.~\ref{fig:mN2mdiffplane}, where we show our profile likelihood result in the $(\mneu{2}-\mneu{1}, \mneu{1}$) plane.\footnote{The origin of the grey areas in Fig.~\ref{fig:mN2mdiffplane} is the lower cut on neutralino masses of 62.5\,GeV used in our model scan. Except for very low mass differences ($\lesssim 5$ GeV) this low mass neutralino region is far inside the region excluded due to the production and decay of heavier electroweakinos.} 
We will return to discuss this excess in more detail below in Sec.~\ref{sec:excesses}.

In Region 2, with mass splitting $\mneu{2} - \mneu{1} \lesssim 200$\,GeV, the least excluded points typically have a bino-dominated \neu{1}, but with a significant Higgsino fraction (Fig.~\ref{fig:mN2mN1planecomp}, top left and top right), or points with a mostly wino \neu{1} (Fig.~\ref{fig:mN2mN1planecomp}, top middle). The \neu{2} for these least constrained points can be wino- or Higgsino-dominated (bulk of the  profiled parameter points), or a more even mix of both (towards higher \neu{2} masses), as shown in Fig.~\ref{fig:mN2mN1planecomp}, bottom middle and bottom right, or the \neu{2} can be bino-dominated (Fig.~\ref{fig:mN2mN1planecomp}, bottom left), corresponding to points where the \neu{1} is mostly wino. In short, a range of different theory scenarios survive current LHC constraints for these neutralino masses. 

The searches for simplified models that have the strongest sensitivity to constrain the lower-mass part of Region 2 include \texttt{ATLAS\_2LEPJETS\_EW\_139}, \texttt{ATLAS\_2OR3LEP\_139}, \texttt{ATLAS\_3LEP\_139}, \texttt{ATLAS\_3LEP\_eRJR\_139}, \texttt{CMS\_0LEP\_chargino\_VV\_VH\_137}, \texttt{CMS\_2OSLEP\_EW\_Production\_137} and \texttt{CMS\_MultiLEP\_3LEP\_137}. The signal regions in the above searches typically target fairly ``clean'' scenarios, e.g.\ with production of either pure-wino \neu{2}/\cha{1} or pure-Higgsino \neu{2}/\neu{3}/\cha{1}, decaying down to a bino \neu{1}, and where the assumed decays, e.g.\ $\neu{2} \rightarrow Z \neu{1}$ or $\neu{2} \rightarrow h \neu{1}$ and $\cha{1} \rightarrow W \neu{1}$, have 100\% branching ratio. In our scan, we find that these searches indeed exclude large parts of parameter space predicting such phenomenology. The parameter points that our scan then identifies as the least constrained points in Region 2 include points with a more complicated phenomenology, e.g.\ points where $|\mu| \sim M_2 \gtrsim |M_1|$, meaning that all five heavier electroweakinos can be relevant for the production processes, that their typical mass differences are not what would be predicted in pure-wino or pure-Higgsino scenarios, and that their branching ratios can be diluted across a larger set of possible decay processes.

Region 3 of the ($\mneu{2}, \mneu{1}$) plane is characterised by a large mass gap, $\mneu{2} - \mneu{1} \gtrsim 200$\,GeV. Focusing on the case where \neu{1} is almost fully bino, each of the heavier states will give rise to a decay chain that produces at least one on-shell $Z$, $h$ or $W$ from the final decay step down to the \neu{1}. Thus, any electroweakino pair production (beyond the mostly negligible bino \neu{1}\neu{1} production) leads to final states with at least two on-shell bosons. 

The different theory scenarios $|M_1| < M_2  < |\mu|$,  $|M_1| < |\mu| < M_2$, and $|M_1| < M_2 \sim |\mu|$ can all map to such mass scenarios with a ``lonely'' bino \neu{1} at the bottom of the spectrum, but they give different predictions in terms of production cross-sections and preferred decay channels, depending also on $\tan\beta$. 
Consequently, we find that many of the searches for leptons and missing energy, e.g.
\texttt{ATLAS\_3LEP\_139}, \texttt{ATLAS\_2LEPJETS\_EW\_139}, \texttt{ATLAS\_2OR3LEP\_139} and \texttt{CMS\_2OSLEP\_EW\_Production\_137}, 
as well as the searches for hadronically decaying bosons and missing energy, 
\texttt{ATLAS\_2BoostedBosons\_139} and \texttt{CMS\_0LEP\_chargino\_VV\_VH\_137}, constrain different parts of these parameter regions.

When combined, the surviving parts of Region 3 that our scan picks out as currently least constrained come mainly from scenarios with $|M_1| < M_2 < |\mu|$. This can be seen by comparing the surviving region in Fig.~\ref{fig:mN2mN1plane} (left panel) with the bottom middle and bottom right panels of Fig.~\ref{fig:mN2mN1planecomp}, which show that for the least constrained points in Region 3, \neu{2} is mostly wino-dominated. Thus, there will typically also be a wino-dominated chargino with $\mcha{1} \approx \mneu{2}$. The least constrained points mostly also have $\tan\beta \lesssim 5$ and $\text{BR}(\neu{2} \rightarrow h \neu{1}) \gtrsim 80\%$. Hence, the dominant signals for these points come from on-shell $\neu{2} \cha{1} \rightarrow (h \neu{1}) (W \neu{1})$ and $\cha{1} \cha{1} \rightarrow (W \neu{1}) (W \neu{1})$, and the profile likelihood pattern we see in Region 3 is largely dominated by the contributions from the $Wh$ signal regions of \texttt{CMS\_0LEP\_chargino\_VV\_VH\_137} and \texttt{ATLAS\_1Lep2b\_139}. We note that our implementation of \texttt{ATLAS\_2BoostedBosons\_139} does not include the $b$-jet signal regions, due to challenges with reproducing the $b$-tagging of small-radius jets. If these signal regions were included, this search would likely also have contributed to shaping the profile likelihood surface we see in Region 3 similarly to \texttt{CMS\_0LEP\_chargino\_VV\_VH\_137}. The CMS search largely disfavours the lower-mass scenarios, and shows a weak preference for the selected scenarios at $\mneu{2} \gtrsim 1$\,TeV, due to a couple of slight data excesses in some of their $Wh$ signal regions. The ATLAS search exhibits a mild preference for most of the kinematically accessible points selected across Region 3, due to several small data excesses observed in that search.

As we look to higher \mneu{2} in the bottom middle and bottom right panels of Fig.~\ref{fig:mN2mN1planecomp}, we see that the preferred \neu{2} mixing is a more even wino-Higgsino blend. This is largely just a consequence of our scan constraints; as any point where \mneu{2} is approaching the upper mass bound must necessarily be a scenario where either $|\mu| \sim M_2$ or $|\mu| \sim M_1$, with a mixed composition of some or all of \neu{2}, \neu{3} and \neu{4}.

At the qualitative level, our results broadly align with the findings presented by the SModelS group~\cite{Altakach:2023tsd, Constantin:2025bqp}. Ref.~\cite{Altakach:2023tsd} performs an electroweak MSSM fit up to masses of 3\,TeV, and excludes \cha{1} up to $\sim900$\,GeV, as compared to $\sim760$\,GeV in our results. With a different set of analyses, different approach to exploring the parameter space (differential evolution versus random scan in the SModelS study), and the use of NLO cross-sections in the SModelS study, this difference in exclusion limit fits with expectations. Based on the samples provided by the ATLAS study of electroweakinos in the pMSSM after Run 2 of the LHC~\cite{ATLAS:2024qmx}, Ref.~\cite{Constantin:2025bqp} studies the pMSSM and  finds four regions where excesses can be fit by their model (see Fig.~10 of Ref.~\cite{Constantin:2025bqp}). We have been able to identify two of these regions appearing within our profile likelihood surface, despite the differences in theoretical scenarios. These two preferred regions with highly degenerate masses and where $m_{\cha{1}} > 800$ GeV, correspond to our Regions 1 and 3.

\subsubsection{Excesses}
\label{sec:excesses}
In recent years, there have been various claims of significant excesses in LHC Run 2 searches for weakly-produced particles, including final states with jets plus missing energy and soft leptons with missing energy~\cite{ATLAS:2019lng,ATLAS:2021moa,CMS:2021edw,ATLAS:2021kxv,CMS:2021far,Agin:2023yoq,Agin:2024yfs,Goodsell:2024aig,Fuks:2024qdt}. Most recently, Ref.~\cite{Agin:2025vgn} interpreted these excesses within the context of two MSSM-inspired simplified models that are frequently used to present LHC search results:

\begin{itemize}[itemsep=\medskipamount]
\item \textbf{Higgsino scenario}: In this scenario, it is assumed that the only light sparticles are a set of pure Higgsino neutralinos and charginos $\{\neu{1},\neu{2},\cha{1}\}$, with $m_{\cha{1}}
= ( \mneu{2} + m_{\neu{1}})/2$. It is assumed that any mixing of heavier particles that splits the masses does not affect the dominantly-Higgsino nature of the sparticle interactions.
\item \textbf{Wino--bino scenario}: In this scenario, it is assumed that the only light sparticles are a pure bino \neu{1}, a pure wino $\neu{2}$ and pure wino \cha{1}, with $m_{\cha{1}}=\mneu{2}$. There are two versions of the scenario depending on the sign of $m_{\neu{1}}\times \mneu{2}$, which changes the interactions of the two lightest neutralinos.\footnote{In other mass relations here we implicitly assume absolute values.}
\end{itemize}

If the mass spectrum is sufficiently compressed, the electroweakino decays are dominated by the production of off-shell gauge bosons, via
\begin{align}
&\neu{2} \;\to\;  (Z^\ast \to f\bar{f}) + \neu{1}
\quad\text{and}\\
&\cha{1} \;\to\;  (W^\ast \to f\bar{f}') + \neu{1}\,.
\end{align}
One can therefore describe the model fully using only two parameters which may be chosen as \mneu{2} and $\Delta m \equiv  \mneu{2} - m_{\neu{1}}$. 

Ref.~\cite{Agin:2025vgn} investigated the interplay between the ATLAS and CMS soft lepton searches, and the CMS monojet search, excluding the zero-lepton multijet searches since these were found in that study not to have sensitivity in the interesting parameter region, whilst also having unknown correlations with the monojet search. For the wino--bino scenario with $m_{\neu{1}}\times \mneu{2}>0$, regions of parameter space were found where the MSSM provides a better fit to the data than the SM, with $\mneu{2}\approx 320$\,GeV and $\Delta m \approx 20$\,GeV. Intriguingly, such scenarios can generate a viable \neu{1} dark matter candidate, and the dark matter properties could be improved further since the simplified model analysis is only an approximation of the richer phenomenology one could generate in a more comprehensive MSSM scenario. 

We see our strongest preference in this region centred around $\mneu{2}\approx 200$\,GeV and a mass difference of $\Delta m \approx 20$\,GeV. The largest contributions to this preference are coming from the ATLAS soft lepton search \texttt{ATLAS\_2LEPsoft\_139} and the CMS two-photon GMSB search \texttt{CMS\_2Photon\_GMSB\_36}. The photons targeted by \texttt{CMS\_2Photon\_GMSB\_36} can arise from initial-state radiation,  chargino final-state radiation or loop-decays of $\neu{2} \rightarrow \gamma + \neu{1}$. Fig.~\ref{fig:low mass excesses} (left) shows the profile likelihood in the EWMSSM model in this region. As we discuss in Appendix~\ref{app:soft_lep_validation}, the validation of the implementation of the CMS soft lepton search \texttt{CMS\_2LEPsoft\_137} in \colliderbit was not deemed sufficient to include in the scanned likelihood. Nevertheless, we show the impact of including this search on the region of interest in the centre panel of Fig.~\ref{fig:low mass excesses}. Our \texttt{CMS\_2LEPsoft\_137} likelihood contours approximately match Fig.~5 of Ref.~\cite{Agin:2025vgn} albeit with a weaker preference over the SM, which matches our expectation given that we generally over-predicted the observed confidence limits (see Fig.~\ref{fig:CMS_softlep_validation}). The total likelihood combination including the CMS soft lepton search disfavours the current best-fit point and brings the new best-fit point to $\mneu{2}\approx 295$\,GeV and $\Delta m \approx 15$\,GeV, which lies closer to the findings from Ref.~\cite{Agin:2025vgn}. Fig.~\ref{fig:low mass excesses} (right) shows the ATLAS and CMS soft lepton contributions to the total likelihood (having included the CMS soft lepton in the combination). As the preferred regions in the ATLAS search are disfavoured by the CMS search, this illustrates how we were not able to fit both excesses from the two in combination with all other searches applied in this study.

We have not included monojet searches in our scan, choosing instead to use the zero-lepton multijet searches, even though monojet searches are available in \colliderbit. The main reason behind this choice is our approach to event simulation. To generate large scale Monte Carlo samples for every point in a large parameter space sample, the event generation must be fast and relatively simple. This means that we cannot afford higher order event generation, nor running jet matching and merging. Instead, we rely on the generation of initial and final state radiation in \pythiaeight. This simplification does relatively well for the generation of extra jets up to the $p_T$ scale of around half the mass of the produced sparticles~\cite{Plehn:2005cq}. Further details on the fidelity of this simulation were discussed in~\cite{GAMBIT:2018gjo}, see in particular Sec.~3.3.2 and Figure~3 therein. Since this approach should be more reliable for the softer jets of multijet searches than the harder jets of the monojet searches, this guides our choice. Given that in Refs.  \cite{Agin:2023yoq, Agin:2024yfs, Agin:2025vgn} the authors found that CMS monojet and ATLAS soft lepton excesses could align in various supersymmetry scenarios, we expect that had we included monojet searches in this study, we would observe an increase in the significance of our best-fit regions for the \EWMSSM model.

The CMS zero-lepton searches (such as \texttt{CMS\_0LEP\_137} and \texttt{CMS\_0LEPStop\_137}) have been shown to be highly relevant for these scenarios~\cite{Buanes:2022wgm}, and in our scans these are the searches that most strongly counteract the preference coming from \texttt{ATLAS\_2LEPsoft\_139}. As these two CMS searches are correlated, we select the \texttt{CMS\_0LEPStop\_137} search for the large majority of parameter points, as it has the stronger expected exclusion. At low masses, the production cross-sections involving the two light neutralinos and the lightest chargino are sufficiently large that the relatively rare cases where initial state radiation and/or compressed hadronic decays from \neu{2}/\cha{1} to \neu{1} produce enough jets and hadronic activity to pass the cuts of \texttt{CMS\_0LEPStop\_137},  contribute some constraining power. On the other hand, we find that the ATLAS zero-lepton search \texttt{ATLAS\_0LEP\_139} is less sensitive to these low-mass scenarios, likely due to the fact that the ``softest'' ATLAS signal regions still require two or more jets where each jet must satisfy $p_T \geq 250$\,GeV. In contrast, the CMS search uses a much softer cut $p_T \geq 30$\,GeV to count jets, but then adds cuts on the scalar sum of jet $p_T$ ($H_T$), requiring $H_T \geq 300$\,GeV for the softest signal regions included in our implementation---see Appendix~\ref{app:searches}. This means that the CMS analysis can accept events with only a single quite hard jet, while contributions from softer jets ensure a jet multiplicity $\geq 2$ and help bring the scalar jet $p_T$ sum up to $H_T \geq 300$\,GeV.

Closing the theoretically well-motivated surviving parameter space at small $\Delta m$ for scenarios with degenerate wino or Higgsino states will be very challenging. While the constant improvement of analysis strategies by the LHC experiments, and the large statistics that will be available from the High-Luminosity LHC may improve current limits, the soft nature of the decay products from the produced electroweakinos limits what is achievable at the LHC.

Future $e^+e^-$ colliders such as the FCC-ee will be able to directly probe the parameter space up to masses at half their centre-of-mass energy scale, for the FCC-ee $\sqrt{s}=365$ GeV. Nevertheless, lessons from LEP show that even here degenerate scenarios will be challenging to probe. However, at the FCC-ee indirect searches through loop corrections to oblique parameters, or other electroweak precision observables such as $m_W$ and $\sin\theta_W$, will offer much stronger 
exclusion bounds from the ultra-high statistics of a Tera-$Z$ run. Recent studies have shown that 
Higgsinos with masses less than 500\,GeV and winos with masses less than 1\,TeV  can be excluded, far surpassing limits from direct searches~\cite{Nagata:2025ycf,Greljo:2025ggc}.

\subsection{Global fit results for the \texorpdfstring{\GEWMSSM}{G-EWMSSM}}
\label{sec:GEWMSSM_results}

\begin{figure*} 
  \centering
  \includegraphics[height=0.8\columnwidth]{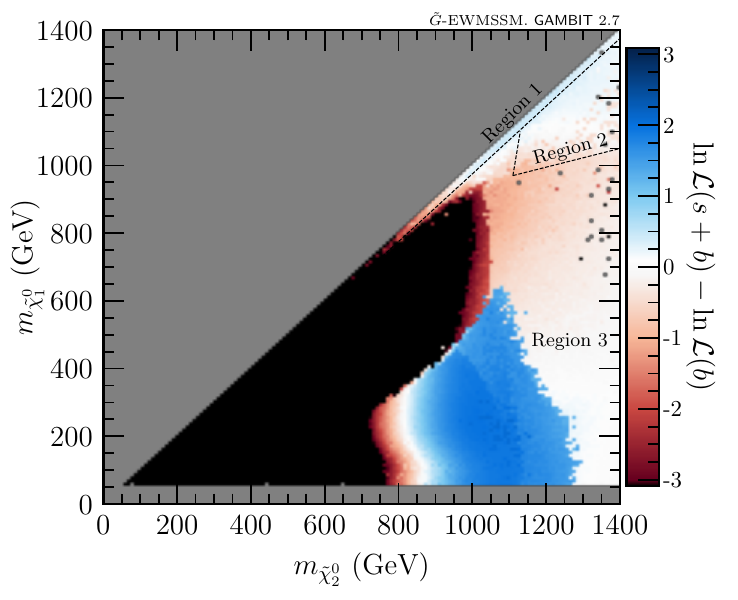}
  \includegraphics[height=0.8\columnwidth]{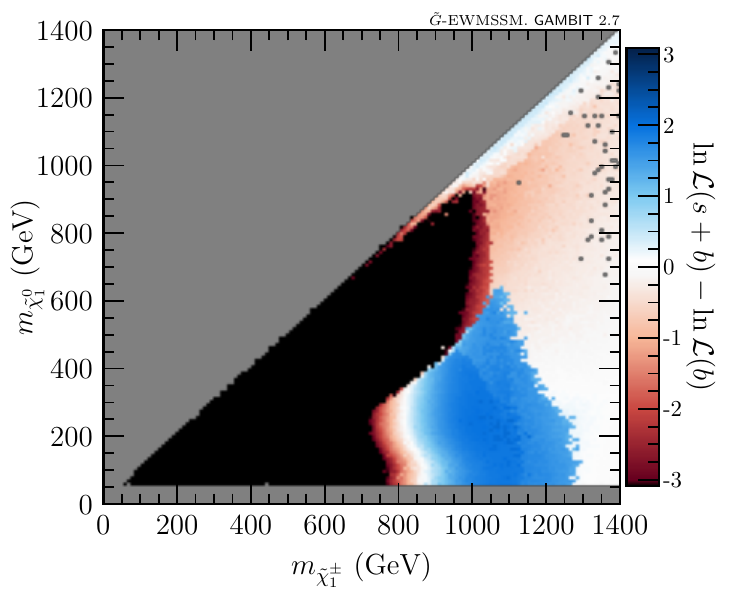}
  \caption{Profile likelihood for the $\tilde{G}$-EWMSSM model in the ($\mneu{2},\mneu{1}$) (left) and ($m_{\cha{1}},\mneu{1}$) (right) planes. Blue (red) regions are favoured (disfavoured) with respect to the background hypothesis, with black disfavoured at or beyond 2$\sigma$. Regions 1--3 as described in Sec.~\ref{sec:GEWMSSM_results} are designated and separated by black lines.
  }
  \label{fig:mN2mN1planeGEWMSSM}
\end{figure*}

\begin{figure*}
    \centering
    \includegraphics[width=0.32\linewidth]{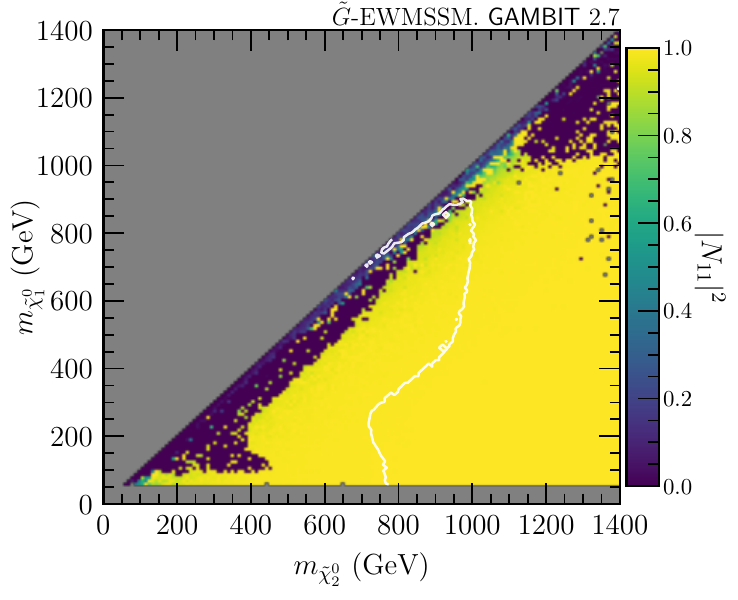}
    \includegraphics[width=0.32\linewidth]{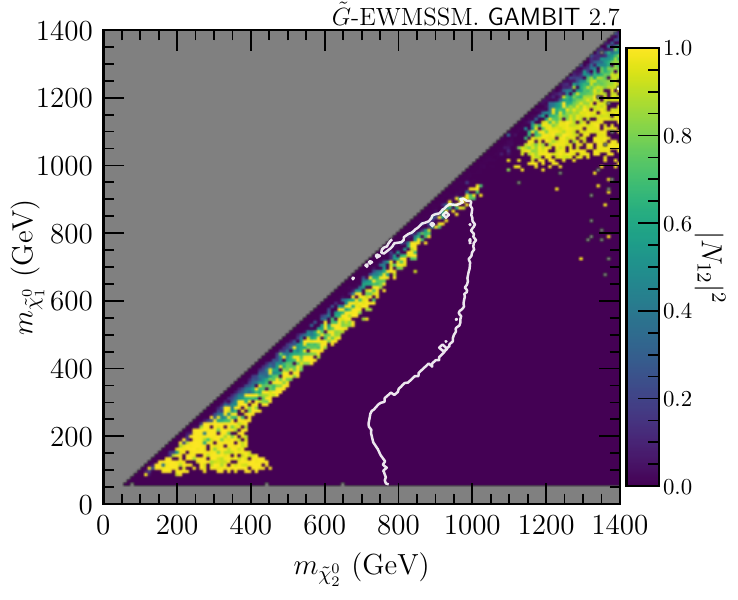}
    \includegraphics[width=0.32\linewidth]{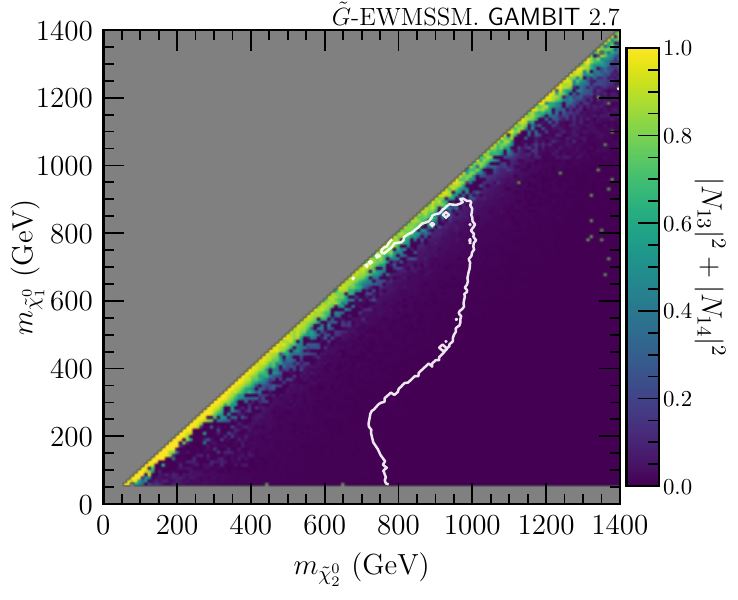}  \\ 
    \includegraphics[width=0.32\linewidth]{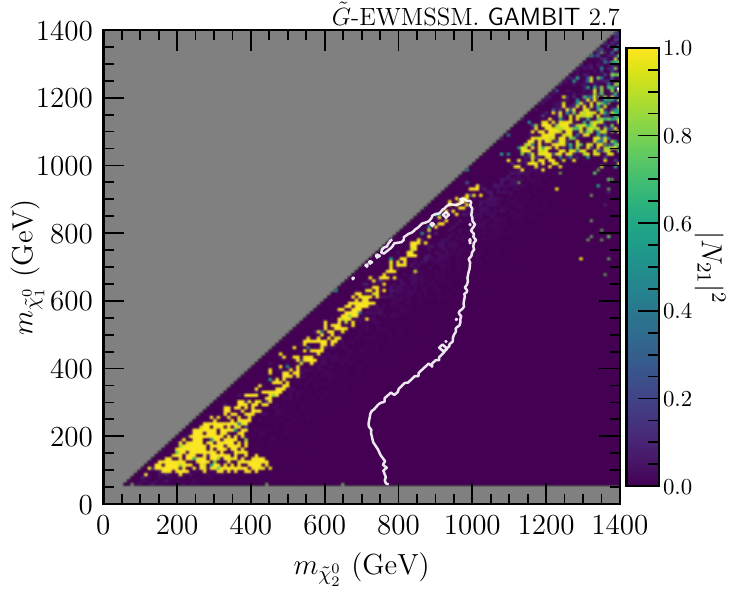}
    \includegraphics[width=0.32\linewidth]{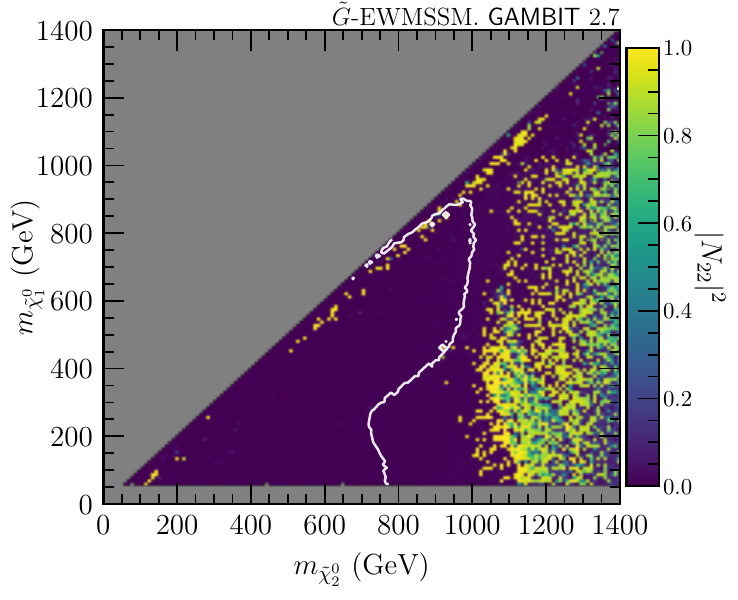}
    \includegraphics[width=0.32\linewidth]{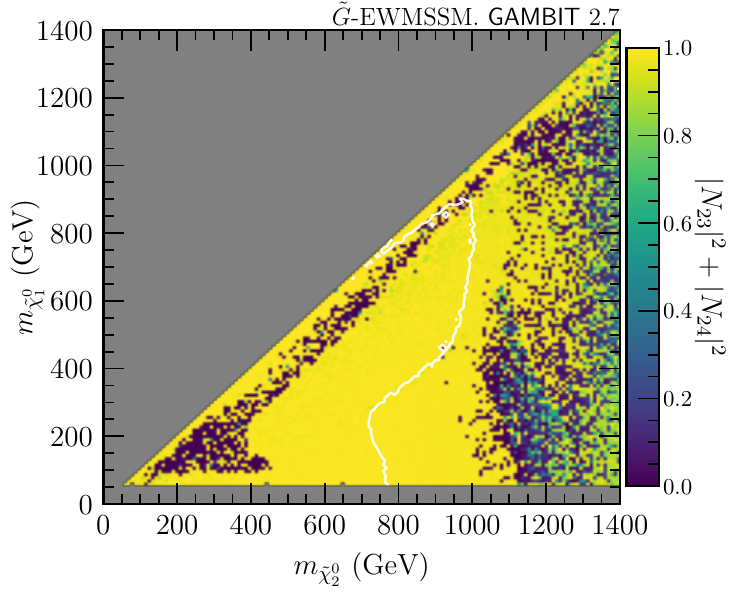}  \\
    \caption{Composition of the lightest neutralino (top) and second lightest neutralino (bottom) for the maximum likelihood $\tilde{G}$-EWMSSM parameter sample in each bin of the ($\mneu{2},\mneu{1}$) plane, i.e.\ the same samples as picked out by the profiling in Fig.~\ref{fig:mN2mN1planeGEWMSSM} (left). The colouring shows the bino fraction (left), wino fraction (middle) and Higgsino fraction (right). The white contour is the border of the 2$\sigma$ excluded region.}
   \label{fig:mN2mN1planecompGEWMSSM}
\end{figure*}

In Fig.~\ref{fig:mN2mN1planeGEWMSSM} we show the 2D profile likelihood map for the \GEWMSSM in the ($\mneu{2},\mneu{1}$) plane (left) and the ($\mcha{1},\mneu{1}$) plane (right). These results can be compared with previous \gambit results~\cite{GAMBIT:2023yih}, although it should be noted that we have extended the upper bound on the scan parameters so that the electroweakino mass parameters in the present results reach 1.5\,TeV, compared to the 1\,TeV upper bound in the previous study.\footnote{See Fig.~3 in Ref.~\cite{GAMBIT:2023yih}.} 

As in the \EWMSSM results, we structure our discussion around which different theory scenarios appear as the currently least constrained in different non-excluded regions of the ($\mneu{2},\mneu{1}$) plane. For this purpose, in Fig.~\ref{fig:mN2mN1planecompGEWMSSM} we show the bino, wino and Higgsino composition of the \neu{1} (top row) and \neu{2} (bottom row), for the same highest-likelihood points as picked out by the likelihood profiling in Fig.~\ref{fig:mN2mN1planeGEWMSSM}. Comparing these neutralino composition plots to the non-excluded region in Fig.~\ref{fig:mN2mN1planeGEWMSSM}, we can again identify three different regions in the ($\mneu{2},\mneu{1}$) plane: 
\begin{itemize}[itemsep=\medskipamount]
    \item \textbf{Region 1}: Along the diagonal, where \neu{1} is dominantly Higgsino, or, towards high masses, a Higgsino-bino mixture. As expected for such a scenario, \neu{2} is also Higgsino in this region.
    \item \textbf{Region 2}: At high masses, a wedge-shaped region at $\mneu{1} \gtrsim 1000$\,GeV and $\mneu{2} \gtrsim 1100$\,GeV, where the least constrained scenarios have wino-dominated \neu{1} and bino-Higgsino-dominated \neu{2}.
    \item \textbf{Region 3}: The remainder of the non-excluded ($\mneu{2},\mneu{1}$) plane, where the currently least constrained scenarios have a bino \neu{1}, with a \neu{2} which is either Higgsino, wino or a Higgsino-wino mix.
\end{itemize}
The nature of these regions also explains why the profile likelihood maps in the ($\mneu{2},\mneu{1}$) and ($\mcha{1},\mneu{1}$) planes (Fig.~\ref{fig:mN2mN1planeGEWMSSM}) are so similar: for almost all the scenarios picked out by the likelihood profiling, \neu{2} is either mostly Higgsino or mostly wino, and in either case the model will predict an accompanying \cha{1} with $\mcha{1} \approx \mneu{2}$.

In Region 1 we see that the lowest-mass points that (barely) survive exclusion in our analysis have $\mneu{1} \approx \mneu{2} \approx 650$\,GeV. This is a considerably stronger limit compared to what we found in our previous study of the $\tilde{G}$-EWMSSM, which showed a preference for scenarios with \mneu{1} and \mneu{2} as low as $\sim 150$\,GeV. This illustrates the considerable progress that has been made by the LHC experiments in targeting such scenarios. 

\begin{figure*}
    \centering
    \includegraphics[width=0.48\linewidth]{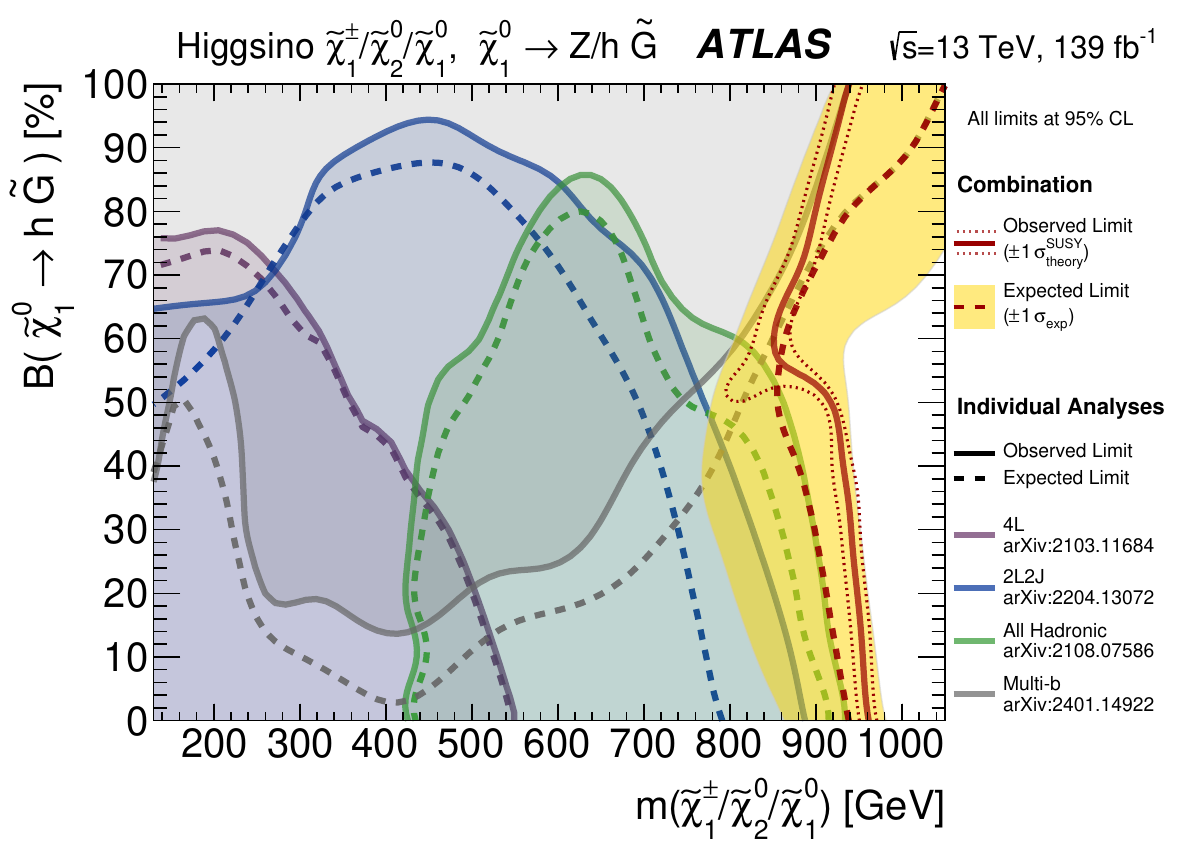}
    \includegraphics[width=0.48\linewidth]{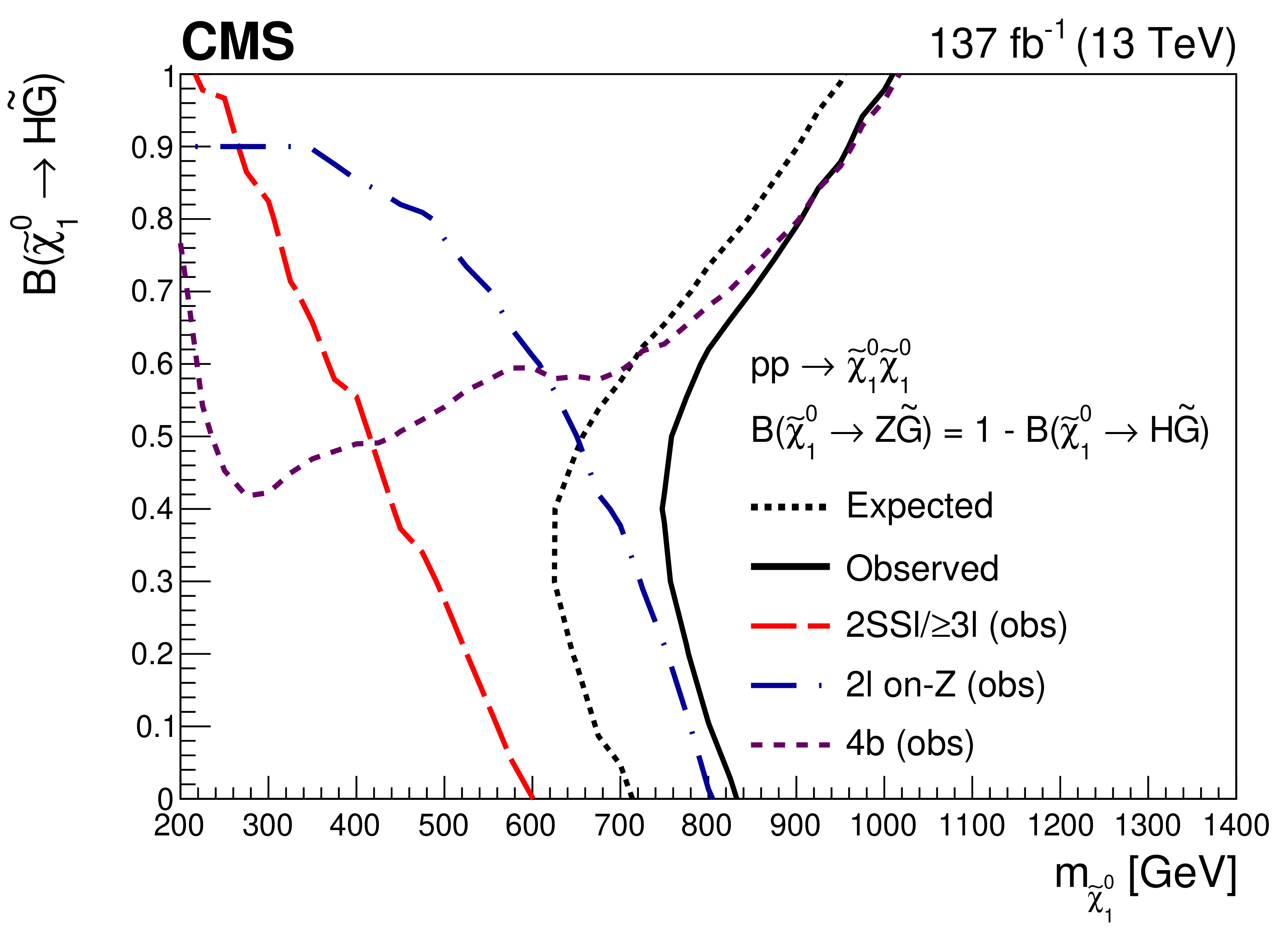} \\
    \includegraphics[width=0.48\linewidth]{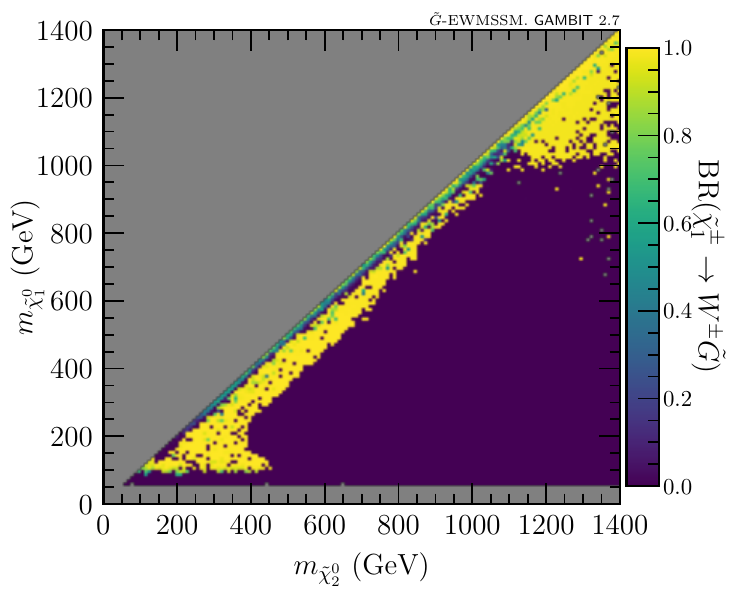}
    \caption{\textit{Top:} ATLAS~\cite{ATLAS:2024lda} (left) CMS~\cite{CMS:2024gyw} (right) simplified model exclusion limits targeting GMSB scenarios. \textit{Bottom:} Branching ratio for direct decay of the lightest chargino to the gravitino LSP, $\text{BR}(\cha{1} \rightarrow W^\pm \gravitino)$, profiled according to the total likelihood for the \GEWMSSM model.}
    \label{fig:GEWMSSM-BRs}
\end{figure*}

As the decay process $\tilde\chi^0 \rightarrow \gamma \gravitino$ at leading order depends on non-zero wino and/or bino components of the neutralino, the limit where the \neu{1} is close to 100\% pure Higgsino (the limit $|\mu| \ll |M_1|, M_2$) predicts that the NLSP \neu{1} will decay through either $\neu{1} \rightarrow Z \gravitino$ or $\neu{1} \rightarrow h \gravitino$. Close in mass to the $\neu{1}$ there will also be a \neu{2} and a \cha{1}, both pure Higgsino. This scenario has inspired ATLAS and CMS searches based on a simplified model where the production is assumed to be that of pure-Higgsino \neu{1}\neu{2}, \neu{1}\cha{1}, \neu{2}\cha{1} and \cha{1}\cha{1} pairs, and where the produced \neu{2} and/or \cha{1} decay via off-shell channels to \neu{1}, resulting in an effective production of \neu{1}\neu{1} pairs, both decaying to the gravitino LSP through either $Z\gravitino$ or $h\gravitino$. In Fig.~\ref{fig:GEWMSSM-BRs} we show summary plots from ATLAS~\cite{ATLAS:2024lda} (left) and CMS~\cite{CMS:2024gyw} (right) with exclusion limits for such a simplified model. The limits are set in the ($\mneu{1}, \text{BR}(\neu{1} \rightarrow h \gravitino)$) plane, assuming $\mneu{2} = \mcha{1} = \mneu{1}$ and $\text{BR}(\neu{1} \rightarrow Z \gravitino) = 1 - \text{BR}(\neu{1} \rightarrow h \gravitino)$. The simplified-model exclusion limits reach between $\sim800$\,GeV and $\sim1000$\,GeV, considerably higher than the limits found here. 

The ATLAS and CMS limits are based on NLO production cross-sections, while our scan uses LO+LL cross-sections. However, the main reason we observe weaker mass limits in our Region 1 compared to the ATLAS/CMS simplified model limits, is that our profiling picks out scenarios that differ from the simplified model in two key aspects. First, in the scenarios preferred by our scan, the \neu{1} will not be 100\% pure Higgsino, but retain small wino/bino components, which means there will often be a non-zero $\text{BR}(\neu{1} \rightarrow \gamma \gravitino)$, growing to around 50\% for $\mneu{1} \gtrsim 1000$\,GeV. Second, even for pure Higgsinos for masses of the near degenerate \cha{1}, \neu{2} and \neu{1} above $\sim300$\,GeV, the direct decays of the $\neu{2}$ and $\cha{1}$ down to the gravitino start competing with the decays via \neu{1}. For \neu{2} it means the decays to $h\gravitino$ and $Z\gravitino$ become competitive, but as these are the same final states as are expected to dominate from \neu{1} decays, the overall contribution is mostly to modify the predicted balance between these signals. However, the \cha{1} can decay to $W\gravitino$---in fact, for $\mcha{1} \gtrsim 700$\,GeV this decay has up to 100\% branching ratio for the scenarios our scan picks out, as can be seen along the diagonal in the bottom panel of Fig.~\ref{fig:GEWMSSM-BRs}. This enables final states $WW\gravitino\gravitino$, $hW\gravitino\gravitino$ and $ZW\gravitino\gravitino$,  rendering searches for four leptons or four $b$-jets plus missing energy less sensitive.\footnote{The \texttt{ATLAS\_4b\_139} analysis will have a reduced impact due to the lack of high-mass signal regions in our implementation. We do, however, include all signal regions for the corresponding CMS search, \texttt{CMS\_2Higgs\_4b\_neutralino\_137}, which does not give overly competitive constraints on this model.}

For the currently least constrained scenarios in Region 1, we therefore find that searches targeting signals with $W$ bosons have the most purchase. \texttt{CMS\_0LEP\_chargino\_VV\_VH\_137} and \texttt{ATLAS\_2LEPJETS\_EW\_139} are the most constraining, followed by \texttt{ATLAS\_2BoostedBosons\_139} and \texttt{CMS\_2OSLEP\_EW\_Production\_137}.\footnote{We recall that our implementation of \texttt{ATLAS\_2BoostedBosons\_139} is lacking the $b$-jet signal regions.} Small excesses in the searches \texttt{ATLAS\_1Lep2b\_139} and \texttt{ATLAS\_4b\_139} contribute to partly counter the constraints on these scenarios.\footnote{The ATLAS electroweakino search in Ref.\ \cite{ATLAS:2025uij}, which targets events with photons, jets and large missing energy, was published after our main scans and analysis were done. We have, however, added a preliminary implementation of this search in \colliderbit and investigated the impact on selected benchmark points from our scans. The ATLAS search is optimised for scenarios with the production of pairs of near-degenerate Higgsino \neu{1}/\neu{2}/\cha{1}, all of which decay via $\neu{1} \rightarrow Z/h/\gravitino$. Our tests on benchmark points indicate that this search would add some constraining power for the scenarios picked out in our Region 1, leading to a somewhat higher mass bound in this region.}

In the high-mass Region 2, the currently least constrained scenarios have wino-dominated \neu{1} and \cha{1}, and hence the dominant production is \neu{1}\cha{1} and \cha{1}\cha{1} pairs. Yet, with all electroweakino masses above $\sim1000$\,GeV, the total production cross-section is limited. The produced \cha{1} decays to $W\gravitino$ with a $\sim100\%$ branching ratio. For the \neu{1}, these scenarios predict $\text{BR}(\neu{1} \rightarrow Z\gravitino) \approx 75\%$ and $\text{BR}(\neu{1} \rightarrow \gamma\gravitino) \approx 25\%$. If produced, the heavier electroweakinos decay via $\neu{1}$ or $\cha{1}$ rather than directly to the \gravitino. In our scan, the little constraining power we find for these scenarios comes from \texttt{ATLAS\_2BoostedBosons\_139} and \texttt{ATLAS\_2LEPJETS\_EW\_139}, but this is offset by a weak preference coming from \texttt{CMS\_0LEP\_chargino\_VV\_VH\_137}.

In Region 3, which constitutes the rest of the non-excluded $(\mneu{2},\mneu{1})$ plane, scenarios identified by our scan as being in best agreement with the combined data have bino-dominated \neu{1}. These will decay dominantly as $\gamma \gravitino$. For the lowest $\mneu{1}$ we have $\text{BR}(\neu{1} \rightarrow \gamma\gravitino) \approx 100\%$. As $\mneu{1}$ increases this drops down to about 80\% for $\mneu{1} \gtrsim 400$\,GeV, with the channel $\neu{1} \rightarrow Z\gravitino$ making up the remaining 20\%. Since the cross-section for directly producing the bino pair $\neu{1}\neu{1}$ is tiny, the scenarios in Region 3 rather predict the production of pairs of heavier electroweakinos, which are mostly Higgsino, wino or a Higgsino/wino mix. None of these heavier electroweakinos decay directly to the gravitino, so all signal events, regardless of production mode, have two $\neu{1} \rightarrow (\gamma/Z)\gravitino$ decays as their last steps. For almost all of Region 3 the mass difference between \neu{1} and the heavier electroweakinos is large, so the decays of the two produced electroweakinos down to $\neu{1}\neu{1}$ give at least two on-shell $Z$/$h$/$W$. So in sum, for Region 3 almost all signal events have at least four on-shell bosons, of which up to two will be photons. 

The most eye-catching feature of Region 3 in Fig.~\ref{fig:mN2mN1planeGEWMSSM} is the dark blue region at $\mneu{1} \lesssim 600$\,GeV. Here, the model fits the data better than the SM background expectation does, due to an excess in the CMS two-photon search \texttt{CMS\_2Photon\_GMSB\_36}. A smaller excess in the similar ATLAS search \texttt{ATLAS\_PhotonGGM\_2Photon\_36} leads to a weaker than expected constraint from this search, which therefore does not offset the preference from the CMS search. Also contributing is a small, positive $\Delta \log \mathcal{L}$ contribution from \texttt{CMS\_1Photon1Lepton\_emu\_combined\_36} across most of Region 3. Moving towards lower $\mneu{2}$, the production cross-section grows and the model eventually overpredicts the observed data and is excluded.

Interestingly, the searches \texttt{CMS\_Photon\_GMSB\_137} and \texttt{ATLAS\_PhotonGGM\_1Photon\_139}, which look for signals with at least one photon, show very limited constraining power for the high-likelihood scenarios at $\mneu{1} \lesssim 600$\,GeV in Region 3, despite these searches using the full Run 2 luminosity.\footnote{As seen in Fig.~\ref{fig:Gravitino analysis correlations}, the \texttt{ATLAS\_PhotonGGM\_1Photon\_139} search is correlated above 5\% with the \texttt{ATLAS\_PhotonGGM\_2Photon\_36} search, which will tend to be the more constraining search. The \texttt{CMS\_Photon\_GMSB\_137} search, however, does not correlate with the corresponding \texttt{CMS\_2Photon\_GMSB\_36} search.} These full Run 2 searches do not have dedicated diphoton signal regions, but rather target events with a single harder photon, and also impose lepton vetos and require larger hadronic activity and/or missing energy.\footnote{Additionally, for the ATLAS search, only the hardest photon is counted towards the hadronic activity, making it even harder for our typical two-photon events to pass the selection.} Overall, the two full Run 2 searches are more aggressively optimised for the production of heavy, strongly produced sparticles, where the photon signal typically originates from the decay of a fairly heavy \neu{1}. 
The $36\,\text{fb}^{-1}$ analyses instead retain dedicated diphoton signal regions with lower photon $p_T$ thresholds, impose no requirement on hadronic activity, and have no (ATLAS) or less strict (CMS) lepton vetoes. These selection requirements better match the events predicted by our preferred Region 3 scenario, where the event energy is typically shared between two $Z$/$h$/$W$ bosons (which will often produce leptons), two gravitinos and two photons, and where the photons and gravitinos come from decays of relatively light \neu{1}. New data analyses using looser requirements on the number of leptons, or looking for leptons and photons, can potentially probe much of this parameter space. Particularly, our Region 3 scenarios predict striking signals such as $4\ell$/$4b$/$2\ell2b$ + $2\gamma$ + MET, which may be detectable despite low signal rates.\footnote{We have tested several benchmark points from the high-likelihood part of Region 3 against our preliminary implementation of the recent $140\,\text{fb}^{-1}$ ATLAS electroweakino search in Ref.~\cite{ATLAS:2025uij}, which was not included in our scans. The tests indicate that this search too has very limited sensitivity to the high-likelihood scenarios in Region 3. This is not unexpected, given that the ATLAS search targets kinematics where the mass of the decaying \neu{1} equals the mass of the produced electroweakinos, in contrast to the two-level mass hierarchy of our high-likelihood scenarios.}

In the remainder of Region 3, we have $\Delta \log \mathcal{L} \lesssim 0$. Here, the expected most sensitive signal region from \texttt{ATLAS\_PhotonGGM\_2Photon\_36} identified in our scan changes to signal regions with no observed excess, so the negative $\Delta \log \mathcal{L}$ contribution from this search and \texttt{ATLAS\_PhotonGGM\_1Photon\_139} counteracts the small preference coming from \texttt{CMS\_2Photon\_GMSB\_36}, as well as from \texttt{CMS\_1Photon1Lepton\_emu\_combined\_36}.

\begin{figure*} 
  \centering
  \includegraphics[height=0.8\columnwidth]{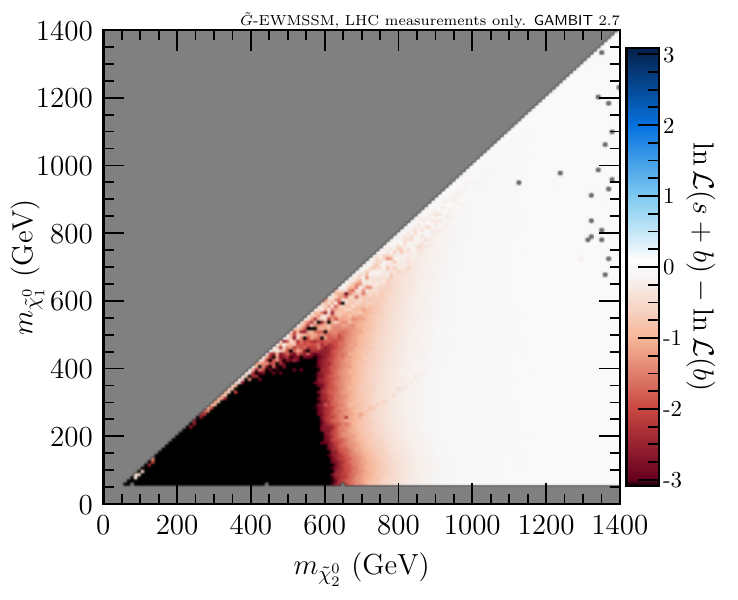}
  \caption{Measurement-likelihood contribution evaluated at the points selected by the total-likelihood profiling for the $\tilde{G}$-EWMSSM model. Points are sorted based on the total likelihood from Fig.~\ref{fig:mN2mN1planeGEWMSSM}. Red regions are disfavoured with respect to the background hypothesis, with black disfavoured at or beyond 2$\sigma$. Note that for the \contur mode used here, it is not possible for a region to be favoured over the background hypothesis.}
  \label{fig:mN2mN1plane_measurements_only}
\end{figure*}

Finally, in Fig.~\ref{fig:mN2mN1plane_measurements_only} we show the $\Delta \log \mathcal{L}$ contribution coming solely from LHC measurements of SM signatures, displayed across the parameter points picked out by the profiling in Fig.~\ref{fig:mN2mN1planeGEWMSSM}. The collection of ``SM measurements'' now constrains much more of the parameter space compared to our previous \gambit study of the \GEWMSSM. Contrast, in particular, our Fig.~\ref{fig:mN2mN1plane_measurements_only} to Figure~6 of Ref.\ \cite{GAMBIT:2018gjo}. In the current study, we used the updated set of \rivet analyses present in \contur~\textsf{3.0.0}~\cite{CONTUR:2025yis}. This includes 19 additional analyses compared to version \textsf{2.3.0}, which we used in Ref.~\cite{GAMBIT:2023yih}. Particularly strong constraints for our \GEWMSSM scenarios come from four \contur likelihood pools: \texttt{ATLAS\_13\_GAMMA}, \texttt{ATLAS\_13\_GAMMA\_MET}, \texttt{ATLAS\_13\_L1L2\_JETS} and \texttt{ATLAS\_13\_LL\_GAMMA}, which contain measurements from Refs.~\cite{ATLAS:2018nci,ATLAS:2019gey,ATLAS:2019iaa,ATLAS:2021mbt,ATLAS:2022fnp,ATLAS:2022wmu,ATLAS:2022wnf,ATLAS:2023gsl}. Unlike the \EWMSSM, whose parameter space is not as strongly probed by photon final states, the \GEWMSSM's much stronger constraints here are driven primarily by the photon measurement pools.

\subsubsection{Dependence on gravitino mass}
\label{subsec:gravitino-mass-dependence}

\begin{figure*} 
  \centering
  \includegraphics[width=0.325\textwidth]{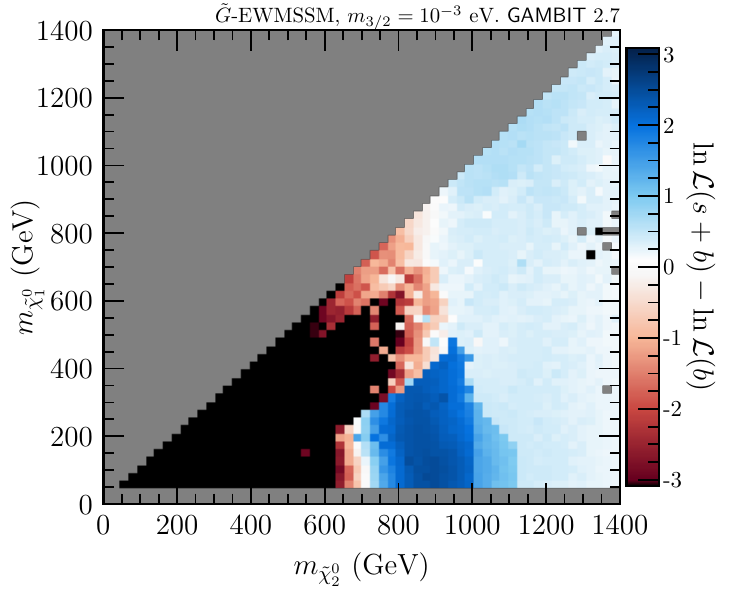} 
  \includegraphics[width=0.325\textwidth]{figures/Gravitino/FINAL_PLOTS/mN2mN1_ProfLogL}
  \includegraphics[width=0.325\textwidth]{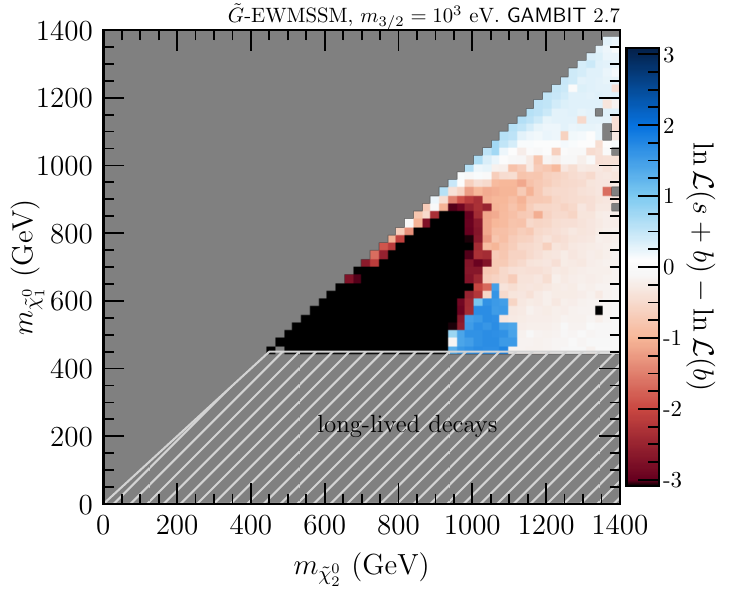}
  \caption{Profile likelihood for the $\tilde{G}$-EWMSSM model with a gravitino mass of $10^{-12}$\,GeV (left) and  $10^{-6}$\,GeV (right). Fig.~\ref{fig:mN2mN1planeGEWMSSM} is repeated here (centre) for ease of comparison. }
  \label{fig:VaryingGravitinoMass}
\end{figure*}

For our main scans of the \GEWMSSM we fixed the gravitino mass at $\mg = 1$\,eV. In Fig.~\ref{fig:VaryingGravitinoMass} we show what happens to our ($\mneu{2},\mneu{1}$) profile likelihood result when the gravitino mass is decreased (left) or increased (right) by a factor of 1000. For comparison, the middle panel shows our main result from Fig.~\ref{fig:mN2mN1planeGEWMSSM}. The results with $\mg = 1$\,meV and $\mg = 1$\,keV were obtained using looser scan settings than the main scans, resulting in roughly a factor of ten fewer parameter samples.\footnote{To be precise, the scan with $\mg = 1$\,meV has 34708 samples, while the scan with  $\mg = 1$\,keV has 30340 samples.} The plots are therefore shown in lower resolution, i.e.\ with larger bins used for the profiling, to avoid overly noisy profile likelihood maps.

For the results with $\mg = 1$\,keV (right) we have removed all parameter samples with long-lived neutralino decays (expected decay length > 1\,cm), as indicated by the hatched region at \mneu{1} below $\sim450$\,GeV. Since our simulations do not include any LHC searches targeting such long-lived scenarios, we do not present any results for this parameter region. The \mneu{1} bound where the \neu{1} starts becoming long-lived is a function of the assumed gravitino mass, coming from the $1/\mg^2$ dependence of the \neu{1} decay widths, as discussed in Sec.~\ref{sec:gravitino_model}. Other than that, the result is very similar to our main $\mg = 1$\,eV result in the middle panel.

When lowering the gravitino mass to $\mg = 1$\,meV (left), the model survives down to lower \mneu{2}, compared to the middle panel. This change is driven by a weakening in the constraints from the photons $+$ MET searches, due to the lighter gravitino mass making the direct decays of the \cha{1} and \neu{2} into gravitinos competitive, thus reducing the predicted photon signals coming from $\neu{1} \rightarrow \gamma\gravitino$. 

We note that our results for $\mg = 1$\,meV and $\mg = 1$\,keV are shown mainly to indicate in which directions our collider results would change when lowering or raising the gravitino mass. We do not consider other, non-collider constraints that can impact these scenarios, and as discussed in Sec.~\ref{sec:gravitino_model}, taking the gravitino mass as low as $\mg = 1$\,meV will be in tension with our assumption that other superpartners are generally heavy.

\section{Conclusions}
\label{sec:conclusions}

In this study, we have used \gambitVer to perform large, collider simulation-based global fits of electroweak supersymmetry scenarios, to assess the combined constraints on the production and decays of neutralinos and charginos at the LHC after Run 2. We have explored two models, \GEWMSSM and \EWMSSM, which describe the MSSM electroweak sector with four neutralinos and two charginos, with and without a light gravitino, respectively, and with all other sparticles decoupled. Our study is restricted to parameter regions where all sparticle decays are prompt.
We performed large-scale parameter scans over the four relevant MSSM parameters, $M_1$, $M_2$, $\mu$ and $\tan \beta$, with additional smaller-scale scans to investigate the dependence on the gravitino mass assumption for the \GEWMSSM.  
Our joint likelihood function is based on a comprehensive set of LHC searches, LHC measurements and LEP cross-section limits, and with a careful treatment of potential event overlap between different LHC searches. This study significantly extends our previous collider-based studies of the \EWMSSM~\cite{GAMBIT:2018gjo} and the \GEWMSSM~\cite{GAMBIT:2023yih}.

For the \EWMSSM, we see substantially larger parameter regions excluded at 2$\sigma$ confidence level compared to our earlier study. In terms of the physical neutralino masses, the exclusion reaches $\mneu{2} \sim 760$\,GeV for a light \neu{1}. Both ATLAS and CMS have observed excesses in searches targeting compressed spectra, and we observe that the \EWMSSM can partly fit these excesses. However, the excesses seen in ATLAS and CMS favour somewhat different regions of the $(\mneu{2}$,$\mneu{1})$ plane in their simplified model, and we did not find that the additional model freedom in either the \EWMSSM or the \GEWMSSM allowed for scenarios that reconciled these excesses completely within the limitations of our recasts.

For the \GEWMSSM, assuming a $\mg = 1$\,eV gravitino mass, our combination of collider results can exclude scenarios up to $(\mneu{1},\mneu{2}) \sim (900,1000)$\,GeV. For scenarios with Higgsino-dominated \neu{1} and \neu{2}, we exclude $\mneu{1} \approx \mneu{2} \lesssim 650$\,GeV---the low-mass Higgsino region preferred in our previous \GEWMSSM study has been disfavoured by the addition of newer and higher-luminosity LHC searches and measurements.

Our mass exclusion limit in this degenerate Higgsino scenario is significantly weaker than the mass limits presented by ATLAS and CMS in Higgsino-inspired simplified models, where exclusion limits reach $\sim800$--$1000$\,GeV. A key reason is that the high-mass Higgsino scenarios we identify predict the chargino decaying directly to the gravitino, $\cha{1} \rightarrow W^\pm\gravitino$, as opposed to decaying via the \neu{1}, as the simplified model used by ATLAS and CMS assumes. This effect depends on the assumed gravitino mass, with lower gravitino masses opening up more allowed parameter space. The range of gravitino masses we consider is bounded on the low side by the requirement of light superpartners through the connection between the gravitino mass and supersymmetry breaking, and on the high side by the onset of long-lived decays, where our current analysis assumes prompt decays of the NLSP.

Our best-fit region for the \GEWMSSM has $|M_1| \ll |\mu|,M_2$ and spans $\mneu{1} \lesssim 650$\,GeV and $\mneu{2} \sim 900$--$1300$\,GeV in the physical masses. The preference arises from the model fitting small excesses, mainly in a $36\,\text{fb}^{-1}$ CMS search for final states with two photons and missing energy~\cite{CMS:2019vzo}.
The preferred scenario predicts the production of two heavy Higgsino/wino electroweakinos, each initiating a cascade decay that always goes via the \neu{1}, which subsequently decays as $\neu{1} \rightarrow \gamma \gravitino$ (80--100\%) or $\neu{1} \rightarrow Z \gravitino$ (0--20\%). The predicted events therefore have at least four on-shell EW bosons (up to two photons with the remainder made up of $W/Z/h$) plus missing energy.
These scenarios avoid constraints from several photon searches with the full Run 2 dataset, in part due to the use of lepton vetoes in those searches, and in part due to the searches targeting different decay kinematics. 
While the production cross-section for heavy Higgsinos/winos is limited, new searches for signatures such as $4\ell$/$4b$/$2\ell{}2b$ + $2\gamma$ + MET could potentially constrain this model space.

Looking ahead to the results from Run 3 of the LHC, while the increase in collision energy to 13.6\,TeV  and the roughly 300\,fb$^{-1}$ integrated luminosity will both improve sensitivity to electroweakino production at higher masses, advances in search strategies have the potential to extend the reach of these limits much further. The analysis of results from Run 3 should help clarify whether the excesses currently seen by ATLAS and CMS in compressed-spectra searches, which favour different regions of the $(\mneu{2}$,$\mneu{1})$ plane, and which our global fits only partially accommodate, are statistical fluctuations or hints of a common underlying signal, particularly as both experiments continue to refine their analysis strategies for these soft, compressed final states. 

The natural future extension of this work is the inclusion of long-lived particle searches, requiring careful propagation of event vertex information from the \pythia event generator through to the \colliderbit analysis implementation. The maturing displaced-vertex and disappearing-track programmes at ATLAS and CMS during Run~3 will be vital for probing the long-lived NLSP scenarios discussed in Sec.~\ref{subsec:gravitino-mass-dependence}. More broadly, as Run 3 searches and measurements become available, repeating this global-fit approach will be the most direct way to track how the preferred and excluded regions for both the \EWMSSM and \GEWMSSM models evolve, and we regard the present study as a baseline against which such future updates can be compared.

Taken together, these results provide a comprehensive state of the art status of light electroweakinos at the LHC. All parameter samples, \gambit input files and plotting scripts can be found on the public Zenodo record associated with this publication~\cite{Zenodo_SUSYRun2}.

\begin{acknowledgements}
\end{acknowledgements}

We thank our colleagues in the GAMBIT Community for many helpful discussions and comments. We would also like to thank Jason Lea Oliver for useful conversations. 

AK, AR, CC and TK were supported by the Research Council of Norway through the FRIPRO grant 323985 PLUMBIN’. 
AB's contributions were enabled by the CHIST-ERA OpenMAPP network, grant number EP/Y036360/1. 
The work of AJ is supported by Korea Institute for Advanced Study (KIAS) grant funded by the Korean government. 
The research of CB is supported by the Australian Research Council via DP220100643 and LE250100010.
FM is funded in part by the National Research Agency (ANR) under project no. ANR-21-CE31-0002-01.
The work of R.~RdA was supported by PID2023-148162NB-C22 from the Spanish Ministerio de Ciencia e Innovación and by the PROMETEO/2022/69 from the Spanish GVA. 
MJW is supported by the Australian Research Council grants CE200100008 and DP220100007. 
PA and YZ are supported by the National Natural Science Foundation of China (NSFC) Key Projects grant No.\ 12335005. 
PZ is supported by the Australian Research Council grants CE200100007. 
VH has been primarily supported by the DOE grant DE-SC0007861. 
AF was supported by the National Natural Science Foundation of China (NNSFC) RFIS-II W2432006. 
GM is supported under award No.\ 11458 from the Gordon \& Betty Moore Foundation and funds from Andrea Ghez via the Lauren B.\ Leichtman \& Arthur E. Levine Chair in Astrophysics.

We gratefully acknowledge the EuroHPC Joint Undertaking for awarding this project access to the EuroHPC supercomputer LUMI, hosted by CSC (Finland) and the LUMI consortium, through a EuroHPC Extreme Scale Access call.
The authors also gratefully acknowledge the access to computer resources at MareNostrum and technical support provided by Barcelona Supercomputing Center (RES-FI-2025-2-0043, RES-FI-2025-3-0070 and RES-FI-2026-1-0043), as well as computing resources provided by Sigma2---the National Infrastructure for High-Performance Computing and Data Storage in Norway, through allocation NN9284k.


\begin{appendices}
\renewcommand \thetocsection {\Alph{section}}

\section{Profile likelihood maps for the input parameters} \label{app:input_params}

\begin{figure*} 
  \centering
  \includegraphics[width=0.32\linewidth]{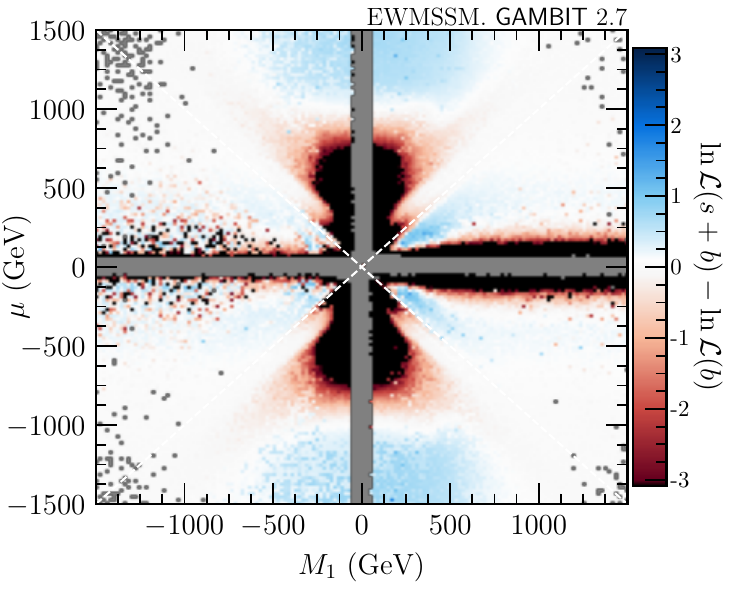}
  \includegraphics[width=0.32\linewidth]{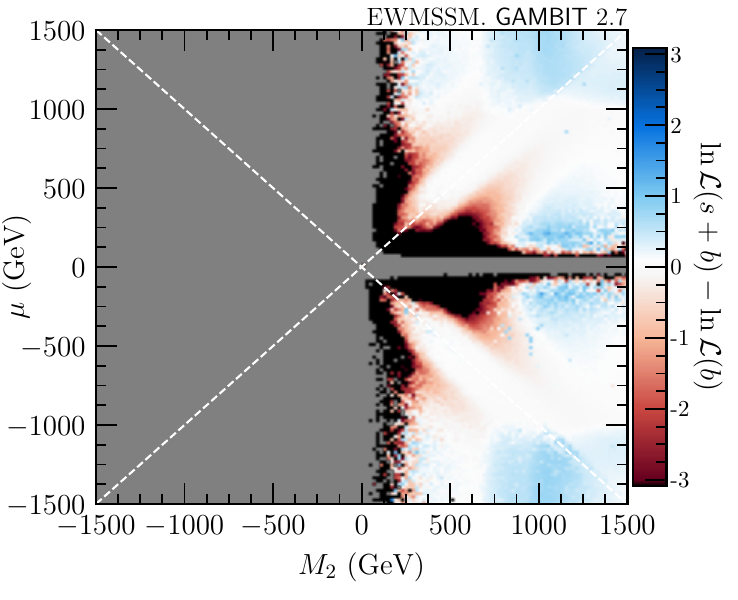}
  \includegraphics[width=0.32\linewidth]{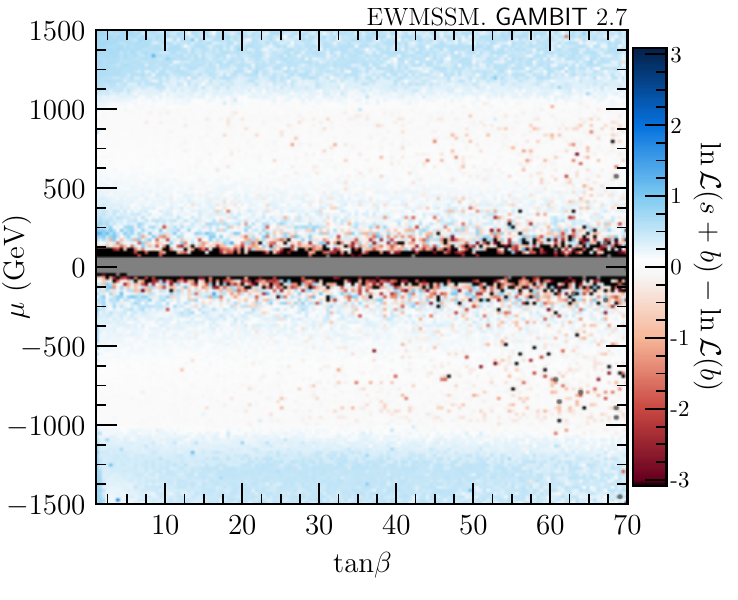}
  \caption{\label{fig:EWMSSM_pars}Profile likelihoods in the $(M_1, \mu)$, $(M_2, \mu)$, and $(\tan\beta, \mu)$ planes for the EWMSSM model. Blue (red) regions are favoured (disfavoured) with respect to the background hypothesis, with black indicating exclusion at or beyond $2\sigma$. }
\end{figure*}

\begin{figure*} 
  \centering
  \includegraphics[width=0.32\linewidth]{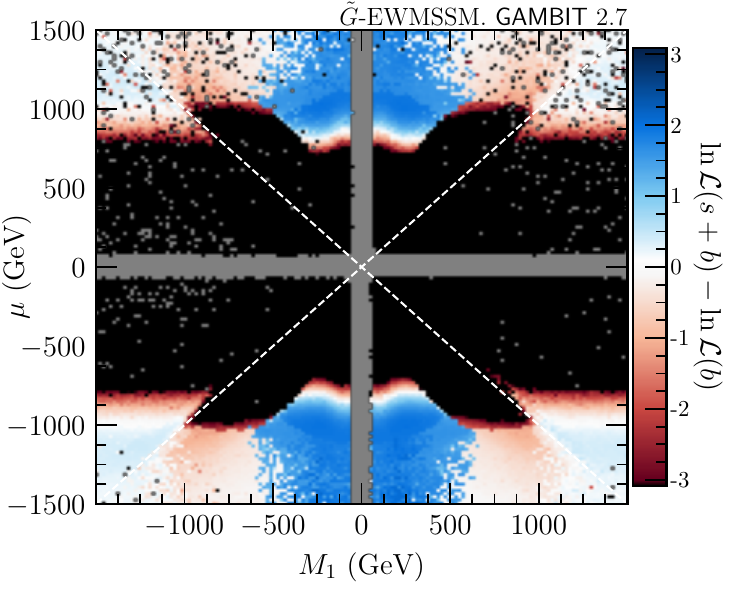}
  \includegraphics[width=0.32\linewidth]{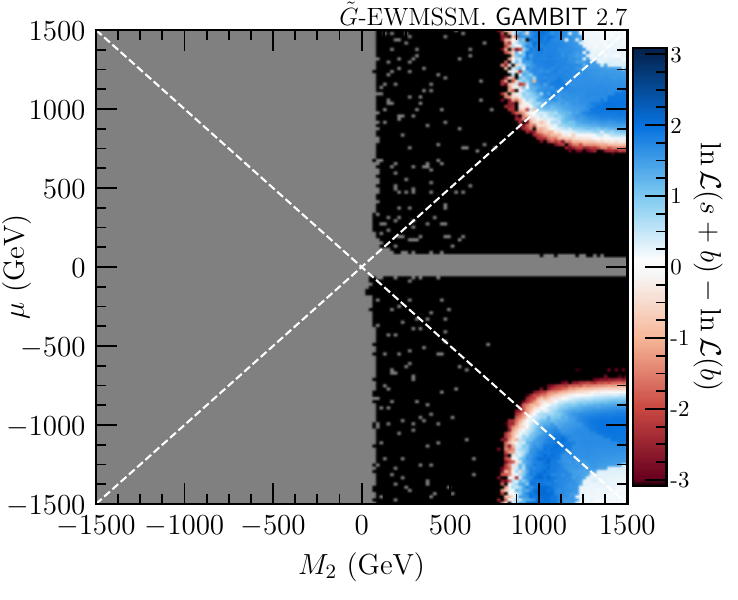}
  \includegraphics[width=0.32\linewidth]{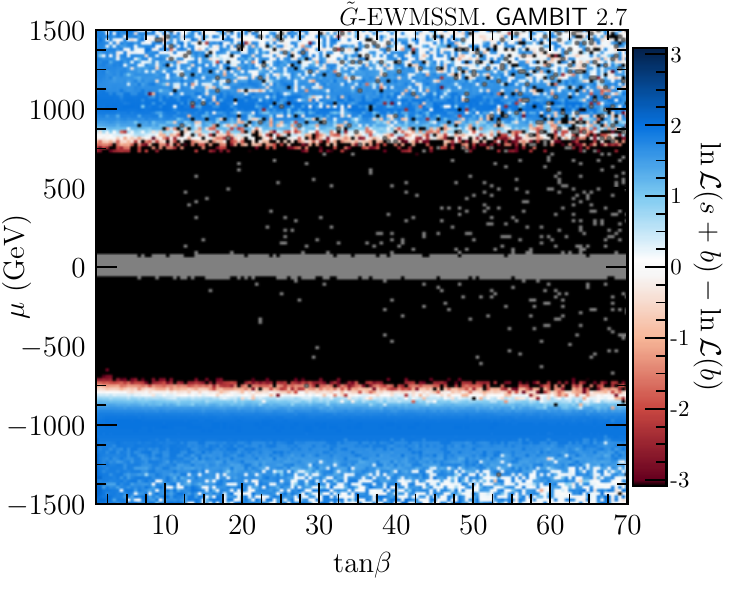}
  \caption{\label{fig:GEWMSSM_pars}Profile likelihoods in the ($M_1$, $\mu$), ($M_2$, $\mu$), and ($\tan{\beta}$, $\mu$) planes for the $\tilde{G}$-EWMSSM model. Blue (red) regions indicate parameter space favoured (disfavoured) relative to the background hypothesis, while black regions are disfavoured at or beyond the 2$\sigma$ level.
  }
\end{figure*}

In Fig.~\ref{fig:EWMSSM_pars}, we show profile likelihood results for three different planes of the \EWMSSM Lagrangian parameters. The different regions we introduced when discussing the profile likelihood map for the $(\mneu{2},\mneu{1})$ plane in Sec.~\ref{sec:results} can be connected with the parameter maps in Fig.~\ref{fig:EWMSSM_pars}. Region 1, where the two lightest neutralinos are near mass degenerate, have $|\mu| < |M_1|, M_2$. The highest-likelihood points in this region can be seen as the preferred blue region at low $|\mu|$ in all three panels of Fig.~\ref{fig:EWMSSM_pars}.
Several different parameter regions map to scenarios picked out in our Region 2 of the $(\mneu{2},\mneu{1})$ plane, where the mass splitting is $\mneu{2} - \mneu{1} < 200$\,GeV. For instance, many of the lowest-mass scenarios in Region~2 correspond to points with $|M_1| < |\mu| < M_2$, and can be seen in the left panel of Fig.~\ref{fig:EWMSSM_pars} as the four white/red regions with $|\mu| \approx |M_1| + 100$\,GeV, extending inwards towards low $|\mu|$ and $|M_1|$.
Region 3, where $\mneu{2} - m_{\neu{1}} > 200$\,GeV, and the lightest neutralino is almost entirely bino, corresponds to parameter regions with $|M_1| \ll |\mu|, M_2$, e.g.\ the upper and lower regions of the left-hand panel.

Fig.~\ref{fig:GEWMSSM_pars} shows the corresponding profile likelihood maps for the Lagrangian parameters of the \GEWMSSM. The exclusion is much stronger than in our previous \GEWMSSM study \cite{GAMBIT:2023yih}. In the previous study, the best-fit region had $|\mu| < |M_1|, M_2$, with $\mu \sim -150$\,GeV and $\tan\beta \sim 1$. In our current combination the preferred region has $|M_1| \ll |\mu|, M_2$, with no preference toward any particular $\tan \beta$ values. We exclude $|M_2| \lesssim 750$\,GeV and $|\mu| \lesssim 700$\,GeV at the 2$\sigma$ confidence level. 

Again we can connect these parameter planes to the three regions we used in Sec.~\ref{sec:GEWMSSM_results} to discuss the results for the $(\mneu{2},\mneu{1})$ plane of the \GEWMSSM. Our Region 1, along the diagonal of the $(\mneu{2},\mneu{1})$ plane, corresponds to the four regions with $|\mu| \lesssim |M_1|$ in the left-hand panel. For the high-mass Region 2, the contributing parameter points mainly have $M_2 < |M_1|, |\mu|$, with $M_2 \gtrsim 1000$\,GeV. Due to having $\Delta \log \mathcal{L} \sim 0$, these parameter points are largely hidden by the profiling in the $(M_2,\mu)$ panel (middle). In the $(M_1,\mu)$ panel (left), points from the outer corners contribute to Region 2. Finally, Region 3, including the dark blue region of positive $\Delta \log \mathcal{L}$, corresponds to where $|M_1| \ll |\mu|, M_2$, e.g.\ the upper and lower parts of the left-hand panel.

\section{Updates to \colliderbit} \label{app:CB_updates}
In addition to the new implementations of LHC searches from Run 2 presented in Appendix~\ref{app:searches}, we detail here some other changes that have been made to the \colliderbit module in \gambitVer released with this paper, as well as additional information on the validation of some more challenging searches.

\subsection{Interface to \texorpdfstring{\pythiaeight}{Pythia 8}} \label{app:pythia_8_3_update}
The three body decays of electroweakinos into fermion pairs and a lighter electroweakino proceeding through a (potentially off-shell) vector boson: $\tilde\chi^0_i\to f\bar f\tilde\chi^0_j$, $\tilde\chi^\pm_i\to f\bar f'\tilde\chi^0_j$, $\tilde\chi^0_i\to f\bar f'\tilde\chi^\pm_j$, and $\tilde\chi^\pm_i\to f\bar f\tilde\chi^\pm_j$, will have correlations between the kinematics of the decay products due to the spin structure. These decays are currently being modelled as isotropic decays in \pythiaeight. This can have important phenomenological consequences, as pointed out in Ref.~\cite{Agin:2024yfs}, since it affects distributions like the invariant mass of the final state leptons, and thus directly impacts the cuts of some searches. In particular this is important for the searches discussed below in Appendix~\ref{app:soft_lep_validation}.

Starting with the version of \colliderbit found in \gambitVer, we update the version of \pythia used from \textsf{8.212} to \textsf{8.312}, and implement a patch to its decay routines that includes the application of the full leading-order matrix element for all four electroweakino decay channels. The only approximation used is that the final state SM fermions are kept massless. We have checked that this reproduces the expected invariant mass distributions of SM leptons in the electroweakino decays for both the Higgsino and wino/bino scenarios. This functionality is turned on by, in the steering \textsf{YAML} file \yaml{pythia_settings}, adding  \yaml{- SUSYResonance:3BodyMatrixElement = on}. This completely replaces the current effect of that setting in \pythia.

\subsection{Soft lepton validation} \label{app:soft_lep_validation}

\begin{figure*} 
  \centering
  \includegraphics[width=0.49\textwidth]{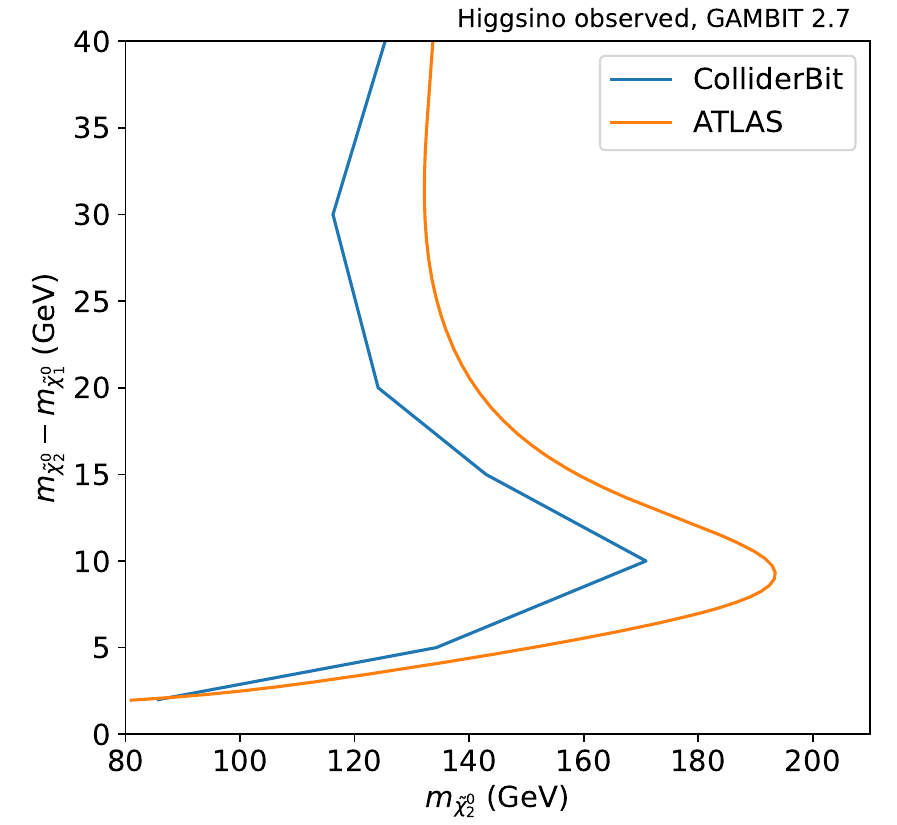}
  \includegraphics[width=0.49\textwidth]{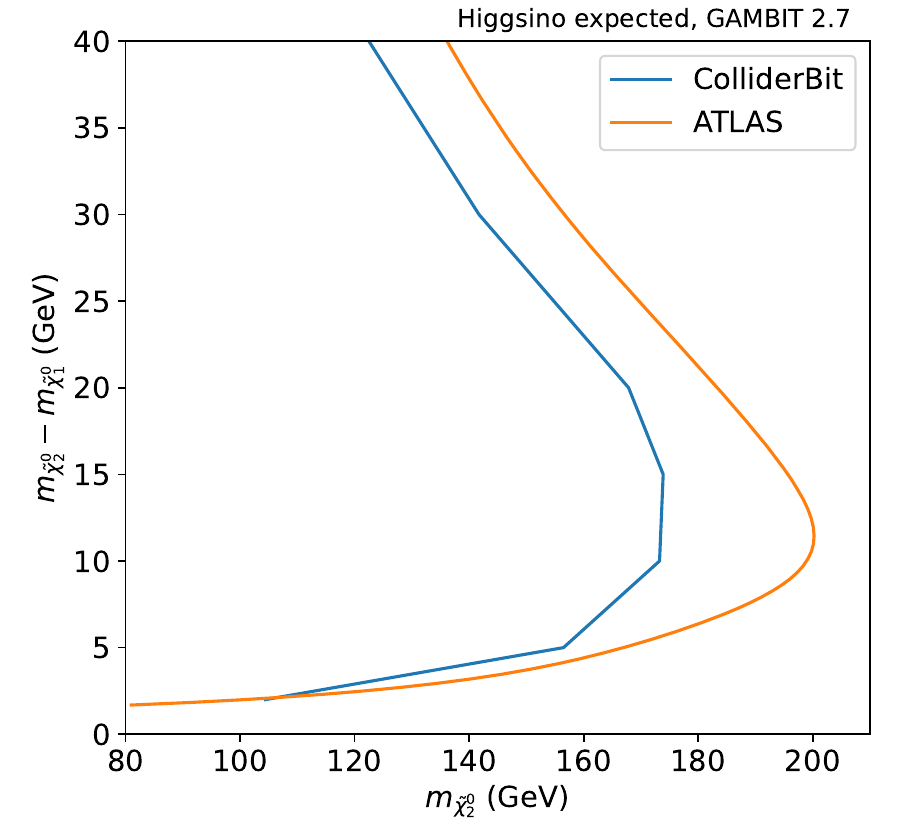}
  \caption{
  Comparison of the 95\% observed (left) and expected (right) confidence upper limit on the mass of the second lightest neutralino between ATLAS~\cite{ATLAS:2019lng} (orange) and \colliderbit's \texttt{ATLAS\_2LEPsoft\_139} implementation (blue) for the Higgsino signal model.
  }
  \label{fig:ATLAS_softlep_validation}
\end{figure*}

Fig.~\ref{fig:ATLAS_softlep_validation} shows the bound on the mass of the second lightest neutralino when the mass difference between the two lightest neutralinos is small, using the \textsf{ATLAS-SUSY-2018-16} search for a compressed spectrum with two soft leptons~\cite{ATLAS:2019lng}. The limits are presented for a direct Higgsino production simplified signal model where all the three lightest electroweakinos are purely Higgsino-like. This comparison relies on pre-computed precision electroweak cross-sections provided by the LHC SUSY cross-section working group~\cite{Fuks:2012qx, Fuks:2013vua}. \colliderbit's implementation of this search, using the modifications to the electroweakino decays detailed in Appendix~\ref{app:pythia_8_3_update}, manages to reproduce the general trend, albeit with a consistently more conservative limit. Given the differences in detector simulation and event generation between the ATLAS result and ours we do not expect to exactly reproduce their limits, and so the level of agreement seen in our validation against the ATLAS results was deemed sufficient the purpose of validating our implementation.

\begin{figure*} 
  \centering
  \includegraphics[width=0.49\textwidth]{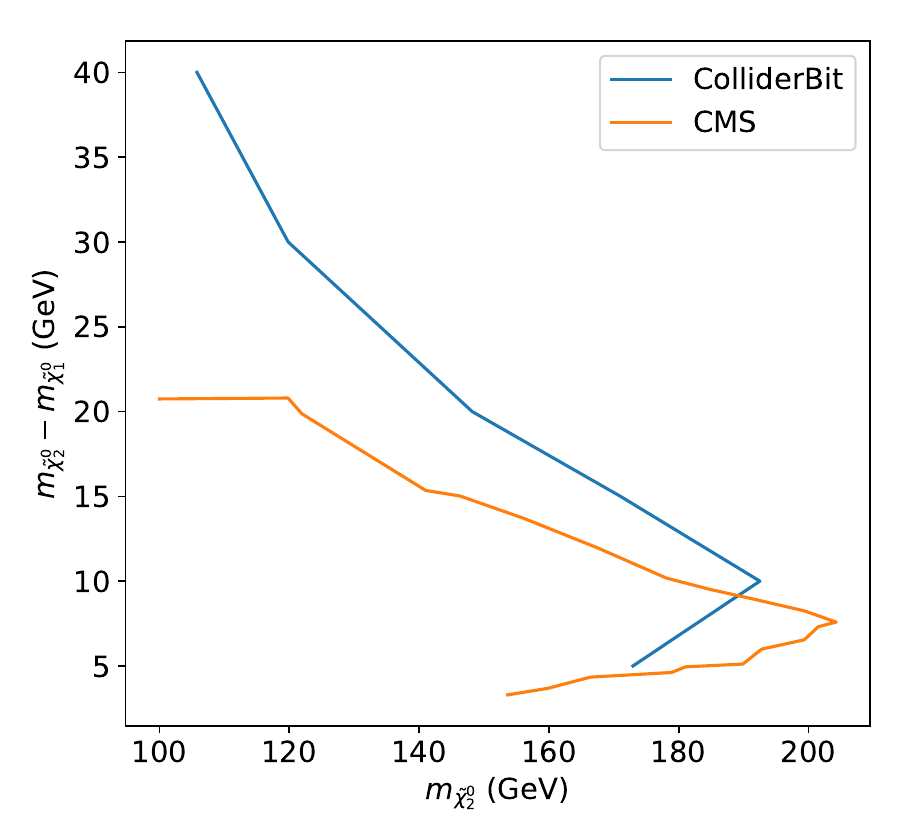}
  \includegraphics[width=0.49\textwidth]{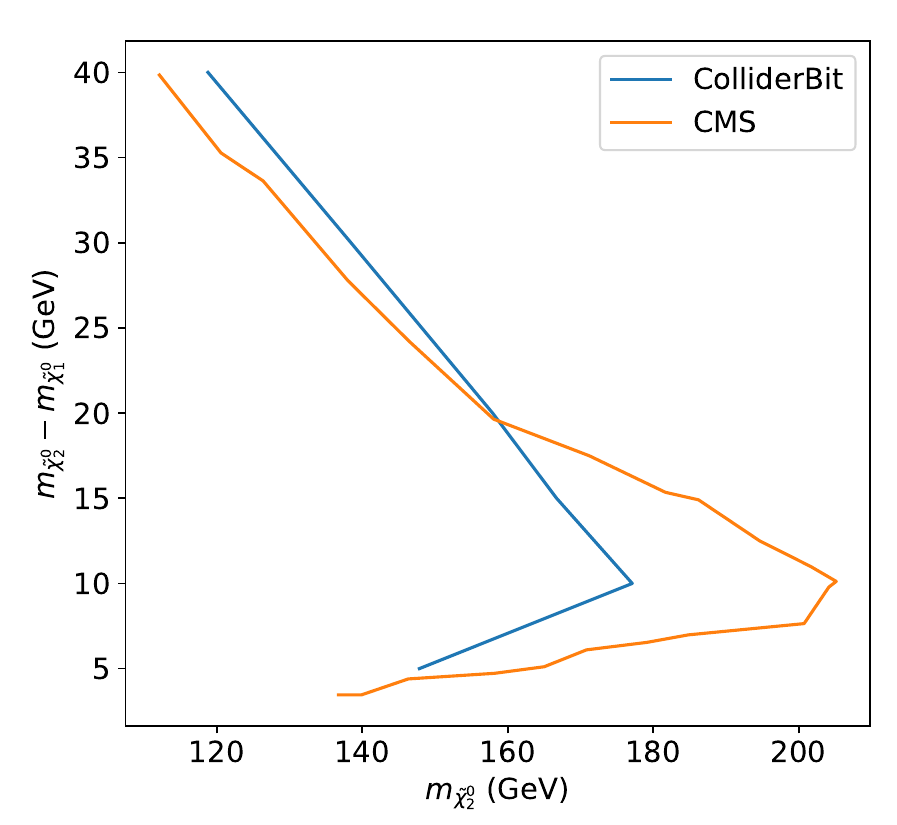}
  \caption{
  Comparison of the 95\% observed (left) and expected (right) confidence upper limit on the mass of the second lightest neutralino between CMS~\cite{CMS:2021edw} (orange) and  \colliderbit's \texttt{CMS\_2LEPsoft\_137} implementation (blue) for the Higgsino signal model.
  }
  \label{fig:CMS_softlep_validation}
\end{figure*}

This level of agreement was not found with the corresponding \texttt{CMS\_2LEPsoft\_137} analysis for two soft leptons~\cite{CMS:2021edw}. Fig.~\ref{fig:CMS_softlep_validation} shows the equivalent plots to Fig.~\ref{fig:ATLAS_softlep_validation}, for the CMS results. Here we predicted much stronger observed limits, in particular for mass differences larger than 20\,GeV, where the CMS limits weaken. Given the uncertainty in our predictions, we would prefer to lie on the side of conservative limits. We thus decided that our validation was not sufficiently accurate to include the \texttt{CMS\_2LEPsoft\_137} search in the total likelihood used to drive the scanner. We did simulate it alongside other collider likelihoods, and the results are contained in our publicly released simulated samples~\cite{Zenodo_SUSYRun2}. Fig.~\ref{fig:mN2mdiffplane} shows the impact that including this would have on the \EWMSSM model results.

\subsection{Multiple jet collections} \label{app:multiple_jet_collections}

\colliderbit and the version of \heputils that it includes have been extended to allow the storage of multiple jet collections with different jet algorithms/settings accessible to every analysis. Previously, this was possible only by performing the reclustering of jets inside of a particular \colliderbit analysis class, as it processes each event. Allowing multiple jet clusterings to run and be stored in the event record provides a speed-up in cases where two or more jet settings are each used within multiple \colliderbit analyses.

To allow for this, the \textsf{YAML} settings for event simulation have been updated, and an example setting choice for the \textsf{HardScatteringSim} capability is shown below using two anti-kt jet clustering algorithms with different radii. The name of each jet collection in the \textsf{YAML}, e.g.\ \yaml{antikt_R04}, should match those expected by the analyses. Inside of a given \colliderbit analysis, jets are accessed with a tag for the jet collection using the pointer to the event object, in our example \cpp{event->jets("antikt_R04")}. In the event that the required jet collection has not been run, \colliderbit  will throw a runtime exception. In addition to the \yaml{jet_collections} setting, an additional required setting \yaml{jet_collection_taus} has been added. This is required in order to determine which jet collection will be used for tau tagged jets.

\begin{lstyaml}
- capability:  HardScatteringSim
  type: Py8Collider_defaultversion
  function: getPythia
  options:
    LHC_13TeV:
    xsec_veto: 0.007 # corresponds to ~1 expected event at L = 139 fb^-1.
    jet_collections:
      antikt_R04:
        algorithm: antikt
        R: 0.4
        recombination_scheme: E_scheme
        strategy: Best
      antikt_R08:
        algorithm: antikt
        R: 0.8
        recombination_scheme: E_scheme
        strategy: Best
    jet_collection_taus: antikt_R04
\end{lstyaml}

\subsection{Signal region overlap} \label{app:signal_region_overlap}

\colliderbit has been extended to drop CSV files (one for ATLAS analyses and one for CMS analyses) which contain an account of which signal regions are filled by each event for every analysis. The purpose of this is to allow the user to measure the level of overlap between signal regions, and determine the correlation between different signal regions or analyses. A utility script \textsf{ColliderBit/scripts/generate\_correlation\_plots.py} is provided by \gambit, which allows the user to generate correlation estimates such as Figs.~\ref{fig:EWMSSM analysis correlations} and ~\ref{fig:Gravitino analysis correlations}. Unlike final signal sums, keeping a permanent record of every event may cause prohibitively large disk usage when run during a large global fit. It is for this reason that we recommend that this setting only be used when running at most a handful of parameter points. The \textsf{YAML} rule required to instruct \colliderbit to drop these files is:

\begin{lstyaml}
# Tell ColliderBit to drop accepted events CSV for all signal regions
  - if:
      capability: any
    then:
      options:
        drop_accepted_events_file: true
\end{lstyaml}

\subsection{Neural nets and BDT implementations} \label{app:NN_BDT_implemenations}

In the implementation of the \texttt{ATLAS\_2LEP0JET\_EW\_139} analysis (describing the \textsf{ATLAS-SUSY-2019-02} search), \colliderbit has been extended to incorporate externally trained Boosted Decision Trees (BDTs) provided by ATLAS in the form of ROOT weight files. During the initialisation of the analysis, these weight files are registered and linked to internal TMVA readers~\cite{TMVA:2007ngy}, which allow the trained classifiers to be accessed by name within the analysis. For each event, the relevant kinematic variables (such as lepton transverse momenta, missing transverse energy, dilepton invariant mass, angular separations, etc.) are computed and combined into feature vectors that match the training setup of the ATLAS BDTs. These feature vectors are then passed to the TMVA readers, which evaluate the BDTs and return classifier scores. The output scores are compared against thresholds defined by ATLAS to determine the signal region(s) into which an event falls. In this way, the trained BDT weights are applied directly within \colliderbit to reproduce the multivariate selections of the experimental analysis.

\subsection{Addition of new LEP analysis} 
\label{app:lep_extension}
We extend the existing LEP searches in \colliderbit by the L3 multi-photon and missing energy search~\cite{L3:2003yon}. For light enough neutralinos this places stringent constraints on the cross section for the process $e^+ e^- \to \neu{1} \neu{1} \to \gravitino \gravitino \gamma \gamma$. We apply the $95\%$ limits shown in Figure~6c of \cite{L3:2003yon} following the treatment described in~\cite{ColliderBit}.

\newpage

\section{LHC searches}
\label{app:searches}

\begin{table*}
\setlength\tabcolsep{10pt}\renewcommand{\arraystretch}{1.0}
\centering
\resizebox{\linewidth}{!}{
\begin{tabular}{rllcl}
\toprule
No. &    Analysis label in \colliderbit & Report Number  & $\mathcal{L}$ [${\rm fb}^{-1}$]  & Source \\ \midrule
\rownumber&   \texttt{ATLAS\_13TeV\_1Lep2b\_139invfb}    & \textsf{ATLAS-SUSY-2019-08} & 139 & ATLAS $Wh$ search~\cite{ATLAS:2020pgy} \\
\arrayrulecolor{lightgray}\hline
\rownumber&   \texttt{ATLAS\_13TeV\_2LEP0JET\_EW\_139invfb}    & \textsf{ATLAS-SUSY-2019-02} & 139 & ATLAS 2 OS lepton search~\cite{ATLAS:2022hbt} \\ 
\hline  
\rownumber&    \texttt{ATLAS\_13TeV\_2LEPJETS\_EW\_139invfb} & \multirow{2}{*}{\textsf{ATLAS-SUSY-2018-05}} & \multirow{2}{*}{139} & \multirow{2}{*}{ATLAS 2 lepton, jets search~\cite{ATLAS:2022zwa}} \\
&    \texttt{ATLAS\_13TeV\_2LEPJETS\_RJR\_139invfb} \\
\hline
\rownumber&    \texttt{ATLAS\_13TeV\_3LEP\_eRJR\_139invfb} & \textsf{ATLAS-SUSY-2018-06} & 139 & ATLAS 3 lepton search using eRJR technique~\cite{ATLAS:2019wgx} \\ 
\hline    
\rownumber&    \texttt{ATLAS\_13TeV\_2OR3LEP\_139invfb} & \textsf{ATLAS-SUSY-2019-22} & 139 & ATLAS 2 or 3 leptons search~\cite{ATLAS:2023lfr} \\
\hline
\rownumber&    \texttt{ATLAS\_13TeV\_bTaus\_StopStau\_139invfb} & \textsf{ATLAS-SUSY-2019-18} & 139 & ATLAS $\tau$-lepton, $b$-jets search for stop~\cite{ATLAS:2021jyv}  \\
\hline
\rownumber&    \texttt{ATLAS\_13TeV\_1OR3LEP\_StopHZ\_139invfb} & \textsf{ATLAS-SUSY-2018-21}     & 139   & ATLAS Higgs or Z boson search for stop~\cite{ATLAS:2020aci} \\ 
\hline
\rownumber&    \texttt{ATLAS\_13TeV\_4b\_139invfb} & \textsf{ATLAS-SUSY-2020-16}     & 139   & ATLAS 2 Higgs boson search, using 4 $b$-jets~\cite{ATLAS:2024tqe}     \\ 
\hline
\rownumber&    \texttt{ATLAS\_13TeV\_2LEPsoft\_139invfb} & \textsf{ATLAS-SUSY-2018-16}       & 139   & ATLAS 2 soft-leptons search for compressed spectrum~\cite{ATLAS:2019lng} \\ 
\hline
\rownumber&    \texttt{ATLAS\_13TeV\_0LEP\_139invfb}   & \textsf{ATLAS-SUSY-2018-22}       & 139   & ATLAS $0\ell$, jets search for squark and gluino~\cite{ATLAS:2020syg} \\ 
\hline 
\rownumber&    \texttt{ATLAS\_13TeV\_2BoostedBosons\_139invfb} & \textsf{ATLAS-SUSY-2018-41} & 139   & ATLAS 2 boosted bosons search~\cite{ATLAS:2021yqv} \\
\hline
\rownumber&    \texttt{ATLAS\_13TeV\_2LEPStop\_inclusive\_139invfb}    & \textsf{ATLAS-SUSY-2018-08} & 139  & ATLAS search for stop in the di-lepton channel~\cite{ATLAS:2021hza} \\
\hline
\rownumber&    \texttt{ATLAS\_13TeV\_2OSLEP\_chargino\_inclusive\_139invfb}    & \textsf{ATLAS-SUSY-2018-32}  &   139 &   ATLAS 2 lepton chargino search~\cite{ATLAS:2019lff} \\ 
\hline
\rownumber&    \texttt{ATLAS\_13TeV\_3LEP\_139invfb}   & \textsf{ATLAS-SUSY-2019-09} &   139 & ATLAS 3 lepton search for chargino-neutralino pair~\cite{ATLAS:2021moa} \\
\hline
\rownumber&    \texttt{ATLAS\_13TeV\_4LEP\_139invfb}   & \textsf{ATLAS-SUSY-2018-02}  & 139   & ATLAS 4 lepton search~\cite{ATLAS:2021yyr} \\
\hline 
\rownumber&    \texttt{ATLAS\_13TeV\_2bMET\_36invfb}   & \textsf{ATLAS-SUSY-2016-28} & 36.1    & ATLAS 2 $b$-jets search~\cite{ATLAS:2017avc}   \\
\hline
\rownumber&    \texttt{ATLAS\_13TeV\_MultiLEP\_strong\_139invfb}   & \textsf{ATLAS-SUSY-2018-09}   & 139   & ATLAS same-sign leptons, jets search for squark and gluino~\cite{ATLAS:2019fag} \\
\hline
\rownumber&    \texttt{ATLAS\_13TeV\_PhotonGGM\_2Photon\_36invfb}  & \textsf{ATLAS-SUSY-2016-27}   & 36.1  & ATLAS photons search~\cite{ATLAS:2018nud} \\
\hline
\rownumber&    \texttt{ATLAS\_13TeV\_PhotonGGM\_1Photon\_139invfb} & \textsf{ATLAS-SUSY-2018-11} & 139   & ATLAS 1 photon, jets search for SUSY GGM~\cite{ATLAS:2022ckd} \\ 
\hline
\rownumber&    \texttt{ATLAS\_13TeV\_ZGammaGrav\_CONFNOTE\_80invfb}    & \textsf{ATLAS-CONF-2018-019}  & 79.8  & ATLAS 1 or more photon search for Higgs exotic decay~\cite{ATLAS:2018vzq} \\ 
\arrayrulecolor{black}\midrule 
\rownumber&    \texttt{CMS\_13TeV\_1LEPbb\_137invfb}   & \textsf{CMS-SUS-20-003} &   137 & CMS 1 lepton and 2 $b$-jets search~\cite{CMS:2021few} \\ 
\arrayrulecolor{lightgray}\hline
\rownumber&    \texttt{CMS\_13TeV\_2Higgs\_4b\_neutralino\_137invfb}   & \textsf{CMS-SUS-20-004}  & 137  & CMS 4 $b$-jets search~\cite{CMS:2022vpy} \\
\hline
\rownumber&    \texttt{CMS\_13TeV\_0LEPStop\_137invfb} & \textsf{CMS-SUS-19-010}   & 137   & CMS 0 lepton search for stop~\cite{CMS:2021beq} \\ 
\hline
\rownumber&    \texttt{CMS\_13TeV\_0LEP\_chargino\_VV\_VH\_137invfb} & \textsf{CMS-SUS-21-002} & 137   & CMS $WW$, $WZ$ or $Wh$ search in fully hadronic decays~\cite{CMS:2022sfi} \\
\hline 
\rownumber&    \texttt{CMS\_13TeV\_Photon\_GMSB\_137invfb} & \textsf{CMS-SUS-21-009}   & 137   & CMS 1 photon, multi-jets search~\cite{CMS:2023xlp}   \\ 
\hline
\rownumber&    \texttt{CMS\_13TeV\_2OSLEP\_EW\_Production\_137invfb}   & \textsf{CMS-SUS-20-001}  & 137  & CMS 2 oppositely charged same-flavour leptons search~\cite{CMS:2020bfa}  \\ 
\hline
\rownumber&    \texttt{CMS\_13TeV\_2OSLEP\_for\_chargino\_36invfb}     & \multirow{2}{*}{\textsf{CMS-SUS-17-010}}  & \multirow{2}{*}{35.9}  & \multirow{2}{*}{CMS 2 opposite charged lepton search~\cite{CMS:2018xqw}} \\  
&    \texttt{CMS\_13TeV\_2OSLEP\_for\_stop\_36invfb}  \\ \hline
\rownumber&    \texttt{CMS\_13TeV\_0LEP\_137invfb} & \textsf{CMS-SUS-19-006}  & 137  & CMS 0 lepton, multi-jet search~\cite{CMS:2019zmd}  \\
\hline
\rownumber&    \texttt{CMS\_13TeV\_1LEPStop\_36invfb}  & \textsf{CMS-SUS-16-051}  & 35.9  & CMS 1 lepton search for top squark~\cite{CMS:2017gbz}   \\ 
\hline
\rownumber&    \texttt{CMS\_13TeV\_1Photon1Lepton\_emu\_combined\_36invfb} & \textsf{CMS-SUS-17-012}   & 35.9  & CMS 1 photon, 1 lepton search~\cite{CMS:2018fon}  \\ 
\hline
\rownumber&    \texttt{CMS\_13TeV\_2LEPStop\_36invfb}  & \textsf{CMS-SUS-17-001}  & 35.9  & CMS 2 OS lepton search for stop~\cite{CMS:2017jrd}   \\ 
\hline
\rownumber&    \texttt{CMS\_13TeV\_2Photon\_GMSB\_36invfb} & \textsf{CMS-SUS-17-011}  & 35.9  & CMS 2 photon search~\cite{CMS:2019vzo}  \\ 
\hline
\rownumber&   \texttt{CMS\_13TeV\_2SSLEP\_Stop\_137invfb}     & \textsf{CMS-SUS-19-008}  & 137  & CMS 2 SS lepton or 3 lepton, jets search~\cite{CMS:2020cpy}   \\ 
\hline
\rownumber&    \texttt{CMS\_13TeV\_MultiLEP\_2LEP\_137invfb}   & \multirow{3}{*}{\textsf{CMS-SUS-19-012}}  & \multirow{3}{*}{137}  & \multirow{3}{*}{CMS 3 or 4 lepton search~\cite{cms:2021cox-fix}}  \\ 
&    \texttt{CMS\_13TeV\_MultiLEP\_3LEP\_137invfb}   &   &   &   \\ 
&    \texttt{CMS\_13TeV\_MultiLEP\_4LEP\_137invfb}   &   &   &   \\ 
\arrayrulecolor{black} 
\bottomrule
\end{tabular}
}
\caption{\label{apptab:analist} Summary of LHC analyses included in this work, detailing the search channels, corresponding report numbers, integrated luminosities, and relevant references.
}
\end{table*}

In this section, we briefly introduce the 13\,TeV  LHC searches used in our study and provide concise descriptions of each signal region (SR) used in our Monte Carlo simulations. All searches, along with their corresponding labels in \colliderbit, are summarised in Table~\ref{apptab:analist}.

\begin{enumerate}
    \item \textbf{ATLAS search in final states with one lepton and two $b$-jets}~\cite{ATLAS:2020pgy}. This search targets the electroweakino pair production process where the chargino decays into a $W$ boson and the lightest neutralino \neu{1}. Meanwhile, the heavier neutralino \neu{2} decays into the Standard Model's $125$\,GeV Higgs boson $h$ and \neu{1}. Signal events require precisely one lepton from the leptonic decay of a $W$ boson, two $b$-tagged jets with an invariant mass $m_{bb}$ in the range of $100$ to $140$\,GeV from $h$ boson decays, and significant missing transverse momentum, $E_{\rm T}^{\rm miss} > 240$\,GeV, due to LSPs. To reduce the background from $W+$jets and semi-leptonic $t\bar{t}$ events, the transverse mass of the lepton and the missing transverse momentum, $m_{\rm T}(p_{\rm T}^{\ell}, p_{\rm T}^{\rm miss})$, must exceed $100\,{\rm GeV}$. The contransverse mass of the two $b$-jets $m_{\rm CT}(b, b)$~\cite{Polesello:2009rn, Tovey:2008ui} is required to be larger than $180\,{\rm GeV}$ to suppress the $t\bar{t}$ background. Three distinct classes of SRs for discovery, labelled as \textsf{SR-LM}, \textsf{SR-MM}, and \textsf{SR-HM}, are defined based on the targeted mass splitting between the heavier electroweakino states, $\cha{1} / \neu{2}$, and the lightest supersymmetric particle (LSP), \neu{1}. These classes correspond to low (\textsf{LM}), medium (\textsf{MM}), and high (\textsf{HM}) mass differences, respectively. To enhance the SUSY sensitivity, each of the three SRs is binned in three $m_{\rm CT}$ regions for a simultaneous two-dimensional fit in $m_{\rm CT}$ and $m_{\rm T}$. In order to further suppress $t\bar{t}$ and single-top background events, the \textsf{SR-HM} also requires the invariant mass of the lepton and the leading $b$-jet $m_{\ell, b_1}$ to exceed $120\,{\rm GeV}$.  Modest excesses are observed in all three SRs, with significance $Z$~\cite{Cousins:2007yta} values ranging from 1.30 to 1.88. Masses of $\cha{1} / \neu{2}$ are excluded up to $740\,{\rm GeV}$ for a massless \neu{1}. For \neu{1} with a mass exclusion limit of up to $250\,{\rm GeV}$, the corresponding exclusion for $\cha{1} / \neu{2}$ is up to $600\,{\rm GeV}$. 

	\item \textbf{ATLAS search for chargino pairs in final states with two leptons}~\cite{ATLAS:2022hbt}. This analysis targets the electroweak production of charged slepton or chargino pairs, which decay into two-lepton final states accompanied by missing transverse momentum. Signal events require two oppositely charged leptons and large $E_{\rm T}^{\rm miss}$ significance. SRs are defined using kinematic variables of the dilepton system, including the magnitude of the vector sum of the di-lepton and missing momentum (denoted as $\vec{p}_{\rm T, boost}^{\ell\ell}$), stransverse mass $m_{\rm T2} (\ell_1, \ell_2, \vec{p}_{\rm T}^{\rm miss})$, the angle and azimuthal angle between the two leptons $\cos{\theta_{\ell\ell}}$ and $\Delta \phi (\ell, \ell)$, and between the leptons and the missing momentum $\Delta \phi (\ell, p_{\rm T}^{\rm miss})$. In the chargino production model, SRs  are categorised into same-flavour and different-flavour groups. These signals are further refined using fine bins based on boosted decision tree scores. Chargino masses up to 140\,GeV are excluded, as well as mass-splittings between the chargino and neutralino as low as 100\,GeV.

	\item \textbf{ATLAS search in final states with two leptons and multi-jets}~\cite{ATLAS:2022zwa}. This search targets SUSY signals in final states containing exactly one opposite-sign, same-flavour (OSSF) lepton pair plus at least two jets. The search strategies are optimised for various signal processes, and we recast the electroweakino pair (EWK) SRs and the SRs using the recursive jigsaw reconstruction (RJR) variables~\cite{Jackson:2017gcy, Santoni:2017lcl, Jackson:2016mfb} in this work. For the EWK search, 13 orthogonal SRs are defined using various kinematic variables. These include the transverse masses $m_{\rm T}$ of two leptons, the $m_{\rm T2}$ variable, the invariant mass of the dilepton system $m_{\ell\ell}$, the missing transverse energy significance $\mathcal{S}(E_{\rm T}^{\rm miss})$~\cite{ATLAS:2018uid}, the jet system mass $m_X$, $\Delta R_{X}$ and the azimuthal opening angle between the lepton system and the missing momentum $\Delta \phi(p_{\rm T}^{\ell\ell}, p_{\rm T}^{\rm miss})$. The RJR search uses two decay trees to reconstruct the process of chargino/neutralino pair decays to $WZ$ and/or whether the events contain energetic ISR radiation. SRs \textsf{SR2L-Low} and \textsf{SR2L-ISR} are defined targeting chargino/neutralino models with mass-splitting $m_{\cha{1} } - \mneu{1} $ of approximately 100\,GeV as a follow-up check for the two corresponding SRs in search~\cite{ATLAS:2018eui}. Limits are set on the sparticle masses within a mass-degenerate chargino/neutralino model and a GMSB model with a Higgsino NLSP, up to 820\,GeV and 900\,GeV, respectively. 

    \item \textbf{ATLAS search for electroweakinos using an emulated RJR (eRJR) technique}~\cite{ATLAS:2019wgx}. This search is targeting chargino-neutralino pairs with a mass splitting $m_{\neu{2} / \cha{1}}  - \mneu{1}$ near the electroweak scale and charginos/neutralinos decaying into a \neu{1} and either a W or Z gauge boson. Events must include at least one OSSF lepton pair within the $Z$ boson mass window and have 1 -- 3 jets. Two SRs, \textsf{SR-Low} and \textsf{SR-ISR}, are defined to test the two corresponding SRs identified in an earlier report~\cite{ATLAS:2018eui} using additional data and an enhanced RJR technique. This search can probe electroweakino pairs decaying through the on-shell $WZ$ decay mode for masses up to $350\,{\rm GeV}$. 
    
    \item \textbf{ATLAS search for electroweakinos using two or three lepton events}~\cite{ATLAS:2023lfr}. This search is optimised for event topologies of electroweakino production in the $Wh$ decay mode with final states containing two same-sign (SS) leptons. Kinematic observables, such as the $m_{\rm T2}$ variable, the minimum of two $m_{\rm T}$ variables, the missing transverse energy $E_{\rm T}^{\rm miss}$, and its significance, are employed to suppress SM backgrounds such as $WZ$, $Wh$, and $t\bar{t}V$. 14 SRs are defined for $Wh$ decay modes, and no significant excess is observed. This search shows a significant improvement compared to the prior $36.9~{\rm fb}^{-1}$ search. Electroweakino masses are excluded up to approximately 525\,GeV for a massless LSP. This search provides a thorough examination of the scenario discussed in this work, particularly the Higgsino NLSP region explored in our previous study~\cite{GAMBIT:2023yih}.

    \item \textbf{ATLAS search in final state with $\tau$ lepton, $b$-jets and transverse missing energy}~\cite{ATLAS:2021jyv}. This search targets new particles that preferentially decay into third-generation SM particles. SRs are categorised into \textsf{Di-tau SR} and \textsf{Single-tau SR} based on the number of hadronically decayed $\tau$ candidates. The signal events of \textsf{Di-tau SR} require exactly one opposite-sign (OS) $\tau$ pair, at least one $b$-jet, $E_{\rm T}^{\rm miss}$ larger than $280\,{\rm GeV}$, and the stransverse mass $m_{\rm T2}(\tau, \tau)$ exceeding $70\,{\rm GeV}$. The \textsf{Signal-tau SR} requires exactly one $\tau_{\rm had}$ candidate, at least two $b$-jets, and $E_{\rm T}^{\rm miss} > 280\,{\rm GeV}$. Additionally, the sum of the transverse masses $m_{\rm T}$ of the two $b$-jets must exceed $700\,{\rm GeV}$, the transverse mass $m_{\rm T}(\tau)$ must be greater than $300\,{\rm GeV}$, and the scalar sum of the transverse momenta, defined as $s_{\rm T} \equiv p_{\rm T}^{\tau} + p_{\rm T}^{j_1} + p_{\rm T}^{j_2}$, must be over $800\,{\rm GeV}$. \textsf{Signal-tau SR} is further binned and the results showed a $1.37 \sigma$ deviation.

    \item \textbf{ATLAS search for top squark production and decay with a Higgs or $Z$ boson}~\cite{ATLAS:2020aci}. This search focuses on stop pair production, where each stop decays into a \neu{2} and a top quark. Subsequently, the neutralino \neu{2} decays into a LSP and either a Higgs or a $Z$ boson. The events are classified into two distinct categories: four SRs with $3\ell$ for signal events involving a $Z$ boson, and two SRs with $1\ell$ for events involving a Higgs boson. In the $3\ell$ selection, signal events require one OSSF lepton pair whose invariant mass matches the $Z$ boson mass, a large $m_{\rm T}$ of the third lepton, significant $E_{\rm T}^{\rm miss}$, and high jet multiplicity. The selection of $1\ell$ SRs requires two $b$-tagged jets, identified as a Higgs candidate using a neural network. Additionally, a high jet multiplicity, $m_{\rm T} > 150\,{\rm GeV}$, and significant object-based $E_{\rm T}^{\rm miss}$ is required. The model-independent interpretation results indicate that the exclusion limit for the top squark mass exceeds $1.2\,{\rm TeV}$. 

    \item \textbf{ATLAS search for Higgsino pairs in events with 4 $b$-jets}~\cite{ATLAS:2024tqe}. This search focuses on Higgsino pair production, where each Higgsino decays into a Higgs boson and a nearly massless gravitino. Each Higgs boson is assumed to subsequently decay into a $b\bar{b}$ pair. The low-mass channel is specifically targeting Higgsino masses slightly heavier than the Higgs mass. Two Higgs candidates are reconstructed from the four $b$-jets. To suppress the $Wt$ background, a quantity $X_{Wt}$, measuring the compatibility of reconstructing the nominal $W$ and top masses, is required to be larger than $1.8$ for any combination. The signal events are defined by the requirement $X_{hh}^{\rm SR} < 1.6$, where $X_{hh}^{\rm SR}$ quantifies how closely Higgs boson candidates match the Higgs mass. The SRs are binned in $E_{\rm T}^{\rm miss}$ and the effective mass $m_{\rm eff}$, which is defined as the scalar sum of $E_{\rm T}^{\rm miss}$ and the $p_{\rm T}$ of the jets associated with Higgs boson candidates. The high-mass channel is optimised for Higgsino masses above $200\,{\rm GeV}$, focusing on detecting final states with high $E_{\rm T}^{\rm miss}$. The signal events require at least three $b$-jets and no signal leptons. After meeting the preselection requirements, the XGBoost algorithm was employed to distinguish between background and signal. The inputs for this process included the number of jets $N_{\rm jets}$ and $b$-jets $N_{\text{b-jets}}$, $H_{\rm T}$, $E_{\rm T}^{\rm miss}$ and its significance $\mathcal{S}(E_{\rm T}^{\rm miss})$, the azimuthal angles, the minimum transverse mass $m_{\rm T, min}^{b-jets}$, the minimum angular distance $\Delta R_{\rm min}^{bb}$, the scalar sum of the masses of the re-clustered large-radius jets $M_{J}^{\Sigma}$, the masses of the reconstructed Higgs boson candidates $m_{h_i}^{\rm HM}$ and its radius $\Delta R_{h_i}^{\rm HM}$. Higgsinos with masses ranging from $130\,{\rm GeV}$ to $940\,{\rm GeV}$ are excluded under the pure Higgs boson decay hypothesis. This search is particularly relevant to our \GEWMSSM model. Only the low-mass portion of this analysis is implemented in \colliderbit, with the high-mass portion requiring a BDT that was made publicly available shortly following the validation of this implementation.

    \item \textbf{ATLAS search for compressed electroweakino mass spectra in final states with two soft leptons}~\cite{ATLAS:2019lng}. This search is optimised for electroweakino pair \neu{2}\cha{1} production of compressed SUSY spectra, where the mass difference $\Delta m (\neu{2} /\cha{1}, \neu{1})$ is less than $50\,{\rm GeV}$. Signal events are selected with large $E_{\rm T}^{\rm miss}$ and two OSSF low $p_{\rm T}$ leptons ($2\ell$) or one soft lepton plus one track ($1\ell1T$). SRs are optimised for specific SUSY scenarios. The \textsf{SR-E} category is targeting signals of an electroweakino system recoiled by an ISR jet. Special kinematic variables, such as the RJR variable $R_{\rm ISR}$ and the transverse mass $M_{\rm T}^{\rm S}$ of the whole electroweakino system, are used to distinguish SUSY signals. Similarly, \textsf{SR-VBF} and \textsf{SR-S} categories are defined for electroweakinos produced through vector boson fusion (VBF) and for slepton production recoiled by ISR. Some inclusive SRs for electroweakinos observe slight excesses, with a local significance of $2.7\sigma$, while the exclusive SR results indicate that all deviations are less than $2\sigma$. The recast of this analysis in \gambit 2.7 differs slightly from the version used for production scans in these results. Small improvements have been made in the treatment of baseline track objects. The impact of these changes on any results presented are negligible, and all conclusions presented do not depend on any differences between published recast code and those used during scans.

    \item \textbf{ATLAS search for coloured sparticles using events with jets }~\cite{ATLAS:2020syg}. This search focus on the pair production of squarks and gluinos in final states characterised by jets and $E_{\rm T}^{\rm miss}$, without the presence of electrons or muons. Signal events must have $E_{\rm T}^{\rm miss} > 300\,{\rm GeV}$, leading jet $p_{\rm T} > 200\,{\rm GeV}$, sub-leading jet $p_{\rm T} > 50\,{\rm GeV}$, and $m_{\rm eff}$ for all jets > $800\,{\rm GeV}$. Multi-bin SRs \textsf{MB-SSd} are defined for the direct decays of the squark pair, $\tilde{q} \to q \neu{1}$. \textsf{MB-GGd} are defined for direct decays of the gluino pair, $\tilde{g} \to q\bar{q} \neu{1}$. An additional region \textsf{MB-C} targets small mass splittings between the pair-produced squarks or gluinos and the LSP \neu{1}. Eight exclusive SRs are defined using an independently-trained BDT for either gluino direct decay (\textsf{GGd}) or gluino one-step decay (\textsf{GGo}). An example of the latter is the cascade decay process $\tilde{g} \to q q^\prime \cha{1} \to q q^\prime W^\pm \neu{1}$. In addition to multi-bin and BDT searches, ten inclusive model-independent SRs characterised by increasing minimum jet multiplicity are defined with requirements on $m_{\rm eff}$, $N_{\rm jets}$ and $E_{\rm miss}/\sqrt{H_{\rm T}}$. We implement all 10 model-independent SRs in our analysis. 

    \item \textbf{ATLAS search for two boosted hadronically-decaying bosons and $E_{\rm T}^{\rm miss}$}~\cite{ATLAS:2021yqv}. This search targets pair-produced electroweakinos that decay into high-$p_{\rm T}$ weak bosons in the fully hadronic channel. This search is motivated by several signal models, which can be categorised into three physics scenarios: a general EWMSSM spectrum, light Higgsinos decaying into gravitinos, and light Higgsinos with a QCD axion as the LSP. The signal event requires at least two large-$R$ jets, with no selected leptons and $b$-tagged jets. We implement the four inclusive \textsf{4Q} SRs, including \textsf{4Q-WW}, \textsf{4Q-WZ}, \textsf{4Q-ZZ} and \textsf{4Q-VV} to target different boosted candidates. Boson tagging is required for the two leading large-$R$ jets, with an efficiency of approximately $50\%$. Signal events typically exhibit a relatively spherical topology, with the large-$R$ jets generally well separated from the $E_{\rm T}^{\rm miss}$. Consequently, the SRs are defined by the following requirements: $E_{\rm T}^{\rm miss} > 300\,{\rm GeV}$, $m_{\rm eff} > 1300\,{\rm GeV}$, and $\min \Delta\phi (E_{\rm T}^{\rm miss}, j) > 1.0$. The results, interpreted within various simplified models, show that the sensitivities to the \EWMSSM and \GEWMSSM models are significantly enhanced. In \colliderbit, due to challenges in matching the $b$-tagging to the ATLAS results during validation, we do not include the two $b$-jet SRs. For this reason, we do not apply the ATLAS \textit{full likelihood} framework to this analysis. 

    \item \textbf{ATLAS search for events with two OS leptons, jets and $E_{\rm T}^{\rm miss}$}~\cite{ATLAS:2021hza}. This search is optimised for the pair production process $\tilde{t} \to t \neu{1}$. Signal events are required to have exactly two OS leptons with an invariant mass $m_{\ell\ell}$ greater than $20\,{\rm GeV}$ to remove low-mass resonances, as well as at least one $b$-jet. Events with same-flavour lepton pairs are additionally required to have $m_{\ell\ell}$ outside the $Z$ boson mass window. SRs targeting two-body decays of the top squark are defined by the following kinematic requirements: the azimuthal angle $\Delta \phi_{\rm boost}$ between the missing transverse momentum ($p_{\rm T}^{\rm miss}$) and the di-lepton transverse momentum ($p_{\rm T, boost}^{\ell\ell}$) must be less than 1.5; the $E_{\rm T}^{\rm miss}$ significance must be greater than 12; and $m_{\rm T2}^{\ell\ell}$ must exceed $110\,{\rm GeV}$. Various SRs are also optimised for 3-body and 4-body stop decays. In our analysis, we use the seven inclusive SRs for 2-body top quark decays, labeled as \textsf{SR2bInc}, which provide less model-dependent sensitivity.

    \item \textbf{The ATLAS search for charginos and sleptons in events with two leptons and missing transverse momentum}~\cite{ATLAS:2019lff}. This search  focuses on the pair production of charginos and sleptons which decay into two light-flavour leptons. Three physics scenarios are considered: (1) chargino decays into a $W$ boson and an LSP, (2) slepton decays into a lepton and an LSP, and (3) chargino cascade decays via a slepton or sneutrino. Signal events are required to have exactly two OS leptons with an invariant mass $m_{\ell}$ greater than $100\,{\rm GeV}$, no $b$-jets, $E_{\rm T}^{\rm miss} > 110\,{\rm GeV}$ and $E_{\rm T}^{\rm miss}$-significance greater than 10. SRs are defined based on the $m_{\rm T2}$ variable. According to the lepton flavours and the number of non-$b$-tagged jets, SRs is identified into four sets, which are labelled as \textsf{SR-SF-0J}, \textsf{SR-DF-0J}, \textsf{SR-SR-1J} and \textsf{SR-DF-1J}. A total of 16 inclusive SRs were defined to enhance model-independent sensitivity, along with 36 finely binned SRs in the $m_{\rm T2}$ variable to maximise sensitivity to the simplified models. 
    
    \item \textbf{ATLAS search for chargino-neutralino pair production in final states with three leptons and missing transverse momentum}~\cite{ATLAS:2021moa}. This search is optimised for the golden channel of $\cha{1} \neu{2}$ pair production with decays via $WZ$ to $3\ell$ final states. Events selections are designed independently for the targeted models: for the on-shell $WZ$, off-shell $WZ$ or $Wh$ selections. For SRs targeting $WZ$ scenarios, an OSSF lepton pair is required to come from a $Z$ boson candidate, and the remaining lepton is assigned to the $W$ boson candidate. Therefore, the $m_{\rm T}$ variables are always constructed using the $W$ lepton and the $E_{\rm T}^{\rm miss}$. In the $Wh$ selection, an OS lepton pair is assumed to be indirectly produced in the Higgs boson decay and can be either same-flavour or different flavour. A common pre-selection for ${\sf SR}^{\sf WZ}$ and ${\sf SR}^{\sf Wh}_{\sf OSSF}$ requires events to contain exactly three signal leptons, no $b$-tagged jets, and $E_{\rm T}^{\rm miss} > 50\,{\rm GeV}$. For the $Z$ candidate, the invariant mass of the OSSF lepton pair ($m_{\ell\ell}$) must be within $[75, 105]\,{\rm GeV}$, while for the Higgs candidate, $m_{\ell\ell}$ must fall outside this range. Additionally, the invariant mass of the three leptons ($m_{3\ell}$) must not be within the $Z$ mass window. The 20 SR bins are defined for $\sf SR^{WZ}$ using the variables  $m_{\ell\ell}$, $E_{\rm T}^{\rm miss}$, jet multiplicity $n_{\rm jets}$, the jet $p_{\rm T}$ scalar sum $H_{\rm T}$, the scalar $p_{\rm T}$ of the three selected leptons $H_{\rm T}^{\rm lep}$, and $m_{\rm T}$. The same set of variables is used in the selection criteria for the 19 $\sf SR_{DFOS}^{Wh}$ bins. Two regions for ${\sf SR}^{\sf Wh}_{\sf DFOS}$ rely on object resolution variables that are not available in the \colliderbit framework, and are therefore not included in our implementation. This search is sensitive to both the \EWMSSM and \GEWMSSM models considered in this work. For the wino/bino production mode, the exclusion limit reaches $640\,{\rm GeV}$, while for the pure Higgsino mode, the limit is set up to $210\,{\rm GeV}$. 

    \item \textbf{ATLAS search for supersymmetry in events with four or more leptons}~\cite{ATLAS:2021yyr}. This analysis targets both $R$-parity-conserving (RPC) general gauge-mediated (GGM) Higgsino scenarios and $R$-parity-violating (RPV) bino LSP scenarios. The RPC GGM Higgsino model contains an almost mass-degenerate Higgsino triplet, which decays into a nearly massless gravitino $\tilde{G}$ plus a $Z$ or Higgs boson, $\neu{1} \to Z/h + \tilde{G}$. The simplified RPV SUSY scenario considers a bino LSP which decays via an RPV interaction into two charged leptons and a neutrino $\neu{1} \to \ell \ell \nu$. SRs are defined by the multiplicity of light leptons ($N_\ell$) and hadronic $\tau$ candidates ($N_\tau$). Two \textsf{SR0-ZZ} bins require two OSSF lepton pairs as $Z$ candidates, no $b$-tagged jets, and missing transverse energy ($E_{\rm T}^{\rm miss}$) thresholds of $>100$ or $>200\,{\rm GeV}$ for the Higgsino GGM model. Variants without a $b$-jet veto and with looser $E_{\rm T}^{\rm miss}$ requirements are used to reproduce the excess regions reported in Ref.~\cite{ATLAS:2018rns}. A general $4L0T$ SR is defined by applying a $Z$-veto and requiring $m_{\rm eff} > 600\,{\rm GeV}$. Two similar SRs for the RPV scenario are defined with higher $m_{\rm eff}$ thresholds. The \textsf{SR5L} region requires at least five light-flavour leptons, with no additional kinematic requirements. All SRs without hadronic $\tau$ candidates are implemented as described above. Since we do not simulate all SRs in this analysis, we do not apply the ATLAS \textit{full likelihood} framework in this case. 

    \item \textbf{ATLAS search for supersymmetry in events with $b$-tagged jets and $E_{\rm T}^{\rm miss}$}~\cite{ATLAS:2017avc}. This search focuses on bottom and top squark direct pair production, where sbottoms (stop) decay via $\tilde{b}_1 \to b\neu{1}$ and $\tilde{b} \to t \cha{1}$ ($\tilde{t}_1 \to t\neu{1}$ and $\tilde{t} \to b \cha{1}$). In the decay modes involving the chargino \cha{1}, the mass splitting between \cha{1} and the LSP \neu{1} is assumed to be $1\,{\rm GeV}$. As a result, the decay products are too soft to be efficiently reconstructed. SRs can be classified into two sets: the zero-lepton channel for $\tilde{b} \to b \neu{1}$ and the one-lepton channel targeting $\tilde{b} \to t \cha{1}$. The zero-lepton selection requires no baseline leptons and exactly two $b$-tagged jets. Based on the mass splitting between $\tilde{b}$ and the LSP, three SR categories are defined. The \textsf{0L-SRA} regions are optimised for cases where $\Delta m(\tilde{b}, \neu{1})$ is greater than $250\,{\rm GeV}$. Signal events are discriminated by the contransverse mass $m_{\rm CT}$ along with other kinematic variables, including $E_{\rm T}^{\rm miss}$, $m_{\rm eff}$, $H_{\rm T}$, $m_{jj}$, and others. The \textsf{0L-SRB} region targets intermediate $\Delta m(\tilde{b}, \neu{1})$ between $50\,{\rm GeV}$ and $250\,{\rm GeV}$. The selections are based on the minimal transverse mass $m_{\rm T}^{\rm min}({\rm jet}_{1-4}, E_{\rm T}^{\rm miss})$ and azimuthal angles such as $\Delta \phi (b, E_{\rm T}^{\rm miss})$. The \textsf{0L-SRC} region targets the compress spectrum, where an ISR jet with $p_{\rm T} > 500\,{\rm GeV}$ is required to boost bottom squark pairs. We implemented all these zero-lepton SRs, with the one-lepton selection regions relying on variables not available in \colliderbit. 

    \item \textbf{ATLAS search for squarks and gluinos in final states with same-sign leptons and jets}~\cite{ATLAS:2019fag}. This search focuses on processes involving the pair production and cascade decays of coloured sparticles into final states with leptons and jets. Five SRs are built to isolate signals from various hypothetical SUSY processes. These regions are characterised by the multiplicities of leptons ($n_{\ell}$), jets ($n_j$), and $b$-tagged jets ($n_{b}$); the effective mass ($m_{\rm eff}$); the missing transverse energy ($E_{\rm T}^{\rm miss}$) and its ratio to $m_{\rm eff}$; as well as the invariant mass of SS lepton pairs. The \textsf{Rpv2L} region targets gluino pair production in RPV scenarios, so it requires high jet multiplicity $n_j \geq 6$, $m_{\rm eff} > 2600\,{\rm GeV}$, with no requirement on $E_{\rm T}^{\rm miss}$. The SR \textsf{Rpc2L0b} targets RPC gluino pair production with cascade decay chains $\tilde{g} \to q \bar{q}^\prime W Z \neu{1}$. Signal events require a large jet multiplicity $n_{j} \geq 6$ to suppress $WZ$ and other multi-boson backgrounds, as well as $E_{\rm T}^{\rm miss} > 200\,{\rm GeV}$ and $E_{\rm T}^{\rm miss}/m_{\rm eff} > 0.2$. The SRs \textsf{Rpc2L1b} and \textsf{Rpc2L2b} probe scenarios involving $3^{\rm rd}$ generation squarks, such as $\tilde{b} \to t W \neu{1}$. Finally, the SR \textsf{Rpc3LSS1b} provides sensitivity to scenarios with long decay chains but compressed mass spectrum, such as $\tilde{t} \to t\neu{2} \to tWW\neu{1}$. This SR selects events with at least three leptons of the same charge, at least one $b$-tagged jet, and a loose requirement of $E_{\rm T}^{\rm miss}/m_{\rm eff} > 0.14$. Additionally, any pair of same-sign electrons with an invariant mass within the $Z$ boson window is vetoed to suppress backgrounds from $Z\to e^+ e^-$ events where the charge of one electron is mismeasured.

    \item \textbf{ATLAS search for photonic signals of gauge-mediated supersymmetry}~\cite{ATLAS:2018nud}. This search focuses on signatures predicted by models inspired by GGM. In scenarios where the NLSP is bino-like, it predominantly decays via $\neu{1} \to \gamma \tilde{G}$, resulting in final states with two photons and $E_{\rm T}^{\rm miss}$. The signal events are selected by requiring either one photon with $E_{\rm T} > 140\,{\rm GeV}$ for the photon + jets SRs, or two photons with $E_{\rm T}$ thresholds of $35$ and $25\,{\rm GeV}$, respectively, for the di-photon SRs. The di-photon analysis contains four SRs, each targeting different GGM models that feature a purely bino-like \neu{1}. The \textsf{SRaa\_SL} and \textsf{SRaa\_SH} regions are designed to target gluino and squark production, while \textsf{SRaa\_WL} and \textsf{SRaa\_WH} focus on wino production. The photon+jets analysis, defined by three SRs (labels \textsf{SRaj\_L}, \textsf{SRaj\_L200} and \textsf{SRaj\_H}), targeting the NLSP is a mixture of Higgsino and bino, where the branching ratios for $\neu{1} \to \gamma \tilde{G}$ and $\neu{1} \to Z \tilde{G}$ are roughly equal. All of these SRs are defined using kinematic variables such as $H_{\rm T}$, $m_{\rm eff}$, the azimuthal angle differences $\Delta \phi_{\rm min} ({\rm jet}, E_{\rm T}^{\rm miss})$, $\Delta \phi(\gamma, E_{\rm T}^{\rm miss})$, and $\Delta \phi_{\rm min} (\gamma, E_{\rm T}^{\rm miss})$. Additionally, the quantity $R_{\rm T}^4$, defined as the scalar sum of the transverse momenta of the four highest-$p_{\rm T}$ jets divided by the scalar sum of the transverse momenta of all jets in the event, is required to be less than $0.9$ in \textsf{SRaj\_L} and \textsf{SRaj\_H}. All seven SRs described above are implemented in our analysis.

    \item \textbf{ATLAS search for supersymmetry in events with photons, jets and $E_{\rm T}^{\rm miss}$}~\cite{ATLAS:2022ckd}. This search is targeting the simplified GGM model where the neutralino NLSP has large Higgsino or bino components decaying into a gravitino and a photon or a gravitino and either a $Z$ boson or Higgs boson. The analysis targets strong production in three SRs: \textsf{SRL} for large mass splitting between the gluino and neutralino NLSP, \textsf{SRH} for compressed spectra, and \textsf{SRM} for intermediate mass splitting. Signal events in all three regions are required to have at least one high-$p_{\rm T}$ photon, no leptons, at least three jets, and $E_{\rm T}^{\rm miss}$. The SR definitions utilise several kinematic variables, including the scalar sum of transverse momenta ($H_{\rm T}$) of signal jets and leading photons, as well as the angular separations $\Delta \phi({\rm jet}, E_{\rm T}^{\rm miss})$ and $\Delta \phi(\gamma, E_{\rm T}^{\rm miss})$. Additionally, the $R_{\rm T}^4$ variable—defined as the ratio of the scalar sum of the four leading jets to that of all jets—is required in \textsf{SRL} and \textsf{SRH} to suppress SM backgrounds with fewer and softer jets. All three SRs are implemented in our framework.

    \item \textbf{ATLAS search for Higgs exotic decays in final states with photons}~\cite{ATLAS:2018vzq}. This search targets SM Higgs boson decays producing at least one photon and missing transverse energy ($E_{\rm T}^{\rm miss}$), where the photon originates from an exotic Higgs decay to one or two neutralinos, each subsequently decaying into a gravitino and a photon. Signals of Higgs bosons produced in association with a $Z$ boson are considered, where the $Z$ decays into an OSSF lepton pair. The signal selection requires the di-lepton invariant mass $m_{\ell\ell}$ to be within the $Z$ mass window, missing transverse energy $E_{\rm T}^{\rm miss}$ greater than $95\,{\rm GeV}$, a large angular separation between the $Z$ and Higgs systems ($\Delta \phi (\ell\ell, E_{\rm T}^{\rm miss}) > 2.8$), and a small angular separation between the leptons ($\Delta \phi (\ell\ell) < 1.4$). To improve the signal-to-background ratio, events are required to satisfy ${\rm Bal}_{p_{\rm T}} < 0.2$, where ${\rm Bal}_{p_{\rm T}}$ is defined as the transverse momentum difference between the dilepton system and the $\gamma E_{\rm T}^{\rm miss}$ system, divided by the transverse momentum of the $\gamma E_{\rm T}^{\rm miss}$ system. This requirement ensures good signal balance in the transverse plane. This search is sensitive to the \GEWMSSM scenario when the NLSP \neu{1} mass is below the Higgs mass.

    \item \textbf{CMS search for supersymmetry in events with Higgs and $W$ bosons}~\cite{CMS:2021few}. This search targets a simplified SUSY model featuring chargino-neutralino production, where the subsequent decays are $\cha{1} \to W^\pm \neu{1}$ and $\neu{2} \to h \neu{1}$. The final state is characterised by a charged lepton from the $W$ boson decay, two $b$-jets from the $h \to b\bar{b}$ decay, and $E_{\rm T}^{\rm miss}$. Signal events require one signal lepton, two or three jets with exactly two $b$-tagged jets, and $E_{\rm T}^{\rm miss} > 150\,{\rm GeV}$. The baseline selection requires the invariant mass $m_{b\bar{b}}$ to lie in the Higgs mass window $[90, 150]\,{\rm GeV}$, the $m_{\rm T}$ of signal lepton to be larger than $150\,{\rm GeV}$, and the cotransverse mass variable $m_{\rm CT}$ of the two $b$-tagged jets to be larger than $200\,{\rm GeV}$. To improve the signal sensitivity, a collection of ``large-$R$'' AK8 jets with distance parameter $R=0.8$ are defined as Higgs candidates. A total of 12 non-overlapping SRs are categorised in the number of Higgs candidates $N_H$, jet multiplicity $N_{\rm jets}$ and $E_{\rm T}^{\rm miss}$. This search is sensitive to both the \GEWMSSM and \EWMSSM scenarios in this work, and we implement all SRs in \colliderbit. 

    \item \textbf{CMS search for Higgsinos decaying into two Higgs bosons and $E_{\rm T}^{\rm miss}$}~\cite{CMS:2022vpy}. This search targets SUSY processes that produce two Higgs bosons, each decaying via $h \to b\bar{b}$, accompanied by additional jets and large ($E_{\rm T}^{\rm miss}$). Three typical simplified models can lead to this signature: 1. The GGM process, $\neu{1} \neu{1} \to h\tilde{G}~ h\tilde{G}$, where the \neu{1} NLSPs are produced via cascade decays of heavier charginos and neutralinos; 2. Pair production of neutral Higgsinos, which decay into bino LSPs: $\neu{2} \neu{3} \to h \neu{1}~ h \neu{1}$; 3. Gluino pair production, where each gluino decays into a \neu{2} and a quark–antiquark pair, followed by the \neu{2} decaying into a Higgs boson and a \neu{1}. Two types of signatures are considered based on the momentum of the Higgs boson. The resolved signature, which targets Higgs bosons with lower momentum, requires exactly four individually-resolved $b$-tagged jets. In contrast, the boosted signature targets high-momentum Higgs bosons and requires two merged jets, each double $b$-tagged, as Higgs candidates. The baseline selection for both signatures requires a cut on $E_{\rm T}^{\rm miss}$, no charged leptons, and no jets with a small azimuthal separation from $\vec{p}_{\rm T}^{\rm miss}$ in the transverse plane. Signal events for resolved signatures require either 4 or 5 AK4 jets (defined with distance parameter $R = 0.4$). The four jets with the highest $b$-tag discriminator values are selected to form two Higgs boson candidates, and the mass difference between the two candidates is required to be smaller than $40\,{\rm GeV}$. Sixteen SRs are defined based on the $b$-counting variable, which utilises three $b$-tagging working points, $E_{\rm T}^{\rm miss}$, and the maximum angular separation between the $b$-tagged jets within each Higgs boson candidate $\Delta R_{\rm max}$. The boosted signature is characterised by at least two energetic AK8 jets with $p_{\rm T} > 300\,{\rm GeV}$ and jet masses in the range $60 < m_J < 260\,{\rm GeV}$. The Higgs candidate counting variable is defined using the double-$b$ tagging discriminator $D_{bb}$. The six SRs are then categorised based on the variables $N_H$ and $E_{\rm T}^{\rm miss}$. In this work, we implement all 22 SRs. 

    \item \textbf{CMS search for top squark production in fully-hadronic final states}~\cite{CMS:2021beq}. This search is optimised for top squark pair production in events characterised by multiple jets, no leptons, and large $E_{\rm T}^{\rm miss}$. It targets various SUSY simplified models, including $\tilde{t} \to t^{(*)} \neu{1}$, $\tilde{t} \to b \cha{1}$ (with $\cha{1} \to W^+ \neu{1}$), and $\tilde{t} \to c \neu{1}$. In particular, the mass difference $\Delta m = m_{\tilde{t}} - \mneu{1}$ significantly affects not only the top squark decay modes, but also the kinematic distributions of the final-state particles. AK8 jets are also used to identify the boosted high-$p_{\rm T}$ top quark and $W$ boson candidates based on the jet $p_{\rm T}$ and soft-drop mass using the ``DeepAK8'' algorithm. The ``DeepResolved'' algorithm is also used to form top candidates from the combination of three small-$R$ jets. A typical preselection requires at least two small-$R$ jets, $H_{\rm T} > 300\,{\rm GeV}$, $E_{\rm T}^{\rm miss} > 250\,{\rm GeV}$, and no leptons, $\tau_{\rm had}$ candidates, or charged track objects. The low $\Delta m$ search regions additionally require the presence of a high-energy ISR jet recoiling against the stop pair system, no top-tagged, $W$-tagged, or trijet candidates, and $E_{\rm T}^{\rm miss}/\sqrt{H_{\rm T}} > 10~\sqrt{\rm GeV}$. A total of 53 SRs are defined based on the following variables: the number of small-$R$ jets ($N_j$), the number of $b$-jets ($N_{b}$), the number of reconstructed secondary vertices $N_{\rm SV}$, the transverse mass of the $b$-jet $m_{\rm T}^{b}$, the transverse momentum of the ISR jet $p_{\rm T}^{\rm ISR}$, the transverse momentum of the $b$-jet ($p_{\rm T}^{b}$), and $E_{\rm T}^{\rm miss}$. The high $\Delta m$ search region is optimised for stop signals with $\Delta m > m_{W}$. It requires at least five small-$R$ jets, at least one $b$-tagged jet, and $\Delta \phi(j_{1-4}, \vec{p}_{\rm T}^{\rm miss}) > 0.5$. A total of 130 SR bins are defined based on various observables, including $m_{\rm T}^{b}$, $N_{j}$, $N_{b}$, the number of top-tagged jets ($N_{t}$), the number of $W$-tagged jets ($N_W$), the number of tagged trijet candidates ($N_{\rm res}$), $H_{\rm T}$, and $E_{\rm T}^{\rm miss}$. All SRs for both the low $\Delta m$ region (labelled \textsf{lowDM}) and the high $\Delta m$ region (labelled \textsf{highDM}) are implemented in \colliderbit.

    \item \textbf{CMS search for charginos and neutralinos in final states containing hadronic decays of $WW$, $WZ$ or $Wh$}~\cite{CMS:2022sfi}. Unlike most chargino-neutralino searches by ATLAS and CMS, which focus on events containing at least one lepton, this search targets fully hadronic final states. The results are interpreted using simplified models of either chargino pair production ($\cha{1} \tilde{\chi}_1^\mp$) or chargino-neutralino pair production ($\cha{1} \neu{2}$). In these models, the chargino always decays into a $W$ boson and the LSP \neu{1}, while the neutralino decays into either a $Z$ boson or a Higgs boson accompanied by a \neu{1}. Three non-exclusive jet taggers are used to categorise AK8 jets based on their corresponding DNN scores and jet mass requirements: the $W$ tagger identifies $W(q\bar{q}^\prime)$ decays with $65 < m_{J} < 105\,{\rm GeV}$; the $V$ tagger identifies either $W(q\bar{q}^\prime)$ or $Z(q\bar{q})$ decays within the same mass range, $65 < m_{J} < 105\,{\rm GeV}$; and the $b\bar{b}$ tagger identifies $Z(b\bar{b})$ or $h(b\bar{b})$ decays with $75 < m_{J} < 140\,{\rm GeV}$. The baseline selection requires signal events to contain at least two and at most six AK4 jets, at least one AK8 jet with $p_{\rm T} > 200\,{\rm GeV}$, and a scalar sum of all signal AK4 jet transverse momenta ($H_{\rm T}$) greater than $300\,{\rm GeV}$. Additionally, events must have $E_{\rm T}^{\rm miss}$ greater than $200\,{\rm GeV}$, and satisfy a threshold requirement on the azimuthal angle ($\Delta \phi$) between $\vec{p}_{\rm T}^{\rm miss}$ and each of the four highest $p_{\rm T}$ AK4 jets. The $b$-veto SRs (labelled \textsf{0b}) targeting $WW$ or $WZ$ pairs require zero $b$-tagged AK4 jets and at least two AK8 jets with $V$-tagged or $W$-tagged. Three SR categories are defined for events containing at least one $b$-tagged AK4 jet. The \textsf{WH} SR category requires at least one $W$-tagged candidate and at least one $b\bar{b}$-tagged candidate. The \textsf{W} SR category includes events with at least one $W$-tagged candidate and no $b\bar{b}$-tagged candidate, while the \textsf{H} SR category requires no $W$-tagged jet and at least one $b\bar{b}$-tagged candidate. All four categories are finely binned using the $E_{\rm T}^{\rm miss}$ variable, and we implement all the SRs described above. 

    \item \textbf{CMS search for new physics events consisting of at least one photon and multiple jets}~\cite{CMS:2023xlp}. This analysis investigates SUSY signatures produced through strong or electroweak interactions, focusing on final states with at least one photon, high jet multiplicity, and large $E_{\rm T}^{\rm miss}$. The strong production simplified model considers either gluino pair production or stop pair production. In this model, each gluino decays into a neutralino \neu{1} and a pair of quarks, while each top squark decays into a top quark and a neutralino \neu{1}. The electroweak production simplified model includes various electroweakino pair production processes, such as chargino-neutralino $\cha{1} \neu{1}$ and chargino pair $\cha{1} \tilde{\chi}_1^\mp$ production, where the heavier electroweakino decays into the lightest neutralino \neu{1} and a gauge boson. In all scenarios, the lightest neutralino subsequently decays into a gravitino $\tilde{G}$ accompanied by a photon, a $Z$ boson, or a Higgs boson. Signal events are required to satisfy the following criteria: missing transverse energy $E_{\rm T}^{\rm miss} > 300\,{\rm GeV}$, at least two AK4 jets, at least one energetic photon with $p_{\rm T}^{\gamma} > 100\,{\rm GeV}$, a scalar sum of the transverse momenta of jets and the photon ($S_{\rm T}$) greater than $300\,{\rm GeV}$, $\Delta \phi$ between the missing transverse momentum and each of the two highest-$p_{\rm T}$ jets greater than $0.3$, and no signal leptons or isolated charged tracks. Two sets of SRs are defined to target electroweak (EW) and strong production (SP) modes. The EW SRs include events with $2 \leq N_{\rm jets} \leq 6$ and at least one V- or Higgs-tagged jet. The SP SRs comprise all baseline events that do not meet the EW SR selection criteria. We incorporate all 37 SR bins in \colliderbit. 

    \item \textbf{CMS search for supersymmetry in events with two OSSF leptons}~\cite{CMS:2020bfa}. This analysis examines three SUSY signature modes: (i) events where the dilepton invariant mass is consistent with the $Z$ boson mass (on-$Z$); (ii) a kinematic edge in the dilepton invariant mass distribution (edge); and (iii) nonresonant dilepton production from slepton pair decays (slepton). Strong-production simplified models consider gluino or squark pair production, where the heavier neutralino (\neu{2}) decays to a $Z$ boson and \neu{1}, or via an intermediate slepton. Electroweak simplified models include $\cha{1} \neu{2}$ production, with $Z$, $W$, or $H$ bosons in the final state, as well as direct slepton pair production ($\tilde{\ell}^+\tilde{\ell}^- \to \ell^+\ell^- \neu{1} \neu{1}$). Signal events are required to contain two isolated OSSF leptons with $p_{\rm T} > 25,20\,{\rm GeV}$ and missing transverse momentum $E_{\rm T}^{\rm miss} > 50\,{\rm GeV}$, and to satisfy additional selections on the $M_{T2}$ variable, jet multiplicity $n_{\rm j}$, $b$-tag multiplicity $n_{b}$, $H_{\rm T}$, and the dilepton system invariant mass $m_{\ell\ell}$. Multiple SR categories are defined for different signal modes: on-$Z$ SRs (including \textsf{SRA}, \textsf{SRB}, and \textsf{SRC} categories) target gluino/squark and electroweakino production; edge SRs (including \textsf{SRBoostedVZ}, \textsf{SRResolvedVZ}, and \textsf{SRHZ} categories) probe off-shell $Z$ or slepton cascade production modes; and slepton SRs include both jet-less (\textsf{SRoffZ0j}) and jet-inclusive (\textsf{SRoffZj}) categories. We implement all SRs, but only use the edge SRs in this work.

    \item \textbf{CMS search for chargino and top-squark pair production in opposite-sign dilepton final states}~\cite{CMS:2018xqw}. This analysis focuses on SUSY scenarios that produce two oppositely charged leptons and $E_{\rm T}^{\rm miss}$. Two classes of simplified models are investigated. For electroweak production, the process $\tilde{\chi}_1^+\tilde{\chi}_1^-$ is considered, with two possible decay modes: either via slepton or sneutrino mediation, $\cha{1} \to \tilde{\ell}\nu/\tilde{\nu}\ell \to \ell\nu\neu{1}$ with $m_{\tilde{\ell}} = \tfrac{1}{2}(m_{\cha{1}} + \mneu{1})$, or via direct $W$-boson emission, $\cha{1} \to W^\pm \neu{1}$. For strong production, the pair production of top squarks with two decay scenarios are considered: $\tilde{t}_1 \to t \neu{1}$, optimised for compressed mass spectra ($m_W < \Delta m(\tilde{t}_1,\neu{1}) < m_t$), and $\tilde{t}_1 \to b \cha{1} \to b W^\pm\neu{1}$, with $m_{\cha{1}} = \tfrac{1}{2}(m_{\tilde{t}_1} + \mneu{1})$. The baseline selection requires two isolated OS leptons with $m_{\ell\ell} > 20~\text{GeV}$ and a $Z$ window veto for same-flavour pairs, $E_{\rm T}^{\rm miss} \ge 140~\text{GeV}$, and no additional leptons. For chargino searches (labels \textsf{SR-chargino}), the SRs require no $b$-tagged jets and are then finely binned according to $E_{\rm T}^{\rm miss}$, jet multiplicity $N_{\rm jets}$, lepton pair flavour (same-flavour or different-flavour, SF/DF), and the main discriminator, $m_{\rm T2}(\ell\ell)$.  For the top-squark search (labels \textsf{SR-stop}), SRs are categorised by the number of $b$-tagged jets: $N_b = 0$ for compressed-like scenarios and $N_b \geq 1$ for less compressed cases. At high $E_{\rm T}^{\rm miss} \geq 300\,{\rm GeV}$ regions, an ISR topology is required, defined by a leading jet with $p_{\rm T} > 150\,{\rm GeV}$ and $\Delta\phi(j_1, p_{\rm T}^{\rm miss}) > 2.5$. All SRs are implemented in \colliderbit and utilised in this work.
    
    \item \textbf{CMS search for supersymmetry in final states with jets and missing transverse momentum}~\cite{CMS:2019zmd}. This search targets the pair production of gluinos and squarks, which typically decay through a sequence of processes resulting in jets and large $E_{\rm T}^{\rm miss}$. This analysis defines 174 exclusive SRs in a four-dimensional space specified by key variables: the number of jets $N_{\rm jets}$, the number of $b$-tagged jets $N_{b\text{-jets}}$, the $p_{\rm T}$ sum of all jet  $H_{\rm T}$, and the magnitude of the vector $p_{\rm T}$ sum of the jets $H_{\rm T}^{\rm miss}$. Signal events are required to satisfy: $N_{\rm jets} \ge 2$, $H_{\rm T} > 300\,{\rm GeV}$, $H_{\rm T}^{\rm miss} > 300\,{\rm GeV}$, $H_{\rm T}^{\rm miss} < H_{\rm T}$, and no identified isolated leptons, isolated tracks, or isolated photon candidates. Instead of the finely 174 SR bins, we implement the 12 aggregate bins into \colliderbit for simplicity. 

    \item \textbf{CMS search for top squark production using single lepton events}~\cite{CMS:2017gbz}. This search focusses on top squark pair production in events containing exactly one isolated lepton, jets, and $E_{\rm T}^{\rm miss}$. The kinematic distribution of top squark events strongly depends on the mass difference between the stop and the LSP, $\Delta m (\tilde{t}, \neu{1})$. Therefore, two sets of search regions (a total of 31 non-exclusive SRs) are designed to be sensitive to different ranges of $\Delta m (\tilde{t}, \neu{1})$ in parameter space. In order to eliminate the influence of the correlation between signal intervals on the results, we only implement six aggregate SRs into \colliderbit. The signal events have exactly one lepton with $m_{\rm T} > 150\,{\rm GeV}$, and $E_{\rm T}> 450\,{\rm GeV}$. The six aggregate SRs are defined based on the number of jets $N_{\rm J}$, the invariant mass of the lepton and $b$-tagged jet $m_{\ell b}$, $E_{\rm T}^{\rm miss}$, and a modified topness variable $t_{\rm mod}$~\cite{Graesser:2012qy}.
 
    \item \textbf{CMS search for supersymmetry in events with a photon, a lepton, and missing transverse momentum}~\cite{CMS:2018fon}. This search targets the GGM SUSY model, where the NLSP is assumed to be either a bino- or wino-like neutralino, or a wino-like chargino. In this scenario, a neutral NLSP \neu{1} decays into a photon or a $Z$ boson plus a gravitino $\tilde{G}$, while a charged NLSP \cha{1} decays into a $W$ boson plus a gravitino. This search targets events featuring one photon, at least one lepton, and large $E_{\rm T}^{\rm miss}$. Such signatures can arise from gluino pair production $\tilde{g} \to q \bar{q} \neu{1} / \cha{1}$, squark pair production $\tilde{q} \to q \neu{1} / \cha{1}$, and direct electroweak production of $\tilde{\chi}^\pm \tilde{\chi}^0$. Signal events are required to contain at least one isolated photon with $p_{\rm T} > 35\,{\rm GeV}$ and at least one isolated electron or muon with $p_{\rm T} > 25\,{\rm GeV}$. To suppress the effects of electron misidentification and final-state radiation, photon candidates are required to meet the following criteria: they must be separated from any reconstructed electron by $\Delta R > 0.3$, be separated from the highest $p_{\rm T}$ lepton by $\Delta R > 0.8$, and the invariant mass of the $e\gamma$ system must fall outside the $Z$ boson mass window. The SR is defined by requiring $E_{\rm T}^{\rm miss} > 120\,{\rm GeV}$ and a lepton transverse mass $m_{\rm T} > 100\,{\rm GeV}$. This region is further subdivided into 36 search bins (18 for the $\gamma e$ channel, and 18 for the $\gamma \mu $ channel) based on $E_{\rm T}^{\rm miss}$, the photon transverse momentum $p_{\rm T}^{\gamma}$, and $H_{\rm T}$, where $H_{\rm T}$ is the scalar sum of the transverse momenta of all jets separated from both the photon and the lepton by $\Delta R > 0.4$. We implement all 36 SRs in our analysis. 

    \item \textbf{CMS search for top squark production in opposite-charge dilepton final states}~\cite{CMS:2017jrd}. This search focuses on top squark pair production in events containing two OS leptons, $b$-tagged jets, and $E_{\rm T}^{\rm miss}$. The search is interpreted within a typical simplified model describing the top squark pair production with decay mode $\tilde{t} \to t \neu{1} \to bW\neu{1}$ or $\tilde{t} \to b \cha{1} \to bW\neu{1}$, where both $W$ bosons are assumed to decay leptonically. The preselection requires exactly two OS leptons with an invariant mass greater than $20\,{\rm GeV}$ and outside the $Z$ boson mass window, at least two signal jets with at least one being $b$-tagged, $E_{\rm T}^{\rm miss} > 80\,{\rm GeV}$, $E_{\rm T}^{\rm miss} / \sqrt{H_{\rm T}} > 5$, and the azimuthal angular separation between each of the two leading jets and the missing transverse momentum to be $\cos \Delta\phi(p_{\mathrm{T}}^{\mathrm{miss}}, j_{1}) < 0.8$ and  $\cos \Delta\phi(p_{\mathrm{T}}^{\mathrm{miss}}, j_{2}) < 0.96$ respectively. The SRs are divided into different-flavour (\textsf{DF-SR}) and same-flavour (\textsf{SF-SR}) categories. They are further distinguished based on $m_{\rm T2}(\ell\ell)$, $m_{\rm T2}(b\ell\,b\ell)$, and $E_{\rm T}^{\rm miss}$. We implement all SRs into the \colliderbit framework. 

    \item \textbf{CMS search for supersymmetry in events with photons and missing transverse momentum}~\cite{CMS:2019vzo}. This search is optimised for SUSY signals in which the NLSP neutralino decays into a nearly massless gravitino and a photon $\neu{1} \to \gamma \tilde{G}$, resulting in characteristic events with large $E_{\rm T}^{\rm miss}$ and two photons. Two photon candidates are required to satisfy $p_{\rm T} > 40\,{\rm GeV}$, the invariant mass $m_{\gamma \gamma}$ to be larger than $110\,{\rm GeV}$. Six SRs are defined based on $E_{\rm T}^{\rm miss}$, with values greater than $100\,{\rm GeV}$.
 
    \item \textbf{CMS search for supersymmetry in events with jets and two same-sign or at least three leptons}~\cite{CMS:2020cpy}. This search targets SUSY processes with decay chains containing multiple $W$ or $Z$ bosons, which can potentially produce at least one pair of same-sign $W$ bosons. Typical SUSY processes include the production of gluino or squark pairs, as well as top and bottom squark pair production. The event selection requires at least two jets and at least two leptons, among which is an SS pair. Exclusive categories are defined in this search based on the $p_{\rm T}$ of the SS leptons. The \textsf{HH} (High-High) category requires both leptons to have $p_{\rm T} > 25\,{\rm GeV}$, while the \textsf{LL} (Low-Low) category requires both leptons to have $p_{\rm T} < 25\,{\rm GeV}$. The \textsf{LM} (Low $E_{\rm T}^{\rm miss}$) category is defined for events where both leptons have $p_{\rm T} > 25\,{\rm GeV}$ and $E_{\rm T}^{\rm miss} < 50\,{\rm GeV}$. The \textsf{ML} (Multilepton) category includes events with at least three leptons, where at least two form an OSSF pair. A total of 168 finely SR bins are defined. For this study, we implemented 17 inclusive SRs (labelled \textsf{ISR}) designed to facilitate reinterpretation. These are based on events with large $H_{\rm T}$, high $E_{\rm T}^{\rm miss}$, the minimum transverse mass of the same-sign leptons ($m_{\rm T}^{\rm min}$), the number of $b$-jets ($N_{b}$), and/or the total number of jets ($N_{\rm jet}$).

    \item \textbf{CMS search for chargino-neutralino production in events with three or four leptons}~\cite{cms:2021cox-fix}. This search targets the direct production of charginos and neutralinos $\cha{1} \neu{2}$ in final states with multiple leptons $\ell$ and up to two hadronically decayed $\tau$ candidates $\tau_{\rm h}$. Various simplified models are considered: $\cha{1} \neu{2}$ decay via intermediate sleptons or sneutrinos; $\cha{1} \neu{2}$ decay via Higgs, $W$ or $Z$ bosons; and a GGM SUSY breaking model of $\neu{1} \neu{1}$ decay via a Higgs or $Z$ boson $\neu{1} \to h/Z \tilde{G}$. A total of 12 categories are used to classify events based on lepton flavours and charges, allowing for a focused analysis of various signal hypotheses. The \textsf{SRSS} category requires two SS leptons and is further divided into 20 bins based on kinematic variables: $m_{\rm T2}(\ell)$, the dilepton transverse momentum $p_{\rm T}^{\ell\ell}$, and $E_{\rm T}^{\rm miss}$. Six three-lepton search categories (labelled \textsf{SRA} through \textsf{SRF}) require exactly three leptons, with or without one OSSF pair and with or without $\tau_{\rm h}$ candidates. They are further divided into 113 SR bins based on the kinematic variables: the transverse mass of the third non-paired lepton $m_{\rm T}$, $E_{\rm T}^{\rm miss}$, the scalar sum of the $p_{\rm T}$ of all jets $H_{\rm T}$, the OSSF pair invariant mass $m_{\ell\ell}$, the minimum $\Delta R$ between any two leptons $\min (\Delta R (\ell, \ell))$, the $m_{\rm T2}$ variable, and the transverse mass of the combined dilepton system and $m_{\rm T}^{2\ell}$. Five four-lepton search categories (labelled \textsf{SRG} through \textsf{SRK}) are defined based on the number of OSSF pairs and the number of $\tau_{\rm h}$ candidates. The leptons in signal events are then paired into $Z$ boson candidates based on the combination with an invariant mass closest to $m_{Z}$. \textsf{SRG} consists of events with two $Z$ candidates, and 5 SRs are defined based on the $m_{\rm T2}$ variable computed with both $Z$ boson candidates $m_{\rm T2}(ZZ)$, and the mass of the second $Z$ candidate $m_{Z2}$. The remaining events fall into categories \textsf{SRH} to \textsf{SRJ}, binned based on the OSSF pair of mass closest to the $Z$ boson mass (with invariant mass $m_{Z1}$) and the $\Delta R$ between two remaining leptons $\Delta R^{\rm H}$. We implemented all 12 search categories, comprising a total of 150 SR bins, into \colliderbit.

\end{enumerate}


\end{appendices}


\bibliography{refs_1, refs_2, refs_contur_analyses}
\end{document}